\documentstyle[12pt,epsfig,subfigure,amsfonts]{article}
\newcommand{\be}{\begin{equation}}
\newcommand{\ee}{\end{equation}}
\newcommand{\beq}{\begin{equation}}
\newcommand{\eeq}{\end{equation}}
\newcommand{\ber}{\begin{equation}}
\newcommand{\eer}{\end{equation}}
\newcommand{\ba}{\begin{eqnarray}}
\newcommand{\ea}{\end{eqnarray}}
\newcommand{\bea}{\begin{eqnarray}}
\newcommand{\eea}{\end{eqnarray}}
\newcommand{\nn}{\nonumber}
\newcommand{\e}{\epsilon}
\newcommand{\w}{\omega}

\renewcommand{\arraystretch}{1.4}
\begin{document}
\baselineskip=15.5pt \pagestyle{plain} \setcounter{page}{1}
%
\begin{titlepage}

\vskip 0.8cm

\begin{center}

{\Large \bf Deep inelastic scattering cross sections from the\\
gauge/string duality}

\vskip 1.cm

{\large {{\bf Ezequiel Koile}{\footnote{\tt
koile@fisica.unlp.edu.ar}}, {\bf Nicolas Kovensky}{\footnote{\tt
nico.koven@fisica.unlp.edu.ar}}, {\bf and Martin
Schvellinger}{\footnote{\tt martin@fisica.unlp.edu.ar}}}}

\vskip 1.cm

{\it IFLP-CCT-La Plata, CONICET and Departamento  de F\'{\i}sica,
Universidad Nacional de La Plata.  Calle 49 y 115, C.C. 67, (1900)
La Plata,  Buenos Aires, Argentina.}

\vspace{1.cm}

{\bf Abstract}

\vspace{1.cm}

\end{center}

Differential cross sections of deep inelastic scattering of charged
leptons from hadrons are investigated by using the gauge/string
duality. We consider vector mesons derived from different
holographic dual models obtaining a general expression. We focus on
the strongly coupled regime of dual gauge theories for different
values of the Bjorken parameter. We find new predictions which are
particularly interesting for differential scattering cross sections
of polarized leptons scattered off polarized vector mesons. We also
carry out a detailed comparison of the moments of the structure
functions with lattice QCD results.

\noindent

\end{titlepage}

\tableofcontents
\newpage

\section{Introduction}

In a recent series of papers
\cite{Koile:2011aa,Koile:2013hba,Koile:2014vca} we have studied deep
inelastic scattering (DIS) of charged leptons from scalar mesons and
from polarized vector mesons at strong coupling by considering
different holographic dual models based on flavor Dp-branes in the
probe approximation. In those papers we have studied the planar
limit of dual gauge theories corresponding to D3D7-,
D4D6$\mathrm{\overline{D6}}$- and D4D8$\mathrm{\overline{D8}}$-brane
models (\cite{Kruczenski:2003be}, \cite{Kruczenski:2003uq} and
\cite{Sakai:2004cn}, respectively)\footnote{The calculations carried
out in \cite{Koile:2011aa,Koile:2013hba,Koile:2014vca} are inspired
in the methods developed in reference \cite{Polchinski:2002jw},
where DIS of charged leptons from glueballs in the planar limit of
$N=1^*$ SYM theory has been considered \cite{Polchinski:2000uf}.}.
In \cite{Koile:2011aa,Koile:2013hba,Koile:2014vca} we have
investigated the hadronic tensor and, for polarized vector mesons,
we have obtained the corresponding eight structure functions. We
have derived general relations, {\it i.e.} model-independent
relations, of the Callan-Gross type between different structure
functions. This is very interesting because from that one can infer
a sort of universal behavior which would also be expected for QCD in
the planar limit. This universal property holds because, for all the
probe flavor Dp-brane models we have considered, mesons are
described as fluctuations of flavor Dp-branes in terms of the
corresponding Dirac-Born-Infeld action. In particular, in
\cite{Koile:2011aa} we have studied DIS from dynamical holographic
scalar and polarized vector mesons with one flavor, by considering
the flavored Dp-brane models of references
\cite{Kruczenski:2003be,Kruczenski:2003uq,Sakai:2004cn}. Moreover,
in \cite{Koile:2013hba} we have carried out a non-trivial extension
of these results for multi-flavored mesons in the planar limit and
with the condition $N_f \ll N$, where $N$ denotes the number of
colors and $N_f$ the number of flavors. In that paper we also have
worked out the corresponding next-to-leading order Lagrangians in
the $1/N$ and $N_f/N$ expansions. In
\cite{Koile:2011aa,Koile:2013hba} we have focused on the Bjorken
parameter regime where $1/\sqrt{\lambda}\ll x<1$, and therefore the
calculations have been done by using the gauge/string duality in
terms of the type IIA and type IIB supergravity dual descriptions of
the gauge theories in the large $N$ limit and at strong 't Hooft
coupling $\lambda$.

In a more recent work \cite{Koile:2014vca} we have studied DIS of
charged leptons from hadrons in the $e^{-\sqrt{\lambda}} \ll x \ll
1/\sqrt{\lambda}$ and $x \sim e^{-\sqrt{\lambda}}$ regimes.
In \cite{Koile:2014vca} we have focused on single-flavored scalar
and vector mesons in the planar limit. This has been investigated in
terms of different holographic dual models with flavor Dp-branes in
type IIA and type IIB superstring theories
\cite{Kruczenski:2003be,Kruczenski:2003uq,Sakai:2004cn}. We have
calculated the hadronic tensor and the structure functions for
scalar and polarized vector mesons. In particular, for polarized
vector mesons we have obtained the eight structure functions at
small values of the Bjorken parameter.

In the present work we further extend our previous results by
studying DIS cross sections of charged leptons from polarized vector
mesons and carrying out a qualitative comparison between moments of
structure functions calculated by using gauge/string duality methods
and available lattice QCD data. We emphasize that the character of
the comparison which we perform is {\it qualitative} since these
structure functions have been obtained in the context of the
large-$N$ and large-$\lambda$ limits of confining gauge theories
derived from their string theory dual models, which in fact do not
lead to real QCD. At this point we can briefly describe some
important differences between QCD and the holographic dual models we
consider. We can start from the D3D7-brane model, which is the
holographic dual description corresponding to the planar limit of
the strongly coupled $SU(N)$ ${\cal {N}}=2$ supersymmetric
Yang-Mills theory with fundamental quarks \cite{Kruczenski:2003be}.
Obviously, this is not the large $N$ limit of QCD but it is related
to it. On the other hand, we study the
D4D8$\mathrm{\overline{D8}}$-brane model developed in
\cite{Sakai:2004cn} which at low energy is expected to be within the
same universality class as the planar limit of QCD. At high energy
this model behaves in a different way compared with QCD, since in
that regime the Kaluza-Klein modes generated by the compatification
of the $x^4$-direction of the $N$ D4-branes along the $S^1$ become
relevant. Therefore these additional degrees of freedom, which are
obviously absent in QCD, play a role in the dynamics of this dual
gauge theory at high energy. In addition, this model
\cite{Sakai:2004cn} enjoys an $SO(5)$ global symmetry, which is not
present in QCD. Thus, one can say that these Dp-brane models do not
lead to holographic dual gauge theories with exactly the same
properties as QCD in the planar limit. However, these holographic
dual models represent infrared confining gauge theories and share
dynamical properties with QCD at low energy. For instance, the
Sakai-Sugimoto model realizes spontaneous chiral symmetry breaking.
Also, the chiral Lagrangian has been derived from this model
\cite{Sakai:2004cn}.

We also ought to mention that the comparison we carry out between
our results and phenomenology unfortunately has the disadvantage
that our calculations and the existing phenomenology and lattice QCD
calculations belong to different parametric regimes. On the one
hand, in our previous papers
\cite{Koile:2011aa,Koile:2013hba,Koile:2014vca} we have done
calculations within the Regge region. The asymptotic Regge region is
relevant since it dominates total cross sections and differential
cross sections when considering small angle. However, in the
asymptotic Regge regime it is not possible to perform quantum field
theory (QFT) calculations, neither one can use lattice QCD methods
in order to obtain QCD amplitudes \cite{Brower:2006ea}. From this
discussion one infers that it is difficult to make any quantitative
comparison of these models and $SU(3)$ QCD data. In this way our aim
is to compare qualitatively certain aspects of the phenomenology
derived from these dual models with QCD. At the end, by taking into
account all these caveats, we would like to see how our results
compare with lattice QCD data. Indeed, this kind of comparison has
proved to be qualitatively (and even in certain cases
quantitatively) interesting in several situations as we shall
explain in the last section of this paper. In fact, from a
theoretical perspective it has given some important insights about
the behavior of gauge theories in general, in the strong coupling
regime where perturbative QFT methods cannot be used. This assertion
is indeed very important because given a certain solution of
supergravity and/or superstring theory, in principle, the
gauge/string duality can be used to {\it define} a strongly coupled
holographic dual $SU(N)$ gauge theory in the large $N$ limit.
Suppose in addition that such a QFT has composite states. A natural
way to study these states is in terms of the DIS calculations as
described in
\cite{Polchinski:2002jw,Brower:2006ea,Koile:2011aa,Koile:2013hba,Koile:2014vca}.
Therefore, the interest of these methods goes beyond the large $N$
limit of QCD itself, {\it i.e.} it is useful for any other confining
gauge theory with a string theory holographic dual model.

This paper is organized as follows. In Section 2 we introduce and
derive a number of relevant expressions necessary to construct the
DIS cross sections of charged leptons from hadrons of spin $s=0,
\frac{1}{2}$ and $1$ in a generic form. This is a fully general
expression for the DIS differential cross section which holds for
any infrared confining gauge theory at any value of the rank of the
gauge group $N$ ({\it i.e.} beyond the planar limit), and at any
value of the coupling. In Section 3 we focus on the case of
polarized vector mesons and we present these derivations in detail.
Thus, in this section we derive the DIS cross section from leptons
off vector mesons which are unpolarized, longitudinally polarized,
transversally polarized and with partial polarization. We also study
the corresponding helicity amplitudes of the forward Compton
scattering directly related to the DIS process and discuss on
physical implications.

In Section 4 we derive general ({\it i.e.} model independent)
expressions for the DIS cross sections at large 't Hooft coupling,
in the three mentioned regimes of the Bjorken parameter. In general
terms the hadronic tensor can be derived from the operator product
expansion (OPE) of two electromagnetic currents inside the hadron.
At weak 't Hooft coupling the OPE is dominated by approximate
twist-two operators, and it corresponds to the scattering of a
lepton from a weakly interacting parton. However, in this section we
consider the strong coupling limit. In this case the OPE which leads
to the hadronic tensor is dominated by double-trace operators, and
it corresponds to the scattering of a lepton from the entire hadron.
Within this strongly coupled regime we consider the parametric range
$1/\sqrt{\lambda}\ll x<1$ where supergravity describes DIS.
Furthermore, we consider the case where $e^{-\sqrt{\lambda}} \ll x
\ll 1/\sqrt{\lambda}$, in which excited strings are produced. Also,
we study the case for $x \sim e^{-\sqrt{\lambda}}$, where the
produced excited strings have a size comparable to the scale of the
AdS space.

In Section 5 we discuss our results derived in the previous sections
such as differential cross sections in terms of the Bjorken
parameter and the fractional energy loss of the lepton, as well as
the structure functions and their moments, and carry out a
comparison with moments of structure functions for the pion and the
$\rho$ meson obtained from lattice QCD. It turns out that our
results of the first three moments of the $F_2$-structure function
of the pion and the $F_1$-structure function of the $\rho$ meson are
in very good quantitative agreement with the corresponding ones from
lattice QCD.

\section{Deep inelastic scattering cross section}

In this section we first introduce some definitions of kinematical
variables which will be useful for the rest of the work. Then, we
introduce the leptonic and hadronic tensors and give an expression
for the DIS differential cross section which holds for any value of
the coupling.

\begin{figure}
\centering
\includegraphics[scale=0.3]{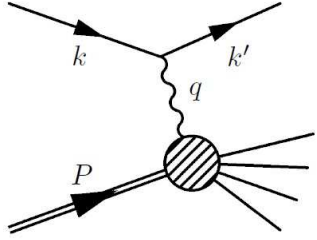}
\caption{\small  Schematic representation of DIS. A lepton with
momentum $k$ interacts with a hadron with momentum $P$ through the
exchange of a virtual photon with momentum $q$.} \label{diagramDIS}
\end{figure}

The DIS process is schematically represented in figure
\ref{diagramDIS}.

We consider the following definitions which are the same as in
reference \cite{Manohar:1992tz} \footnote{Notice however that we use
the mostly-plus metric $\eta^{\mu\nu}=$diag$(-,+,+,+)$.}
\begin{itemize}
\item $M$ is the mass of the hadron, being the on-shell condition $M^2=-P^2$.
\item $E$ is the energy of the incident lepton.
\item $k$ is the momentum of the incident lepton $k^\mu=(E, 0, 0, E)$.
\item $E'$ is the energy of the scattered lepton.
\item $k'$ is the momentum of the scattered lepton $k'^\mu=E'(1,
\sin\theta \, \cos\phi, \sin\theta \, \sin\phi, \cos\theta)$.
\item $P$ is the momentum of the hadron $P^\mu=(M, 0, 0, 0)$
in the rest frame\footnote{Fixed-target experiment.}.
\item $q$ is the momentum transfer or momentum of the virtual photon
$q=k-k'$.
\item $\nu$ is the energy loss of the lepton $\nu=E-E'=-\frac{P \cdot q}{M}$.
\item $y$ is the fractional energy loss of the lepton $y=\frac{\nu}{E}=\frac{P \cdot q}{P \cdot k}$.
\item $x$ is Bjorken parameter $x=-\frac{q^2}{2P \cdot q}$.
\item $t$ is a parameter defined as $t=\frac{P^2}{q^2}$.
\end{itemize}
In what follows we have assumed that the leptons are almost massless.

The DIS differential cross section can be written in terms of the
so-called leptonic and hadronic tensors $l^{\mu \nu}$ and $W^{\mu
\nu}$, respectively. Thus, we have
\ba
\frac{d^2\sigma}{dx \ dy \ d\phi}&=&\frac{e^4}{16 \pi^2 q^4} \ y \
l^{\mu\nu} \ W_{\mu\nu} \, ,
\ea
where $e$ denotes the electron charge. The leptonic tensor depends
on the incident lepton beam and it can be easily derived from
perturbative QED, leading to
\ba
l^{\mu\nu}_{\frac{1}{2}}&=&[2(k^\mu k'^\nu+k^\nu k'^\mu)-2\eta^{\mu\nu}k\cdot k']
+[-2i\epsilon^{\mu\nu\alpha\beta}q_\alpha s_{l\beta}] \quad,\quad m_l^2\sim 0\\
&=& l^{\mu\nu}_{sym}+l^{\mu\nu}_{ant} \, , \nn
\ea
where
$l^{\mu\nu}_{sym}$ and $l^{\mu\nu}_{ant}$ denote the symmetric and
antisymmetric parts, respectively. $s_{l}^{\mu}$ is the lepton spin
four-vector defined by boosting $(0,\vec{s}_{l})$ from the rest
frame, or equivalently by using the spinor states $u$ and
$\overline{u}$ which leads to
\ba
s_{l}^{\mu} = \overline{u}(k,s_l)\gamma^{\mu}\gamma_5 u(k,s_l) \, .
\ea

On the other hand, the hadronic tensor describes the internal
structure of the hadron. It cannot be calculated by using
perturbative QCD because there are soft processes involved. In this
paper we focus on spin-one hadrons since they have a very rich
structure given in terms of eight structure functions which follow
from the most general Lorentz-tensor decomposition of the hadronic
tensor. We also write the results of the structure functions for the
case of scalar mesons since we shall discuss them in section 5.

In general terms, the hadronic tensors for a spin $s=1, \frac{1}{2}$
and $0$ can be written as follows
\ba
W_{\mu\nu}^{1}&=&W_{\mu\nu}^{SF} + W_{\mu\nu}^{Sb}+W_{\mu\nu}^{Ag} \, , \\
W_{\mu\nu}^{\frac{1}{2}}&=&W_{\mu\nu}^{SF} + W_{\mu\nu}^{Ag} \, , \\
W_{\mu\nu}^{0}&=&W_{\mu\nu}^{SF} \, ,
\ea
where
\ba
W_{\mu\nu}^{SF}&=&F_{1}\eta_{\mu\nu}-\frac{F_{2}}{P \cdot q}P_{\mu}P_{\nu} \, , \\
W_{\mu\nu}^{Sb}&=&b_{1}r_{\mu\nu}-\frac{b_{2}}{6}(s_{\mu\nu}+t_{\mu\nu}+u_{\mu\nu})
- \frac{b_{3}}{2}(s_{\mu\nu}-u_{\mu\nu})
- \frac{b_{4}}{2}(s_{\mu\nu}-t_{\mu\nu}) \, , \\
W_{\mu\nu}^{Ag}&=&-\frac{ig_{1}}{P \cdot
q}\epsilon_{\mu\nu\lambda\sigma}q^{\lambda}s^{\sigma}
-\frac{ig_{2}}{(P \cdot
q)^{2}}\epsilon_{\mu\nu\lambda\sigma}q^{\lambda} (P \cdot q
\:s^{\sigma}-s \cdot q \:P^{\sigma}) \, .
\ea
The superscripts refer to the symmetry properties under the exchange
of Lorentz indices ($S$ and $A$) and to the usual name given to the
structure functions involved in each term ($F_i$'s, $b_i$'s and
$g_i$'s). In addition, we have used the following definitions
\cite{Hoodbhoy:1988am}
\begin{eqnarray}
\label{DIS17} && r_{\mu\nu}\equiv\frac{1}{(P \cdot q)^{2}}\bigg(q
\cdot \zeta^{*}\:q \cdot \zeta-\frac{1}{3}(P \cdot q)^{2}\kappa
\bigg)\eta_{\mu\nu} \, , \\
\label{DIS18} && s_{\mu\nu}\equiv\frac{2}{(P \cdot q)^{3}}\bigg(q
\cdot \zeta^{*}\:q \cdot \zeta-\frac{1}{3}(P \cdot q)^{2}\kappa
\bigg)P_{\mu}P_{\nu} \, , \\
\label{DIS19} && t_{\mu\nu}\equiv\frac{1}{2(P \cdot q)^{2}}
\bigg[ -\frac{4}{3}(P \cdot q) P_{\mu}P_{\nu} + \bigg( (q \cdot \zeta^{*})
(P_{\mu}\zeta_{\nu}+P_{\nu}\zeta_{\mu})+c.c.\bigg) \bigg] \, ,  \\
\label{DIS20} && u_{\mu\nu}\equiv\frac{1}{P \cdot
q}\bigg[(\zeta^{*}_{\mu}\zeta_{\nu}+\zeta^{*}_{\nu}\zeta_{\mu})
-\frac{2}{3}M^{2}\eta_{\mu\nu}-\frac{2}{3}P_{\mu}P_{\nu}\bigg] \, , \\
&& \nonumber \\
\label{DIS21} && s^{\sigma}\equiv\frac{-i}{M^{2}}\epsilon^{\sigma\alpha\beta\rho}
\zeta^{*}_{\alpha}\zeta_{\beta}P_{\rho} \, , \\
&& \nonumber \\
\label{DIS22} && \kappa=1-4x^2t \, ,
\end{eqnarray}
where $\zeta^{\mu}$ is the polarization vector of the hadron. Then,
one can write a general expression for the differential cross
section of the DIS of a charged lepton from a hadron in terms of the
symmetric and antisymmetric parts of these tensors
\ba
\frac{d^2\sigma}{dx dy d\phi}&=&\frac{e^4}{16\pi^2 q^4} \ y \
[l^{\mu\nu}_{sym} \ W_{\mu\nu}^{SF} + l^{\mu\nu}_{ant} \
W_{\mu\nu}^{Ag} + l^{\mu\nu}_{sym} \ W_{\mu\nu}^{Sb}] \, ,
\label{WSFAGSB}
\ea
where $\phi$ is the azimuthal angle. The differential cross section
depends trivially on $\phi$ but non-trivially on $x$ and $y$, which
can be chosen as two independent parameters describing the DIS
process, together with the spin polarizations of the lepton and the
hadron. It is worth to note that in the case of spin-zero hadrons
only the first term in the square bracket of Eq.(\ref{WSFAGSB})
contributes, while for the spin-$1/2$ case the second term also
appears. The last term contributes for spin-one hadrons and contains
four additional $b_i$ structure functions.

\section{DIS differential cross sections and structure functions}

In this section we derive the DIS differential cross sections of
charged leptons scattered from hadrons with spin $s=0$,
$\frac{1}{2}$ and $1$. We consider first the general case, and then
we study different polarization states of the hadron beam.

\subsection{The general case}

Let us begin with the first term of Eq.(\ref{WSFAGSB}), which is
present in hadrons with spin $0$, $\frac{1}{2}$ and $1$. We obtain
the following term expressed in terms of the $F_1$ and $F_2$
structure functions
\ba
\frac{e^4}{16 \pi^2 q^4} \ y \ l^{\mu\nu}_{sym} W_{\mu\nu}^{SF}&=&
\frac{e^4 M E}{4\pi^2 q^4} \bigg\{ \ xy^2 \ F_1 +
\bigg[1-y+x^2y^2t\bigg] \ F_2\bigg\}
\nonumber\\
\label{circle1}&\simeq& \frac{e^4 M E}{4 \pi^2 q^4} \bigg\{ \ xy^2 \
F_1 + [1-y] \ F_2\bigg\}  \, ,
\ea
where in the last step we have approximated $|t| \simeq
0$.\footnote{This is the equation (5.2) of reference
\cite{Manohar:1992tz}. An extra factor $2\pi$ appears in the
denominator because we are calculating $\frac{d^2\sigma}{dx dy
d\phi}$ as in \cite{Manohar:1992tz}. In reference
\cite{Hoodbhoy:1988am}, on the other hand, it has been calculated
$\frac{d^2\sigma}{dxdy}=2\pi\frac{d^2\sigma}{dx dy
d\phi}$.}\label{MvsHJM} The second term, which appears for both
$s=\frac{1}{2}$ and $s=1$ spins is given by
\ba
\frac{e^4}{16 \pi^2 q^4} \ y \ l^{\mu\nu}_{ant}W_{\mu\nu}^{Ag} &=&
\label{circle3}\frac{e^4ME}{4\pi^2 q^4}
\bigg\{ y^2 g_1 \bigg[2x\frac{(s_h\cdot s_l)}{(p\cdot q)}+\frac{(q\cdot s_h)
(q\cdot s_l)}{(p\cdot q)^2}\bigg] \nn \\
&&+ 2 x y^2 g_2 \bigg[ \frac{(s_h\cdot s_l)}{(p\cdot
q)}-\frac{(q\cdot s_h)(p\cdot s_l)}{(p\cdot q)^2} \bigg]\bigg\} \, .
\ea
This term depends on the structure functions $g_1$ and $g_2$. Notice
that in the expression for the first term we contract only symmetric
factors, while in the expression for the second one only
antisymmetric tensors are contracted. In addition, in the last
calculation we have used the identity
$\epsilon^{\mu\nu\alpha\beta}\epsilon_{\mu\nu\lambda\sigma}=
-2\big(\eta^\alpha_\lambda \eta_\sigma^\beta-\eta^\alpha_\sigma
\eta_\lambda^\beta \big)$.

Let us now calculate the most general form of the third term which
only appears for spin-one hadrons, and can be written in terms of
the $b_i$ structure functions. Firstly, we need to calculate the
contractions
\be
l^{\mu\nu}_{sym} r_{\mu\nu} \quad;\quad l^{\mu\nu}_{sym} s_{\mu\nu}
\quad;\quad l^{\mu\nu}_{sym} t_{\mu\nu} \quad;\quad l^{\mu\nu}_{sym}
u_{\mu\nu} \, .
\ee
In order to simplify the notation we define a
pseudo-scalar product $\langle \cdot \cdot \rangle$ given by
\be
\langle k_1k_2\rangle \equiv
\frac{3}{2}\frac{(k_1\cdot\zeta^*)(k_2\cdot\zeta)+c.c.}{(P\cdot
q)^2} \, ,
\ee
where $c.c.$ indicates complex conjugate. This pseudo-scalar product
is linear on both sides and it is also commutative. In addition,
since we take the $|t|\ll 1$ limit we set $\kappa=1$. Then, we
obtain
\ba
l^{\mu\nu}_{sym} r_{\mu\nu}&=&[2(k^\mu k'^\nu+k^\nu k'^\mu)
-2\eta^{\mu\nu}k\cdot k']\frac{1}{(P \cdot q)^{2}}\bigg(q
\cdot \zeta^{*}\:q \cdot \zeta-\frac{1}{3}(P \cdot q)^{2}\bigg)\eta_{\mu\nu}\nn\\
&=& -\frac{4}{3}(k\cdot k') \left[ \langle qq \rangle-1\right]\nn\\
&=& \frac{4}{3}M E x y \left[ \langle qq \rangle-1\right] \, , \\
%
%
%
%
l^{\mu\nu}_{sym} s_{\mu\nu}&=&[2(k^\mu k'^\nu+k^\nu k'^\mu)
-2\eta^{\mu\nu}k\cdot k']\frac{2}{(P \cdot q)^{3}}\bigg(q
\cdot \zeta^{*}\:q \cdot \zeta-\frac{1}{3}(P \cdot q)^{2}\bigg)P_{\mu}P_{\nu}\nn\\
&=& \frac{4}{3}\left[\frac{2(P\cdot k)(P\cdot k')}{(P\cdot q)}
-\frac{P^2(k\cdot k')}{(P\cdot q)}\right]\left[\langle qq \rangle-1\right]\nn\\
&=& -\frac{8}{3}ME \left[ \langle qq
\rangle-1\right]\left[\frac{1}{y}-1+x^2yt\right] \, , \\
%
%
%
%
l^{\mu\nu}_{sym} t_{\mu\nu}&=&[2(k^\mu k'^\nu+k^\nu k'^\mu)
-2\eta^{\mu\nu}k\cdot k']\frac{1}{2(P \cdot q)^{2}} \times \nn
\\
&& \bigg[ -\frac{4}{3}(P \cdot q) P_{\mu}P_{\nu} + \bigg( (q \cdot
\zeta^{*})
(P_{\mu}\zeta_{\nu}+P_{\nu}\zeta_{\mu})+c.c.\bigg) \bigg]\nn \\
&=& \frac{-1}{\langle qq \rangle-1}l^{\mu\nu}_{sym} s_{\mu\nu} +
\frac{4}{3}\langle qk'\rangle (k\cdot P)+
\frac{4}{3}\langle qk\rangle (k'\cdot P)\nn\\
&=& \frac{4}{3}ME \left[2\left(\frac{1}{y}-1+x^2yt\right) -\langle
qk'\rangle - \langle qk\rangle (1-y)\right] \, ,
\ea
and
\ba
l^{\mu\nu}_{sym} u_{\mu\nu}&=&[2(k^\mu k'^\nu+k^\nu k'^\mu)
-2\eta^{\mu\nu}k\cdot k']\frac{1}{(P \cdot
q)}\bigg[(\zeta^{*}_{\mu}\zeta_{\nu}+\zeta^{*}_{\nu}\zeta_{\mu})
-\frac{2}{3}M^{2}\eta_{\mu\nu}-\frac{2}{3}P_{\mu}P_{\nu}\bigg]\nn\\
&=& \left[\frac{4}{3}\langle kk'\rangle(P\cdot q)-2\frac{M^2(k\cdot
k')}{(P\cdot q)}\right] -\frac{2M^2}{(P\cdot q)[\langle qq
\rangle-1]}l^{\mu\nu}_{sym} r_{\mu\nu}
-\frac{1}{\langle qq \rangle-1}l^{\mu\nu}_{sym} s_{\mu\nu}\nn\\
&=& \frac{8}{3} M E \left[-y\langle
kk'\rangle+\left(\frac{1}{y}-1\right)+2x^2yt\right] \, .
\ea
Finally, by writing the cross section contribution from the $b_i$
structure functions as $\alpha_1 b_1+ \alpha_2 b_2+\alpha_3
b_3+\alpha_4 b_4$ one finds the following expressions for the
$\alpha_i$ coefficients
\ba
\alpha_1&=&\frac{e^4}{16 \pi^2 q^2} \ y \
l^{\mu\nu}_{sym} r_{\mu\nu} = \frac{MEe^4}{2^2 3\pi^2 q^4} \ xy^2
\left[ \langle qq \rangle-1\right],\\
\alpha_2&=&-\frac{e^4}{16 \pi^2 q^4} \ \frac{y}{6} \ l^{\mu\nu}_{sym}
\left(s_{\mu\nu}+t_{\mu\nu}+u_{\mu\nu}\right) \nn\\
&=& \frac{M E e^4}{2^3 3^2 \pi^2 q^4}  \times \nn \\
&& \left[ 2\left(\langle qq \rangle-3\right)(1-y)+y\langle
qk'\rangle + y(1-y)\langle qk\rangle +2y^2\langle
kk'\rangle-2\left(4-\langle qq\rangle\right) x^2y^2t\right], \nn \\
&& \\
\alpha_3 &=&-\frac{e^4}{16 \pi^2 q^4} \ \frac{y}{2} \
l^{\mu\nu}_{sym}\left(s_{\mu\nu}-u_{\mu\nu}\right) \nn \\
&=&\frac{M E e^4}{2^2 3 \pi^2 q^4}  \ \left[
\left(1-y\right) \langle qq \rangle -y^2 \langle kk'\rangle +x^2y^2t\left(\langle qq \rangle+1\right)\right], \\
\alpha_4&=&-\frac{e^4}{16 \pi^2 q^4} \ \frac{y}{2} \ l^{\mu\nu}_{sym}\left(s_{\mu\nu}-t_{\mu\nu}\right) \nn\\
&=&-\frac{M E e^4}{2^3 3 \pi^2 q^4} \  \left[
2\left(1-y\right)\langle qq \rangle-y\langle qk'\rangle - \langle
qk\rangle y(1-y)+ 2\langle qq \rangle x^2y^2t\right] \, .
\ea
By using the linearity of $\langle \cdot \cdot \rangle$ we obtain
\ba
&& \frac{e^4}{16\pi^2 q^4} \ y \ l^{\mu\nu}_{sym}W_{\mu\nu}^{Sb}=
\nn \\
&& \frac{M E e^4}{12 \pi^2 q^4} \ \left\{ b_1xy^2[\langle qq
\rangle-1] +\frac{1}{6}\left[ (2-3y)\langle qq
\rangle+y(2-3y)\langle qk
\rangle+2y^2\langle kk \rangle-6(1-y)\right]b_2 \right. \nn\\
&&\left. +\left[ \left(1-y\right) \langle qq \rangle -y^2 \langle
kk\rangle+y^2 \langle qk\rangle \right]b_3+ \frac{1}{2}\left[
\left(2-y\right)\langle qq \rangle+y(-2+y)\langle qk\rangle
\right]b_4 +{\mathcal{O}}(t)  \right\} \, . \nn \\
&&
\ea
Now, in the $\sqrt{-t}\ll 1$ limit we can relate the values of
$\langle qq \rangle$, $\langle qk \rangle$, and $\langle kk
\rangle$, since in this limit $\frac{q^\mu}{q^0}=\frac{k^\mu}{E}$,
where the Lorentz index $\mu$ runs from $0$ to $3$. Then, in the
hadron rest frame we obtain the relations
\ba
\frac{q^\mu q^\nu}{M^2(q^0)^2} &=& \frac{\frac{1}{2}(q^\mu
k^\nu+k^\mu q^\nu)}{M^2q^0E} = \frac{k^\mu k^\nu}{M^2E^2} \nn \, ,
\ea
\ba
2\frac{q^\mu q^\nu}{(P\cdot q)^2}\zeta_\mu\zeta^*_\nu &=&
\frac{(q^\mu k^\nu+k^\mu q^\nu)}{(P\cdot q)(P\cdot
k)}\zeta_\mu\zeta^*_\nu = 2\frac{k^\mu k^\nu}{(P\cdot
k)^2}\zeta_\mu\zeta^*_\nu\nn \, ,
\ea
\ba
\frac{3}{2}\frac{(q\cdot\zeta^*)(q\cdot\zeta)+c.c.}{(P\cdot q)^2}
&=& \frac{(P\cdot q)}{(P\cdot k)}
\frac{3}{2}\frac{(q\cdot\zeta^*)(k\cdot\zeta)+c.c.}{(P\cdot q)^2} =
\frac{(P\cdot q)^2}{(P\cdot k)^2}
\frac{3}{2}\frac{(k\cdot\zeta^*)(k\cdot\zeta)+c.c.}{(P\cdot q)^2}\nn
\, ,
\ea
and
%
%
%
%
%
%
\ba
\langle qq\rangle &=&  y\langle qk\rangle \ = \ y^2\langle kk\rangle
\, \, . \nn
\ea
Thus, for $\sqrt{-t}\ll 1$ the part of the cross section associated
with the $b_i$ structure functions of the hadronic tensor reduces to
a simpler form
\ba
\frac{e^4}{16\pi^2 q^4} \ y \
l^{\mu\nu}_{sym}W_{\mu\nu}^{Sb}&=&\frac{MEe^4}{2^2 3 \pi^2 q^4}
(\langle qq \rangle-1) \ \left[ b_1xy^2+b_2(1-y)
\right]+{\mathcal{O}}\left(\sqrt{-t}\right) \, .
\label{finalform2}
\ea
Notice that in this limit the contributions to the cross section
coming from $b_3$ and $b_4$ are sub-leading in $t$ even if $b_3$ and
$b_4$ are not necessarily sub-leading themselves. Thus, the full DIS
differential cross section from spin-one hadrons to this order
becomes
\ba
\frac{d\sigma_{spin 1}}{dx \, dy \, d\phi}&=&\frac{MEe^4}{4  \pi^2 q^4}
\left\{ \left[ xy^2F_1+(1-y)F_2 \right] + \frac{1}{3}(\langle qq \rangle-1) \
\left[ xy^2b_1+(1-y)b_2 \right]  \right. \nn\\
&& \left. + y^2g_1 \bigg[2x\frac{(s_h\cdot s_l)}{(p\cdot q)}
+\frac{(q\cdot s_h)(q\cdot s_l)}{(p\cdot q)^2}\bigg]
+ 2xy^2g_2\bigg[ \frac{(s_h\cdot s_l)}{(p\cdot q)}
-\frac{(q\cdot s_h)(p\cdot s_l)}{(p\cdot q)^2} \bigg]\right\} \nn\\
&=&\frac{MEe^4}{4  \pi^2 q^4}
\left\{ xy^2 \left[F_1+\frac{1}{3}(\langle qq \rangle-1)b_1\right]
+(1-y)\left[F_2+\frac{1}{3}(\langle qq \rangle-1)b_2\right]  \right. \nn \\
&& \left. + y^2g_1 \bigg[2x\frac{(s_h\cdot s_l)}{(p\cdot q)}
+\frac{(q\cdot s_h)(q\cdot s_l)}{(p\cdot q)^2}\bigg] +
2xy^2g_2\bigg[ \frac{(s_h\cdot s_l)}{(p\cdot q)} -\frac{(q\cdot
s_h)(p\cdot s_l)}{(p\cdot q)^2} \bigg]\right\}. \nn \\
&&
\ea
%

\subsection{Hadron polarizations}

Now, we calculate the dimensionless factor
\be
\label{qq}
\langle qq\rangle =
\frac{3}{2}\frac{(q\cdot\zeta^*)(q\cdot\zeta)+c.c.}{(P\cdot
q)^2}=3\frac{q^\mu q^\nu}{(P\cdot q)^2} \ \zeta_\mu\zeta_\nu^{*} \,
,
\ee
for different polarizations of the hadron in order to obtain the DIS
differential cross section $\frac{d\sigma}{dx\, dy \, d\phi}$ for
the particular cases which may be phenomenologically relevant,
namely: when the hadron beam is unpolarized, longitudinally
polarized, transversally polarized and also we study partial
polarizations of the hadron beam. In all cases the longitudinal or
$\hat{z}$-axis is defined by the spatial components of $k$, in other
words, the direction of the incident lepton beam.

\subsubsection*{Unpolarized hadron beam}

In order to describe unpolarized hadron beams we have to average
over polarizations obtaining
\be
\label{average}
\overline{\zeta_\mu\zeta_\nu^*} = \frac{1}{3}(\eta_{\mu\nu}M^2+P_\mu
P_\nu) \, .
\ee
Then,
\ba
\langle qq\rangle_{unpol}=\langle \overline{qq}\rangle =
\frac{1}{(P\cdot q)^2}[M^2q^2+(P\cdot q)^2] =1-4x^2t=\kappa \simeq 1
\, ,
\ea
thus, we obtain a null contribution from the $b_i$ structure
functions as expected for an unpolarized target
\cite{Hoodbhoy:1988am}. In particular, for a longitudinally
polarized leptonic beam we have $s_l = H_l k$ with $H_l = \pm 1$ for
positive and negative helicity, respectively. Then, we obtain
\ba
\frac{d\sigma}{dx \, dy \, d\phi}&=&\frac{MEe^4}{4  \pi^2 q^4}
\left\{ xy^2 F_1 +(1-y)F_2\right\}. \nn
\ea
%

\subsubsection*{Longitudinal polarization}

Now, we consider the case when the hadron beam is polarized in the
direction of the incident beam and call it $\hat{z}$-axis. Thus, we
can take the polarization to be
\be
\zeta_L=\frac{M}{\sqrt{2}}(0,1,iH_h,0)  \quad \rightarrow
\quad s^\sigma=H_h M \hat{z} \, ,
\ee
where $H_h=\pm 1$ indicates that the hadron polarization is
parallel or anti-parallel to the beam direction, respectively. Then,
we obtain the following expressions
\ba
(q\cdot\zeta_L)&=&-\frac{ME'{\sin}\theta}{\sqrt{2}} \ e^{iH_h\phi} \, , \nn\\
(q\cdot\zeta_L^*)(q\cdot\zeta_L)&=&\frac{M^2E'^2{\sin}^2\theta}{2}
=\frac{q^2M^2}{2}[(1-y)+tx^2y^2] \, , \nn\\
\langle qq \rangle_{LP} &=& \frac{3}{2}\frac{1}{(P\cdot
q)^2}M^2E'^2{\sin}^2\theta = -6x^2t[(1-y)+tx^2y^2] \simeq 0 \, ,
\ea
In this case the factor $\langle qq \rangle -1$ only gives a minus
sign. In particular, for a longitudinally polarized lepton beam we obtain
\ba
\frac{d\sigma}{dx \, dy \, d\phi}&=&\frac{MEe^4}{4  \pi^2 q^4}
\left\{ xy^2 \left[F_1-\frac{1}{3}b_1\right] + (1-y)\left[F_2-\frac{1}{3}b_2\right]
-H_l H_h (2-y)yxg_1+{\mathcal{O}}(t)\right\}.\nn\\
\ea
%

\subsubsection*{Transversal polarization}

Since the system under consideration has azimuthal symmetry we can
choose any polarization for the hadron in the $(\hat{x},
\hat{y})$-plane. For transversal polarization we choose for example
the $\hat{x}$-axis and set
\be
\zeta_T=\frac{M}{\sqrt{2}}(0,0,1,iH_h)  \quad \rightarrow
\quad s^\sigma=H_h M\hat{x} \quad;\quad H_h=\pm 1 \, .
\ee
Then, we obtain
\ba
(q\cdot\zeta_T)&=&\frac{M^2}{\sqrt{2}}\Bigg[\sqrt{\frac{(1-y)}{-t}-x^2y^2}
\ \ {\sin}\phi+iH_h\Big(-\frac{1}{2xt}+xy\Big) \Bigg] \nn\\
&\simeq&\frac{M^2}{\sqrt{2}}\Bigg[\sqrt{\frac{(1-y)}{-t}}\ {\sin}\phi-\frac{iH_h}{2xt}\Bigg] \, , \nn\\
(q\cdot\zeta_T^*)(q\cdot\zeta_T)&=&\frac{M^4}{2t^2}\bigg[\frac{1}{4x^2}-t(1-y){\sin}^2\phi\bigg] \, , \nn \\
\langle qq \rangle_{TP} &=& \frac{3}{2}\frac{1}{(P\cdot q)^2}
\frac{M^4}{t^2}\bigg[\frac{1}{4x^2}-t(1-y){\sin}^2\phi\bigg]\nn\\
&=& 6x^2\bigg[\frac{1}{4x^2}-t(1-y){\sin}^2\phi\bigg] \simeq
\frac{3}{2} \, ,
\ea
and the dimensionless factor becomes $\langle qq \rangle - 1 \approx
\frac{1}{2}$. In particular, for a longitudinally polarized lepton beam we obtain
\ba
\frac{d\sigma}{dx \, dy \, d\phi}&=&\frac{MEe^4}{4  \pi^2 q^4}
\left\{ xy^2 \left[F_1+\frac{1}{6}b_1\right] +(1-y)\left[F_2+\frac{1}{6}b_2\right]\right.\nn\\
&&\left.+H_l H_h 2x^2\sqrt{-t}\sqrt{1-y}(yg_1+2g_2) \cos\phi \right\}.
\ea
%

\subsubsection*{Partial polarizations}

The $b_i$ structure function terms in Eq.(\ref{finalform2}) and in
particular the dimensionless factor $\langle qq \rangle - 1$ can be
written in a more general form
\ba
\label{densmatrix}
\frac{e^4}{16 \pi^2 q^4} \ y \ l^{\mu\nu}_{sym}
W_{\mu\nu}^{Sb}=\frac{M E e^4}{2\sqrt{3} \pi^2 q^4} \ {Tr}(\rho
\cdot \lambda_8) \ \left[ b_1 x y^2+b_2(1-y)
\right]+{\cal{O}}\left(\sqrt{-t}\right) \, .
\ea
where $\lambda_8=\frac{1}{\sqrt{3}}$ diag$(1,1,-2)$ and $\rho$ is
the spin-one density matrix that accounts for the possibility of
having a statistical production of the hadron beam. Notice that
\be
{Tr}(\rho \cdot \lambda_8)=\frac{1}{2\sqrt{3}}\left[\langle
S_x^2\rangle+\langle S_y^2\rangle -2\langle S_z^2\rangle+3\langle
S_z\rangle \right] \, .
\ee
We can study particular cases with partial longitudinal and
transversal polarizations. In the $\hat{z}$-axis basis we have
$\rho_{unpol} = (1/3)$ diag$(1,1,1)$ and
\begin{eqnarray}
\rho_{pLP} =\frac{1}{2}
\left( \begin{array}{ccc}
H_L^2 & 0 & 0 \\
0 & 2(1-H_L^2) & 0\\
0 & 0 & H_L^2
\end{array} \right)
 ; \,\,\,\,
\rho_{pTP} =\frac{1}{4}
\left( \begin{array}{ccc}
2-H_T^2 & 0 & -2+3H_T^2\\
  0 & 2H_T^2 & 0 \\
  -2+3H_T^2 & 0 & 2-H_T^2
\end{array} \right)
\end{eqnarray}
where $H_L$ and $H_T$ are the fraction of longitudinally and
transversally polarized hadrons\footnote{Again, we are writing
transversal polarization along the $\hat{x}$-axis. However, by
choosing the $\hat{y}$-axis instead it would lead to the same
results.}, respectively. Here, $H^2=1$ represents the totally
polarized case and $H^2 = \frac{2}{3}$ describes an unpolarized
hadron beam. In the former notation we identify
\ba
\langle qq \rangle_{pLP}&=&-3\left(\frac{2}{3}
-H_L^2\right)\langle qq \rangle_{LP}+3\left(1-H_L^2\right)\langle qq \rangle_{unpol} \, ,\\
\langle qq \rangle_{pTP}&=&-3\left(\frac{2}{3}-H_T^2\right)\langle
qq \rangle_{TP}+3\left(1-H_T^2\right)\langle qq \rangle_{unpol} \, ,
\ea
where the subindex $p$ stands for partial polarization. Then, the
results for this part of the DIS differential cross sections
corresponding to a partially polarized target are
\be
\frac{e^4}{16\pi^2 q^4} \ y \ l^{\mu\nu}_{sym}W_{\mu\nu}^{Sb} =
\frac{M E e^4}{2^2 \pi^2 q^4}\left(\frac{2}{3}-H_L^2\right)  \
\left[ b_1xy^2+b_2(1-y) \right]+{\mathcal{O}}\left(\sqrt{-t}\right),
\ee for partial longitudinal polarization, and
\be
\frac{e^4}{16\pi^2 q^4} \ y \ l^{\mu\nu}_{sym}W_{\mu\nu}^{Sb} =
\frac{M E e^4}{2^3 \pi^2 q^4}\left(H_T^2-\frac{2}{3}\right)  \
\left[ b_1xy^2+b_2(1-y) \right]+{\mathcal{O}}\left(\sqrt{-t}\right),
\ee
for partial transversal polarization.

These expressions lead to the final results of DIS differential
cross section for partially polarized hadrons as a generalization of
the previous cases. For a partially longitudinally polarized target
$H_L$ we obtain
\ba
\frac{d\sigma}{dx \, dy \, d\phi}&=&\frac{MEe^4}{4  \pi^2 q^4}
\left\{ xy^2 \left[F_1+\left(\frac{2}{3}-H_L^2\right)b_1\right] +
(1-y)\left[F_2+\left(\frac{2}{3}-H_L^2\right)b_2\right] \right.\nn\\
&&\left.-H_l H_h (2-y)yxg_1\right\},
\ea
while for a partially transversally polarized target $H_T$ we obtain
\ba
\frac{d\sigma}{dx \, dy \, d\phi}&=&\frac{MEe^4}{4  \pi^2 q^4}
\left\{ xy^2 \left[F_1-\left(\frac{1}{3}-\frac{1}{2}H_T^2\right)b_1\right] +
(1-y)\left[F_2-\left(\frac{1}{3}-\frac{1}{2}H_T^2\right)b_2\right]\right.\nn\\
&& \left.+H_l H_h 2x^2\sqrt{-t}\sqrt{1-y}(yg_1+2g_2) \cos\phi \right\}.
\ea
We have assumed longitudinally polarized leptonic beams $H_l$ in
both cases. Once again, by assuming that $F_2(x)=2 x F_1(x)$ and
$b_2(x)=2 x b_1(x)$ are satisfied, we recover the formulas shown in
\cite{Hoodbhoy:1988am}.

\subsection{Helicity amplitudes}

Deep inelastic scattering is related to the forward Compton
scattering. Let us briefly describe the forward Compton scattering
by using the virtual photon momentum $q$ and the hadron momentum $p$
in analogy with our DIS notation, and the variables $\lambda$ and
$\lambda'$ which denote the initial and final hadronic helicities.
In the forward Compton scattering both the initial and final states
of the system contain a photon and a hadron.
The related amplitude is a function of a tensor $T_{\mu \nu}$
defined in terms of the matrix elements of two electromagnetic
currents by the formula
\be
\left(T_{\mu \nu}\right)_{\lambda \lambda'} \equiv i \int d^4 x \,
e^{i q\cdot x} \langle p,\lambda'| T
\left(J_{\mu}(x)J_{\nu}(0)\right) | p,\lambda \rangle \, ,
\ee
which has a structure similar to the hadronic tensor $W^{\mu \nu}$
involved in DIS. On the other hand, the definition of $W_{\mu \nu}$
in terms of the currents is given by
\be
\left(W_{\mu \nu}\right)_{\lambda \lambda'} = \frac{1}{4\pi} \int
d^4 x \, e^{i q\cdot x} \langle p,\lambda'| \left[J_{\mu}(x),
J_{\nu}(0)\right] | p,\lambda \rangle .
\ee
Since both $T_{\mu \nu}$ and $W_{\mu \nu}$ have the same Lorentz
symmetry transformation properties, they have the same tensor
structure decomposition. Thus, the information they contain can be
encoded in their structure functions in the same way.

\begin{figure}[h]
\centering
\includegraphics[scale=0.7]{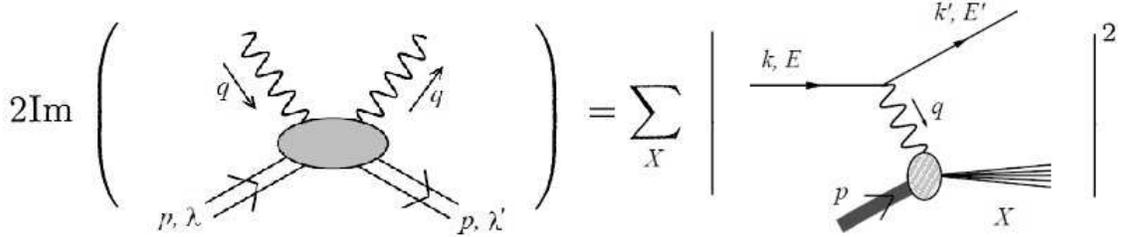}
\caption{\small  The relation between DIS and forward Compton
scattering processes as given by the optical theorem in terms of
Feynman diagrams.} \label{Optical}
\end{figure}

DIS and forward Compton scattering are related to each other by the
optical theorem in quantum field theory as it is schematically shown
in figure \ref{Optical}. It implies that twice the imaginary part of
the forward Compton scattering amplitude leads to the total DIS
amplitude. Therefore, the imaginary part of the $T_{\mu \nu}$
structure functions give the $W_{\mu \nu}$ structure functions times
an extra numerical factor. Actually, this relation between the two
processes is one of the basis of our construction to obtain the
structure functions. As we will see in the next section, it is
possible to describe a holographic dual version of the forward
Compton scattering from the point of view of supergravity and string
theory via the gauge/string duality.

The Compton amplitude can be analyzed for different helicities of
the incoming and out-going particles. The amplitudes in each case
are commonly denoted by $A_{h,H;h',H'}$, where the subindex $h$
corresponds to the helicity of the photon and $H$ labels the
helicity of the hadron. Primed variables indicate helicities of
final states, otherwise labels indicate helicities of initial
states. These helicities can take three different values in the
spin-one case, $-1$, $+1$ and $0$, and they must satisfy the
condition $h + H = h' + H'$. As shown in
\cite{Manohar:1992tz,Hoodbhoy:1988am} they can be calculated as
\be
A_{hH,h'H'} = \epsilon^{\mu \star}_{h}\epsilon_{h'}^{\nu}T_{\mu \nu}
(s_H) \, ,
\ee
where the $\e$'s are the photon polarization vectors and $s_H$ is
the hadron spin. For spin-one hadrons there are eight independent
helicity amplitudes. By considering a general kinematical regime,
and by using the notation from \cite{Hoodbhoy:1988am}, where $a_3
\equiv b_2/3 - b_3$ and $a_4 \equiv b_2/3 - b_4$, these helicity
amplitudes can be written in the following form
\ba
A_{++,++}&=&F_1-\frac{(1-4x^2t ) b_1}{3}-\frac{xt}{3}a_3-g_1-4x^2tg_2 \nn\\
&\simeq& F_1-\frac{b_1}{3}-g_1 \, , \label{A1} \\
A_{+0,+0}&=&F_1+\frac{2(1-4x^2t ) b_1}{3}+\frac{2xt}{3} \ a_3 \nn \\
&\simeq&F_1+\frac{2b_1}{3} \, , \\
A_{+0,0+}&=&\sqrt{-t} \ \big[ 2x(g_1+g_2)+\frac{a_3}{2}+\frac{(1-4x^2t)a_4}{4}\big]\nn \\
&\simeq&\sqrt{-t} \ \big[ 2x(g_1+g_2)+\frac{a_3}{2}+\frac{a_4}{4}\big] \, , \\
A_{+-,+-}&=&F_1-\frac{(1-4x^2t) b_1}{3}-\frac{xt}{3}a_3+g_1+4x^2tg_2\nn \\
&\simeq&F_1-\frac{b_1}{3}+g_1 \, , \\
A_{+-,00}&=&\sqrt{-t} \ \big[ 2x(g_1+g_2)-\frac{a_3}{2}-\frac{(1-4x^2t)a_4}{4}\big] \nn \\
&\simeq&\sqrt{-t} \ \big[ 2x(g_1+g_2)-\frac{a_3}{2}-\frac{a_4}{4}\big] \, , \\
A_{+-,-+}&=&-2xta_3 \simeq 0 \, , \\
A_{0+,0+}&=&-F_1+\frac{(1-4x^2t )F_2}{2x}+\frac{(1-4x^2t ) b_1}{3}-\frac{(3-12x^2t+16x^4t^2)b_2}{18x}\nn\\
&&+\frac{4(-2x^2t)b_3}{3x}+\frac{2xt}{3}(1-4x^2t )b_4 \nn \\
&\simeq&-F_1+\frac{F_2}{2x}+\frac{b_1}{3}-\frac{b_2}{6x} \, , \\
A_{00,00}&=&-F_1+\frac{(1-4x^2t )F_2}{2x}-\frac{2(1-4x^2t ) b_1}{3}+\frac{(3-12x^2t+16x^4t^2)b_2}{9x}\nn\\
&&-\frac{8(-2x^2t)b_3}{3x}-\frac{4xt}{3}(1-4x^2t )b_4 \nn \\
&\simeq&-F_1+\frac{F_2}{2x}-\frac{2 b_1}{3}+\frac{b_2}{3x} \, .
\label{A8}
\ea
Notice that the first line in each equation (\ref{A1})-(\ref{A8})
coincides with the corresponding one in equations (7) in reference
\cite{Hoodbhoy:1988am}\footnote{This is easy to see by writing
$\kappa$ in terms of $x$ and $t$, by taking into account an overall
sign in the definition of the four-dimensional metric, and by noting
that the definition of $\nu=P \cdot q$ in reference
\cite{Hoodbhoy:1988am} differs from the one used in the present work
which is $\nu=-P \cdot q/M$.}, and we write them here for
completeness. On the other hand, the second line in each of these
equations corresponds to the limit $|t|\ll 1$.

These amplitudes are useful tools in order to study the hadron
structure, specially in the cases with $s>1/2$. They provide a
simple way to understand the meaning of the structure functions in
theoretical terms, and give important insights about how they can
affect the scattering amplitudes and how to measure them. This is
the reason why we shall explicitly show the form of these helicity
amplitudes in terms of the structure functions we found by using the
gauge/string duality.

For instance, in the spin-$1/2$ case one has $A_{+\uparrow
,+\uparrow}\sim F_1 - g_1$ and $A_{+\downarrow ,+\downarrow}\sim F_1
+ g_1$, which indicates that $F_1$ is the cross section for a
transverse photon scattered off an unpolarized target, while $g_1$
is the spin asymmetry in the scattering cross section for a
transverse photon \cite{Manohar:1992tz}. One can also say that
$g_1+g_2$ is proportional to the single helicity flip amplitude
$\sim A_{+ \downarrow, 0 \uparrow}$. For spin-one hadrons a similar
analysis can be carried out. The function $F_1 \sim A_{++,++} +
A_{+0,+0} + A_{+-,+-}$ can be interpreted in the same way. Also
$b_1\sim A_{++,++} + A_{+-,+-}-2 A_{+0,+0}$ has an interpretation
from a transverse photon scattered off a polarized spin-one target
\cite{Hoodbhoy:1988am}. The combination $g_1+g_2$ also appears in
the $A_{+-,00}$ amplitude, and at first order in $t$ this is the
only place where $g_2$ is present. The combinations $F_1-b_1/3$ and
$F_2-b_2/3$ define $A_{0+,0+}$ and will be important when we present
our structure functions. Also, the $A_{+-,-+}\sim -b_3+b_2/3$ is an
interesting amplitude (even being sub-leading in $t$) since it
characterizes the double helicity flip. As pointed out in
\cite{Manohar:1992tz}\footnote{In the mentioned paper this is the
so-called $\Delta(x)$ structure function.} this quantity does not
receive contributions from quark interactions at first order in
$\alpha_s$ because quarks cannot change their helicity by
two\footnote{Recall that $\alpha_s=g_{QCD}^2/4\pi$.}.

Finally, all these linear combinations of structure functions that
define the helicity amplitudes are positive, and this is a
consistency check for the structure functions we obtained in our
previous papers \cite{Koile:2011aa,Koile:2013hba,Koile:2014vca} in
terms of the holographic dual models.

\section{DIS differential cross sections from the gauge/string duality}

The present approach is based on a method developed by Polchinski
and Strassler in \cite{Polchinski:2002jw}, where the structure
functions corresponding to glueball states were derived in terms of
the gauge/string duality. They consider the planar limit of the
${\cal{N}}=1^{\star}$ SYM theory obtained from a deformation of the
${\cal{N}}=4$ SYM theory. The ${\cal{N}}=1^{\star}$ SYM theory has a
running coupling and becomes conformal when the energies involved
are higher than a color confinement scale $\Lambda$. Therefore, this
SYM theory and QCD share some important qualitative properties: they
exhibit confinement of the color degrees of freedom and a mass gap
in the infrared. In the context of the gauge/string duality it was
shown that the ${\cal{N}}=1^{\star}$ SYM theory is dual to strings
propagating in a certain deformed $AdS_5\times S^5$ background
\cite{Polchinski:2000uf}. In order to describe DIS, it has been
calculated the scattering amplitude of a process where a particular
graviton perturbation coming from the boundary couples to the
dilaton, which is dual to a glueball state \cite{Polchinski:2002jw}.
In addition, the optical theorem is used to relate DIS and forward
Compton scattering. Therefore, in the gravity dual description the
graviton and the dilaton perturbations are taken to describe both
initial and final states leading to a process which turns out to be
dual to the forward Compton scattering described in the previous
section\footnote{For the large-$x$ case there is also a sum over
intermediate dilaton states.}. It is important to emphasize that in
the large-$x$ case the supergravity description is appropriate to
derive the structure functions, while in the small-$x$ regime the
string theory description is unavoidable.

The calculations of DIS from vector mesons have been presented in
our papers \cite{Koile:2011aa,Koile:2013hba,Koile:2014vca}. We have
obtained the leading order behavior of the eight structure functions
involved in the hadronic tensor $W^{\mu \nu}$ for a spin-one target,
namely $F_1, F_2, b_1, b_2, b_3, b_4, g_1$ and $g_2$ in terms of
$x$, $\frac{\Lambda^2}{q^2}$ and some other parameters which define
the holographic dual description.

The Bjorken parameter $x$ is a very important kinematic variable,
and its physical range $0 \leq x \leq 1 $ can be divided in four
regions\footnote{Recall that $x=1$ corresponds to elastic
scattering.}. In the region $1/\sqrt{\lambda}\ll x<1$ (which we call
the large-$x$ region or range A) the ten-dimensional Mandelstam
variable associated with the intermediate state $\tilde{s}\sim
(\alpha'^2\lambda)^{-1/2}(1/x-1)$ indicates that only massless
string states are exchanged and thus we only have to consider the
low energy limit of string theory, leading to a supergravity
description. The second kinematic region (range B) is reached when
$e^{-\sqrt{\lambda}} \ll x \ll 1/\sqrt{\lambda}$. Thus, in this
region the dynamics of string theory provides the appropriate
description. Fortunately in this region, it is possible to think of
locally interacting strings, in the sense that the characteristic
length scales are smaller than the AdS radius, leading to some
important simplifications \cite{Polchinski:2002jw}. This is the
region we call small-$x$ (but not exponentially small). The third
region (range C) corresponding to $ x \sim e^{-\sqrt{\lambda}}$ is
important because the non-locality of the string scattering is
described by a diffusion operator. This operator leads to the
convergence of the first moments of the structure functions
\cite{Polchinski:2002jw,Brower:2006ea}. For smaller $x$ values,
there is fourth region which is not accessible with the present
approach since it would require to consider strings propagating in a
curved background. Finally, we point out that for actual experiments
large $x$ generally corresponds to $0.1<x<1$, while the small $x$
measurements are supposed to describe the $0<x<0.1$ region.

In the following sections we present our results for the structure
functions in the first two regimes and show how they affect the
differential cross section of the DIS process in the different
polarization cases. We also consider the corresponding helicity
amplitudes.

\subsection{Range A: $1/\sqrt{\lambda}\ll x<1$}

The structure functions found in our previous paper
\cite{Koile:2011aa} in this regime of the Bjorken parameter at
leading order in $t$ are
\begin{eqnarray}\label{strfunctabel}
F_{1}&=& A^{SF}(x)\frac{1}{12x^{3}}(1-x)\,,\nn\\
F_{2}&=& A^{SF}(x)\frac{1}{6x^{3}}(1-x)\,,\nn\\
b_{1}&=& A^{SF}(x)\frac{1}{4x^{3}}(1-x)\,,\nn\\
b_{2}&=& A^{SF}(x)\frac{1}{2x^{3}}(1-x)\,,\nn\\
b_{3}&=& A^{SF}(x)\frac{1}{24x^{3}}(1-4x)\,, \label{61}\\
b_{4}&=& A^{SF}(x)\frac{1}{12x^{3}}(-1+4x)\,,\nn\\
g_{1}&=& A^{SF}(x)\frac{t}{8x^2}(-7+6x)\,,\nn\\
g_{2}&=& A^{SF}(x)\frac{1}{16x^{4}}(3-3x)\,,\nn
\end{eqnarray}
where
\begin{equation}
A^{SF}(x) =
A^{SF}_0\mu_p^2{\mathcal{Q}}^2\alpha'^4R^{2p-6}\bigg(\frac{\Lambda^2}{q^2}\bigg)^{\gamma}
x^{\gamma+2n+5}(1-x)^{\gamma-1}\,,
\label{62}
\label{Axgrande}
\end{equation}
is the only model-dependent factor and it is written in terms of
$q^2$ and the different variables involved in the string theory
calculation: the Dp-brane tension $\mu_p$, the charge of the hadron
${\cal {Q}}$, the confining scale $\Lambda$, the sphere radius $R$,
the string constant $\alpha'$ and the exponents
\be
\gamma^2 = \frac{A^2 + \ell(\ell+p-5)}{B^2} \ , \ A =
\frac{1-2\alpha - \beta (p-3)/2 - p + 4}{2} \ , \ B = \frac{\alpha -
\beta - 2}{2}
\ee
related to the general induced asymptotic metric on the probe
Dp-branes
\beq
ds^2 = \left(\frac{r}{R}\right)^{\alpha} \eta_{\mu \nu} dx^{\mu}
dx^{\nu} + \left(\frac{r}{R}\right)^{\beta} \left[dr^2 + r^2
d\Omega_{p-4}^2\right] \, .
\eeq
Furthermore, we have defined $n=\frac{2+\beta}{4B}$. Notice that
$\ell$ is the usual spherical harmonic index that appears in the
solutions of the scalar and vector mesons, and $A_0^{SF}$ is a
dimensionless normalization constant which depends on the model
\cite{Koile:2011aa,Koile:2013hba}. As explained in
\cite{Koile:2013hba}, for the three considered Dp-brane models the
values of the parameters are shown in Table 1.
\begin{table}
\begin{center}
\centering
\begin{tabular}{|c|c|c|c|}
\hline
Model $/$ Parameter & p & $\alpha$ & $\beta$ \\
\hline
D3D7 & 7 & 2 & -2 \\
\hline
D4D8$\mathrm{\overline{D8}}$ & 8 & $3/2$ & $-3/2$ \\
\hline
D4D6$\mathrm{\overline{D6}}$ & 6 & $3/2$ & $-3/2$ \\
\hline
\end{tabular}
\caption{\small Parameters for each Dp-brane model.}
\end{center}
\end{table}
Using the redefinition $W_i=
\frac{1}{12x^3}A^{SF}(x)\widetilde{W_i}$ we obtain
\begin{eqnarray}
\widetilde{F_{1}}= (1-x)\, \ &,& \ \widetilde{F_{2}}= 2(1-x)\,,\nn\\
\widetilde{b_{1}}= 3(1-x)\, \ &,& \ \widetilde{b_{2}}= 6(1-x)\, \nn \\
\widetilde{b_{3}}= \frac{1}{2}(1-4x)\, \ &,& \ \widetilde{b_{4}}= -(1-4x)\,,\nn\\
\widetilde{g_{2}}= \frac{3}{4x}(3-3x)\, \ &,& \ \widetilde{g_{1}}=
\frac{3xt}{2}(-7+6x) \approx 0 \, .\nn
\end{eqnarray}
This means that in this regime we obtain the relations
\cite{Koile:2011aa,Koile:2013hba}
\ba
F_2=2F_1 \ , \ b_2=2b_1 \ , \ b_1=3F_1 \ , \ g_2=\frac{9}{4x}F_1 \ ,
\ b_4=-2b_3 \nn \, . \label{relationships}
\ea
Note that the first two are typical Callan-Gross type relations,
except for the lack of a factor $x$, which in this regime is
understood as if the lepton is scattered by the entire hadron due to
the strong coupling and the specific kinematic regime. By looking at
the $q^2$-dependence we obtain a power-law decay, which is not
exactly the same as in perturbative QCD. This was already observed
in \cite{Polchinski:2002jw} where it was pointed out that at weak 't
Hooft coupling the leading contribution comes from twist-$2$
operators, while at strong coupling the dominant contribution is
given by double-trace operators.

For completeness we also write the results for the scalar mesons,
which as mentioned only have two structure functions
\begin{equation}
F_1=0 \ , \ F_2 = A_0^{scalar} \mu_p^2 {\mathcal{Q}}^2 \alpha'^4 R^{2p-6}
\left(\frac{\Lambda^2}{q^2}\right)^{\gamma+1}x^{\gamma +3 + 2n}(1-x)^{\gamma},
\label{PiLargex}
\end{equation}
where $A_0^{scalar}= 2^{2\gamma+4}B^{2\gamma+2}\pi^5|c_i|^2|c_X|^2
\frac{\Gamma^2[\gamma+n+2]}{\Gamma^2[n+1]}$, while the constants
$c_i$ and $c_X$ have been defined in \cite{Koile:2011aa}.

By plugging the structure functions from the holographic vector
mesons in the general spin-one DIS differential cross section we
find
\ba
\frac{d\sigma}{dx \, dy \, d\phi}&=&\frac{MEe^4}{4 \pi^2 q^4}
\left\{ \langle qq \rangle\left[ xy^2+2(1-y)\right]\,F_1 +
2xy^2g_2\bigg[ \frac{(s_h\cdot s_l)}{(P\cdot q)}-
\frac{(q\cdot s_h)(P\cdot s_l)}{(P\cdot q)^2} \bigg]\right\} \nn \\
&=&\frac{MEe^4}{4 \pi^2 q^4} F_1\left\{\langle qq \rangle\left[
xy^2+2(1-y)\right] + \frac{9}{2}y^2\bigg[ \frac{(s_h\cdot
s_l)}{(P\cdot q)}-\frac{(q\cdot s_h)(P\cdot s_l)}{(P\cdot q)^2}
\bigg]\right\} \nn
\\
&=&\frac{MEe^4}{4 \pi^2 q^4} \frac{A^{SF}(x)(1-x)}{12x^3} \times \nn \\
&& \left\{\langle qq \rangle\left[ xy^2+2(1-y)\right] +
\frac{9}{2}y^2\bigg[ \frac{(s_h\cdot s_l)}{(P\cdot q)}-
\frac{(q\cdot s_h)(P\cdot s_l)}{(P\cdot q)^2} \bigg]\right\}.
\ea
For a longitudinally polarized lepton beam where $s_l = H_l k$ with
$H_l = \pm 1$ for positive and negative helicity, we can rewrite
this expression in the following form
\ba
\frac{d\sigma}{dx \, dy \, d\phi}&=& \frac{MEe^4A^{SF}(x)}{48 \pi^2 q^4x^3}
(1-x) \left\{\langle qq \rangle\left[ xy^2+2(1-y)\right] - \frac{9H_l\, }{2ME}
s_h^{\mu} \left( y k_{\mu} -q_\mu\right) \right\}. \label{xgrande}
\ea
We can study the behavior of this result for each polarization as
follows.
\begin{itemize}
\item Unpolarized target: $\langle qq \rangle_{unpol} = 1$, $s_h = 0$.
\be
\frac{d\sigma}{dx dy d\phi}\Big|_{unpol} =\frac{MEe^4A^{SF}(x)}{48
\pi^2 q^4x^3}(1-x) \left[xy^2 + 2 (1-y)\right].
\ee
\item Longitudinally polarized target: $\langle qq \rangle_{LP} = 0$,
$s_h = (0, H_h M \hat{z})$ (and using $\cos \theta \approx 1$).
\be
\frac{d\sigma}{dx dy d\phi}\Big|_{LP} = 0 \, .
\ee
This is the result at leading order. It is easy to see why this
happens: since $\langle qq \rangle_{LP}=0$, the $b_i = 3 F_i$
relations for $i=1, 2$ imply that the $b$-$F$ part of the
cross section vanishes. Now, as pointed out in \cite{Manohar:1992tz}
$s_h$ is dotted with $k$ or $q$ both of which have similar $0$ and
$3$ components at leading order. Thus, we replace it by $-H_h p$
\footnote{The minus sign accounts for the difference in the
four-dimensional Minkowski metric components $\eta_{00}$ and
$\eta_{zz}$.} without changing the results. This implies that the
$g_2$-piece becomes sub-leading in this case, and as $g_1 \approx 0$
at first order, therefore the cross section vanishes.
\item Transversally polarized target: $\langle qq \rangle_{TP} = 3/2$, $s_h = (0, H_h M \hat{x})$
\ba
\frac{d\sigma}{dx \, dy \, d\phi}|_{TP}&=& \frac{MEe^4A^{SF}(x)}{48
\pi^2 q^4x^3}(1-x)\left\{\frac{3}{2}\left[ xy^2+2(1-y)\right] -
\frac{9 H_l H_h}{2} \frac{M}{|q|}(1-y) \cos \phi \right\} \nn \\
&&
\ea
Note that the $M/|q| = \sqrt{-t}$ factor makes the spin-dependent
term sub-leading. The reader could think that it is not a reliable
result since in early steps of our calculations we have ignored this
kind of terms. This is not the case because this has been done in
terms involving the $b_i$ functions, which contribute to the leading
term, and mainly because the spin-dependent part is easily
distinguished within experiments.
\end{itemize}

In addition, for the helicity amplitudes in this case we obtain
\ba
A_{++,++}&\simeq& 0+O(t) \, ,\nn \\
A_{+0,+0}&\simeq&3F_1 \, ,\nn \\
A_{+0,0+}&\simeq&\sqrt{-t} \ \big[ \frac{9F_1}{2}+\frac{a_3}{2}+\frac{a_4}{4}\big]=\sqrt{-t} \ 6F_1 \, ,\nn \\
A_{+-,+-}&\simeq& 0+O(t) \, , \\
A_{+-,00}&\simeq&\sqrt{-t} \ \big[ \frac{9F_1}{2}-\frac{a_3}{2}-\frac{a_4}{4}\big]=\sqrt{-t} \ 3\ F_1 \, ,\nn \\
A_{+-,-+}&\simeq& 0 \, ,\nn \\
A_{0+,0+}&\simeq&\frac{4b_3}{3x} \, ,\nn \\
A_{00,00}&\simeq&3\Big(\frac{1}{x}-1\Big)F_1-\frac{8b_3}{3x} \, .\nn
\ea
%

\subsection{Range B: $e^{-\sqrt{\lambda}} \ll x \ll 1/\sqrt{\lambda}$}

The results we have obtained in our previous paper
\cite{Koile:2014vca} in the small-$x$ (but not exponentially small)
regime for the polarized vector meson structure functions can be
summarized in the following equations
\ba
F_1 = \frac{1}{12x^2} I_1 \ &,& \ F_2 = \frac{1}{6x}\left(I_1 + I_0\right), \nonumber \\
b_1 = \frac{1}{4x^2}I_1 \ &,& \ b_2 = \frac{1}{2x}(I_1 + I_0), \label{funcxchico} \\
b_3 = -\frac{1}{3x}(I_1 + I_0) \ &,& \ b_4 = \frac{1}{6x}\left(I_1 + I_0\right), \nonumber \\
g_1 = 0 \ &,& \ g_2 = \frac{1}{8x^2}(I_1 + I_0), \nonumber
\ea
where $I_1$ and $I_0$ are $x$-independent factors given by integrals
of Bessel functions and some constants, plus the $q^2$-dependence.
According to these results, in this regime the Callan-Gross type
relations become
\be
F_2 = 2xF_1 \left(1 + \frac{I_0}{I_1}\right) \ , \ b_2 = 2xb_1
\left(1 + \frac{I_0}{I_1}\right)\ , \ b_4 = \frac{1}{3}b_2 =
-\frac{1}{2}b_3 \, .
\ee
Notice that the first two relations have a Bjorken parameter factor
$x$ with respect to the ones in the regime A described in the
previous section. The pre-factors are given by
\ba
I_1 &=& C \frac{\alpha^2 \Sigma^2 q^4}{
\sqrt{\lambda}\Lambda^4 \Gamma^2 (n+1) B}
\left(\frac{qR}{B}\right)^{(2n+1)-D} I_{n+1,D} \, , \label{I1} \\
I_0 &=& C \frac{\alpha^2 B \Sigma^2 q^2}{
\sqrt{\lambda} \Lambda^4R^2 \Gamma^2 (n+1)}
\left(\frac{qR}{B}\right)^{(2n+3)-D} I_{n,D} \, \label{I2} ,
\ea
where $C$ is a normalization constant and $D = 2n + 1 +
\frac{1}{B}[2\Delta + \alpha + (\frac{\beta}{2}+1)(p-5)]$. We have
used the definition $\Sigma^2=\Delta^2+\Delta(\Delta-2)$. This is
where the $\ell$-dependence enters, since $\Delta = \gamma(\ell) B -
A $. Note that the $x$-dependence of the structure functions is the
same for all models, which was not the case in the previous section.
The Bessel function integrals are defined as \footnote{Recall that
this result is valid provided that the arguments of the gamma
functions are positive.}
\bea
\int_0^\infty d\w \, \w^m \, K^2_n(\w) =  I_{n,m} &=& 2^{m-2}
\frac{\Gamma(\nu+n)\Gamma(\nu-n)\Gamma^2(\nu)}{\Gamma(2\nu)} \ , \
\nu = \frac{1}{2}(m+1).
\eea
In the D4D8$\mathrm{\overline{D8}}$-brane and
D4D6$\mathrm{\overline{D6}}$-brane models $K_n(w)$ functions turn
out to be fractional Bessel functions, however all the integrals are
finite. Again, all these factors are independent of $x$ and $y$
\footnote{We refer the reader to \cite{Koile:2014vca} for a careful
derivation and analysis of these results.}. Finally, the only
model-dependent and $\ell$-dependent numerical factor that enters
the Callan-Gross type relations is
\be
\delta \equiv 1 + \frac{I_0}{I_1} = 1 + \frac{I_{n,D}}{I_{n+1,D}} =
\frac{2D}{D + 2n + 1}.
\ee
Table 2 shows the values of $\delta$ for the three models we
consider. In all the cases we have $1<\delta<2$ since $\Delta$ is
positive. As for the scalar mesons the results are the following
\begin{eqnarray}
F_1 &=& \frac{\pi^2}{16 x^2} \rho_3 |c_i|^2
\left(\frac{\Lambda^2}{q^2}\right)^{\Delta-1}
\frac{1}{\sqrt{4 \pi g_c N}} I_{1,2\Delta+3} \, , \label{PiSmallx1}\\
F_2 &=& \frac{\pi^2}{8 x} \rho_3 |c_i|^2
\left(\frac{\Lambda^2}{q^2}\right)^{\Delta-1}
\frac{1}{\sqrt{4 \pi g_c N}} (I_{0,2\Delta+3}+I_{1,2\Delta+3}) \, \label{PiSmallx2} .
\end{eqnarray}
Equations (\ref{I1}) and (\ref{I2}) show the dependence of our
small-$x$ structure functions with the momentum of the virtual
photon: they always fall-off as a power of $q^2$.  In the Regge
regime the calculations and measurements suggest that a Pomeron is
exchanged\footnote{There are some important differences between hard
and soft Pomerons that we shall not address here, for this see
\cite{Brower:2006ea}.} both at weak and strong 't Hooft coupling.
However, there is an important difference: when $\lambda$ is small
the virtual photon scatters off a parton inside the hadron, leading
to a small (growing) dependence in $q^2$, while at large $\lambda$ a
Pomeron is supposed to collide with the entire hadron. This is what
happens in the present case, since we are using the gauge/string
duality at strong coupling. Thus, we have obtained the same
$q^2$-power fall-off as in the large-$x$ case, and the amplitude is
again dominated by the same double-trace operators
\cite{Polchinski:2002jw}.
\begin{table}
\begin{center}
\centering
\begin{tabular}{|c|c|c|c|}
\hline
Model & D3D7 & D4D8$\mathrm{\overline{D8}}$ & D4D6$\mathrm{\overline{D6}}$ \\
\hline
$\delta$ & $\frac{2\Delta+3}{\Delta+2}$ & $\frac{4 (4\Delta+3)}{8\Delta + 9}$ & $\frac{16(\Delta+1)}{8 \Delta+11}$  \\
\hline
\end{tabular}
\caption{\small Parameter $\delta$ in each model.}
\end{center}
\label{tabla-C}
\end{table}

Plugging our structure functions in the most general DIS
differential cross section for a spin-one target we obtain
\ba
\frac{d\sigma}{dx dy d\phi} &=& \frac{MEe^4}{4\pi^2q^4}
\left\{\langle qq\rangle [xy^2 + 2x(1-y)\delta]F_1 + 2xy^2
g_2\left[\frac{(s_h \cdot s_l)}{(p\cdot q)}
- \frac{(q\cdot s_h)(p\cdot s_l)}{(p\cdot q)^2}\right] \right\} \nn \\
&=& \frac{MEe^4}{4\pi^2q^4} F_1 \left\{\langle qq\rangle [xy^2 +
2x(1-y)\delta] + 3xy^2 \delta \left[\frac{(s_h \cdot s_l)}{(p\cdot
q)}
- \frac{(q\cdot s_h)(p\cdot s_l)}{(p\cdot q)^2}\right] \right\} \nn \\
&=& \frac{MEe^4}{48\pi^2q^4} \frac{I_1}{x} \left\{\langle qq\rangle
[y^2 + 2(1-y)\delta] + 3y^2\delta \left[\frac{(s_h \cdot
s_l)}{(p\cdot q)} - \frac{(q\cdot s_h)(p\cdot s_l)}{(p\cdot
q)^2}\right] \right\} .
\ea
Once again, the experimentally relevant situation occurs when the
lepton is longitudinally polarized, {\it i.e.} when $s_l= H_l k$
with $H_l=\pm 1$. Therefore, the differential cross section takes
the form
\be \frac{d\sigma}{dxdyd\phi} = \frac{MEe^4}{48\pi^2q^4}
\frac{I_1}{x} \left\{\langle qq\rangle [y^2 + 2(1-y)\delta] -
\frac{3\delta}{ME}H_l s_h^{\mu}(y k_{\mu} - q_{\mu}) \right\} \, .
\ee
Then, we can study the behavior of this result for each hadronic
polarization as follows.
\begin{itemize}
\item Unpolarized target: $\langle qq \rangle_{unpol} = 1$, $s_h = 0$.
In this case the spin-dependent term vanishes automatically and we
obtain
\be
\frac{d\sigma}{dxdyd\phi}|_{unpol} = \frac{MEe^4}{48\pi^2q^4}
\frac{I_1}{x}[y^2 + 2(1-y)\delta] \, .
\ee
\item Longitudinally polarized target: $\langle qq \rangle_{LP} = 0$,  $s_h = (0,H_h M \hat{z})$
\be
\frac{d\sigma}{dx dy d\phi}\Big|_{LP} = 0 \, .
\ee
This is just like in the large-$x$ regime. We obtain a vanishing
result at leading order since $g_1 = 0$ (notice that the $g_1$
structure function together with the difference $F_1 -
\frac{1}{3}b_1$ give the main contribution to the longitudinally
polarized cross section), and this is a consistency check of our
results at small $x$ for the polarized vector meson structure
functions.
\item Transversally polarized target:
$\langle qq \rangle_{TP} = 3/2$, $s_h = MH_h \hat{x}$. Now, we have
a vanishing $s_h\cdot k$ but a non-vanishing $s\cdot q = M H_h
k'^{x} = ME' H_h \sin \theta \cos \phi $, and since $E' = E (1-y)$
and $\sin (\theta) \sim \theta \sim M/|q|$ we finally obtain
\be
\frac{d\sigma}{dxdyd\phi}|_{TP} = \frac{MEe^4}{48\pi^2q^4} \frac{I_1}{x}
\left\{\frac{3}{2}[y^2 + 2(1-y)\delta] - 3\delta(1-y)\frac{M}{|q|}
H_l H_h \cos \phi\right\}.
\ee
Since $M/|q| \sim \sqrt{-t}$ the second term also becomes sub-leading in this case.
\end{itemize}

We can also obtain the helicity amplitudes in this regime which
become
\ba
A_{++,++} \simeq 0 \ &,& \ A_{+0,+0} \simeq 3F_1 \, ,  \nn \\
A_{+0,0+} \simeq 0 \ &,& \ A_{+-,+-} \simeq 0 \, ,  \\
A_{+-,00} = O(\sqrt{-t}) \approx 0  \ &,& \ A_{+-,-+} = 0 \, , \nn \\
A_{0+,0+} = 2 (\delta-1)F_1 >0 \ &,& \ A_{00,00} = 3 (\delta-1)F_1
>0 \, . \nonumber
\ea

Note that in this parametric region the factor $a_3=b_2/3-b_3$
defined in the section 3.3 becomes $a_3 = b_2$ at leading order.
This is interesting since it means that the $b_2$ structure function
characterizes the double helicity flip amplitude. The corresponding
$A_{+-,-+}$ helicity amplitude is proportional to $t$, thus it is
always small in the DIS regime, but the $b_2 \sim x^{-1}$ growth
could in principle lead to a non-vanishing helicity amplitude.

\subsection{Range C: $x \sim e^{-\sqrt{\lambda}}$}

This is the exponentially small region of the Bjorken parameter. In
the region B we have neglected the factor ${\tilde s}^{\alpha'
\tilde t/2}$. The effect of this factor is very important within the
parametric region $x \sim \exp{(-\sqrt{\lambda})}$
\cite{Koile:2014vca}. The tildes indicate ten-dimensional Mandelstam
variables. Let us consider the strong coupling regime $1 \ll \lambda
\ll N$ \footnote{In this subsection we focus on the D3D7-brane
model, the other models have a few modifications but the discussion
is very similar.}, and exponentially large values of ${\tilde {s}}$
with $\frac{\log s}{\sqrt{\lambda}}$ fixed. In this region the
interaction cannot be considered local in the AdS background, since
${\tilde {t}}$ becomes a differential operator
\cite{Polchinski:2002jw,Brower:2006ea}, and consequently we have to
replace it by the following operator
\beq
\alpha' \tilde{t} \rightarrow \alpha' \nabla^2 = \alpha' \left(
\frac{R^2t}{r^2} + \nabla^2_{\bot}\right) \, , \label{diffoperator}
\eeq
which takes into account the momentum transfer in the transverse
directions, and it induces the factor
\beq
(\alpha' \tilde{s})^{\alpha'\tilde{t}/2} \sim
x^{-\alpha'\tilde{t}/2} \sim x^{- \alpha' \nabla^2/2} \, .
\label{simeq}
\eeq
The physical meaning of this is that within this region C there is a
Pomeron exchange in the $t$-channel\footnote{In reference
\cite{Koile:2014vca} we have not considered multi-Pomeron
exchange.}. The consequence of this leads to include a factor
$x^{\alpha' \zeta/2}$ multiplying the structure functions. This is
obtained in the region C by replacing the differential operator
$-\nabla^2=-D^2$ by its smallest eigenvalue $\zeta$. Then, the
structure functions become $F_1 \propto x^{-2+\alpha' \zeta/2}$ and
$F_2 \propto x^{-1+\alpha' \zeta/2}$, where the Pomeron intercept is
identified as $2-\alpha' \zeta/2$. This is the expected behavior for
these functions. In particular, for the D3D7-brane model we have
$\zeta = \frac{4}{R^2}(1-\Delta^2)\leq 0$. This implies that in our
results of the previous section 4.2 in the region B we have to
multiply all the structure functions by the factor $x^{\alpha'
\zeta/2}$. This implies that the differential cross sections for
different meson polarizations as well as the helicity amplitudes
must also be multiplied by this factor in the parametric region C.

\section{Results and discussion}

In this section we present our results and discuss them.

\subsection{Structure functions and differential cross sections}

\begin{figure}
\centering \subfigure[$F_1(x)$ function]{
\includegraphics[scale=0.45]{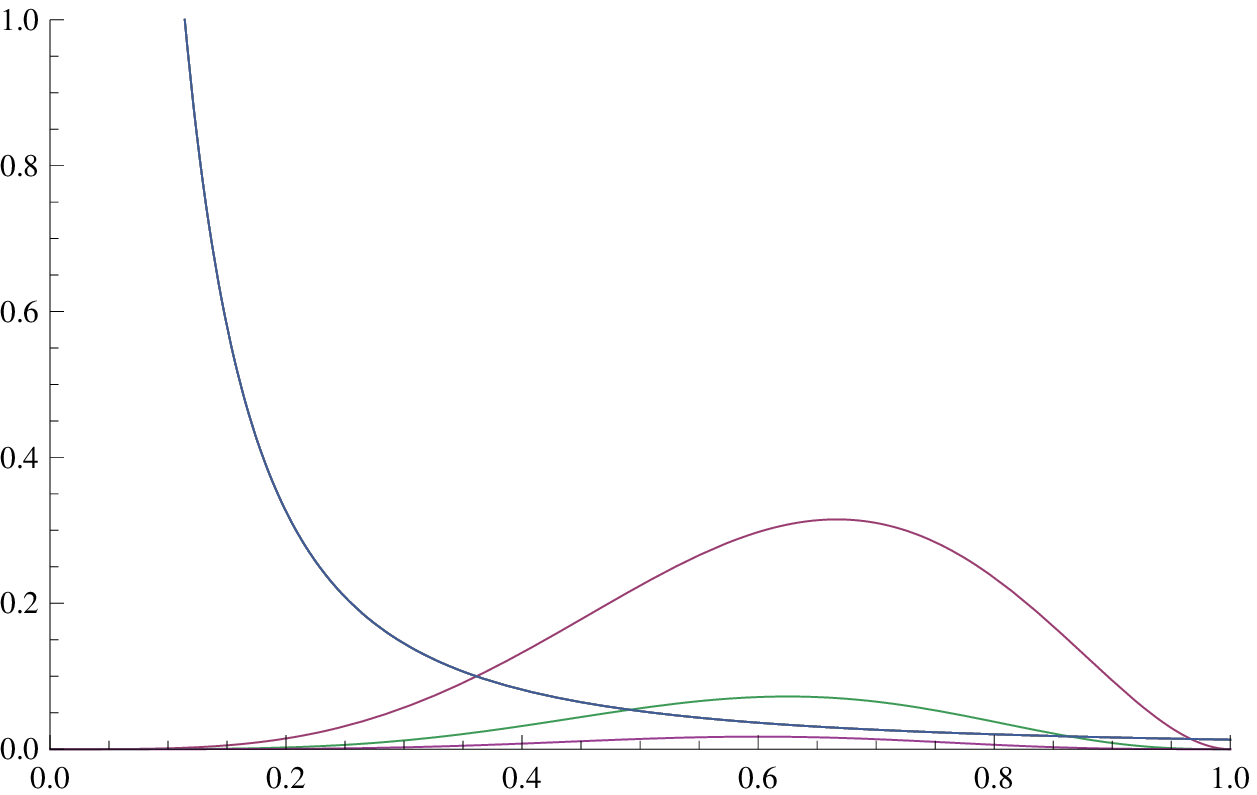}
\label{} } \subfigure[$F_2(x)$ function]{
\includegraphics[scale=0.45]{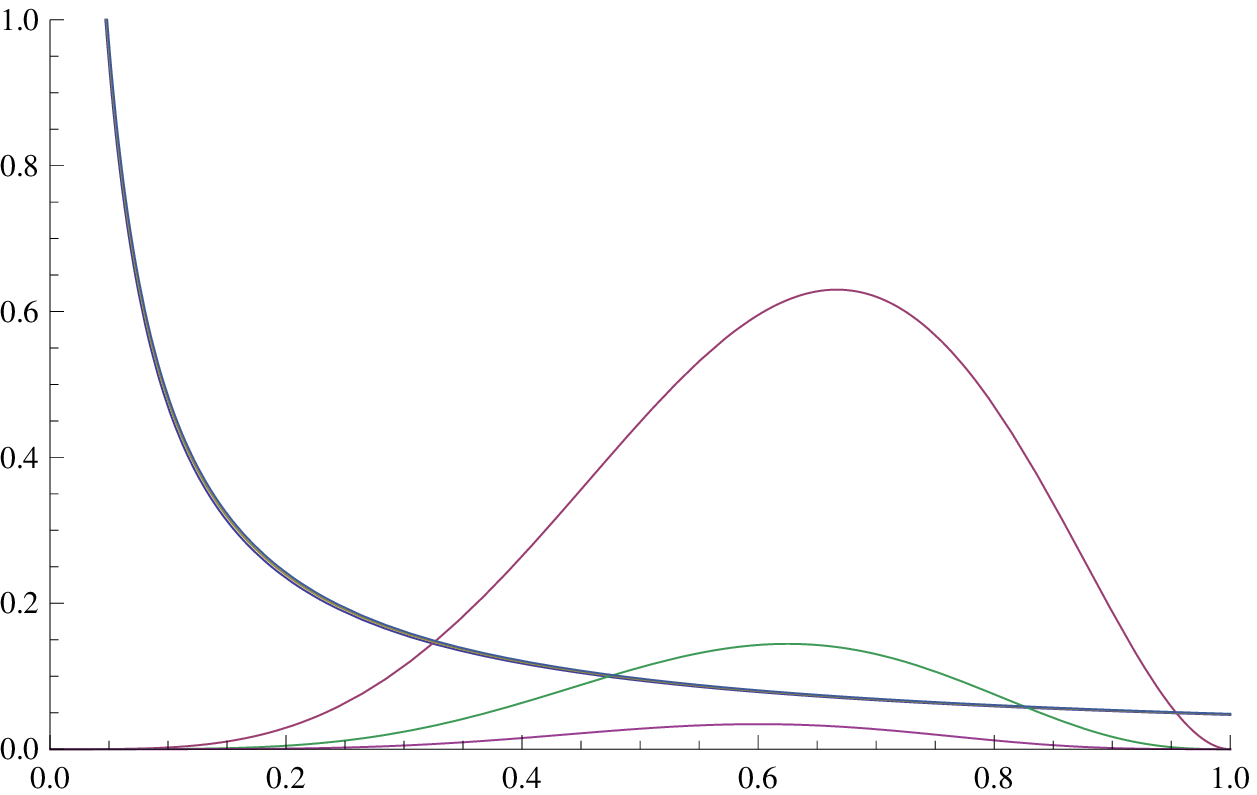}
\label{} } \subfigure[$b_1(x)$ function]{
\includegraphics[scale=0.45]{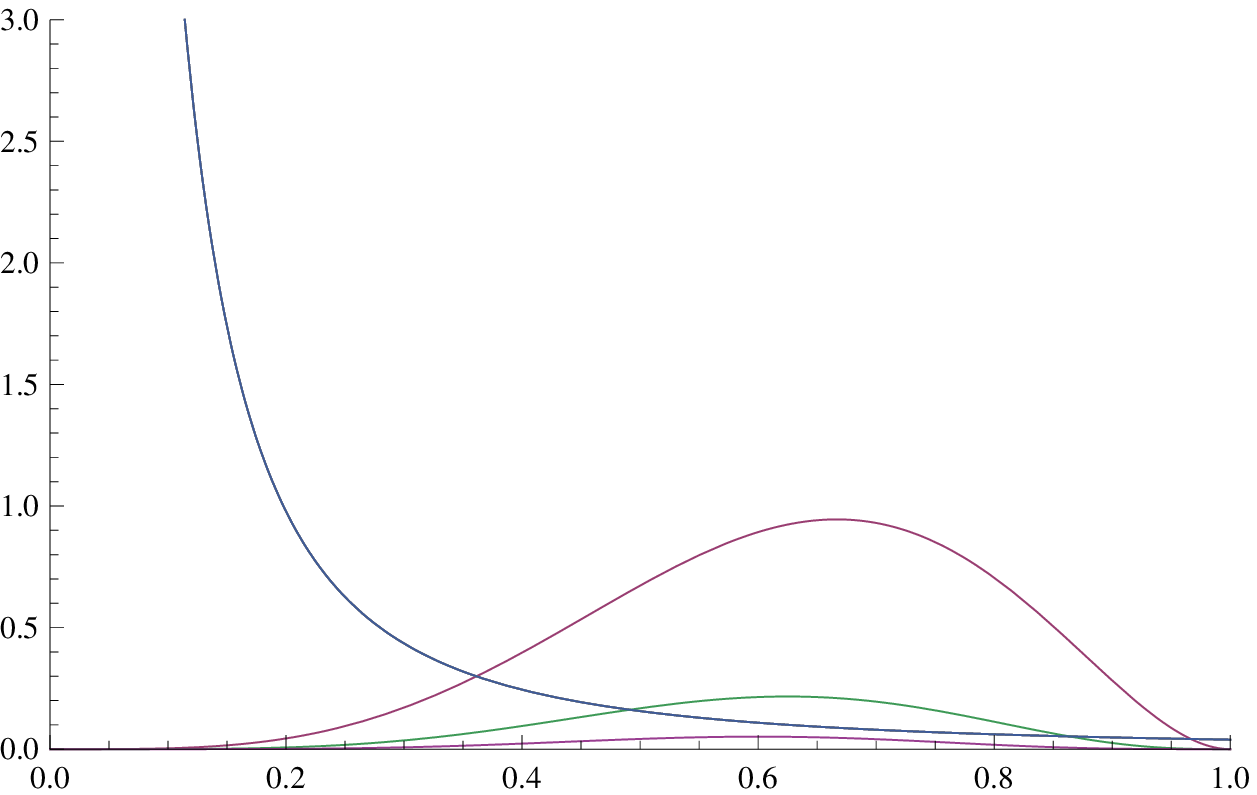}
\label{} } \subfigure[$b_2(x)$ function]{
\includegraphics[scale=0.45]{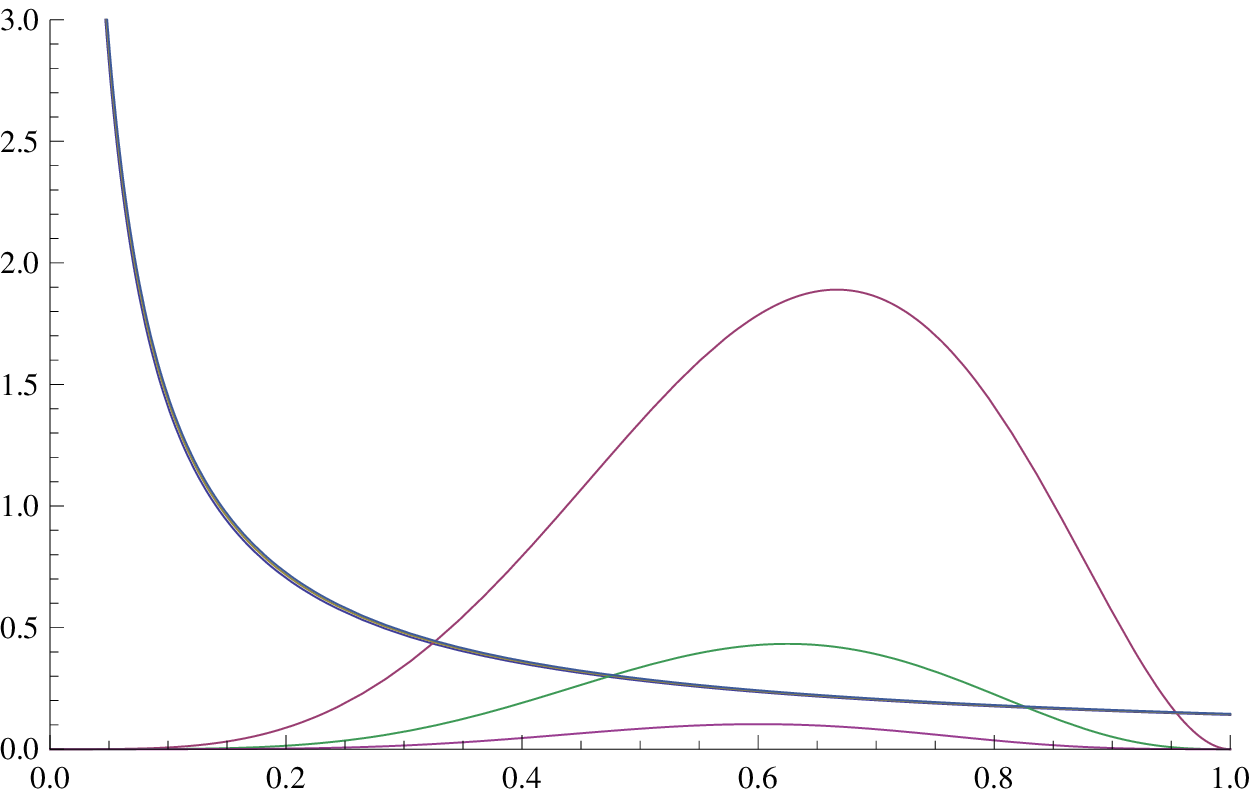}
\label{} } \caption{\small Structure functions $F_1$, $F_2$, $b_1$
and $b_2$ in terms of the Bjorken parameter $x$ for the $\ell=1, 2$
and $3$ vector mesons in the D3D7-brane model. Bell-shaped curves
correspond to the large-$x$ region (height decreases as $\ell$
increases) while the hyperbolic curves correspond to the small-$x$
region.} \label{functions1}
\end{figure}

\begin{figure}
\centering \subfigure[$F_1(x)$ function]{
\includegraphics[scale=0.45]{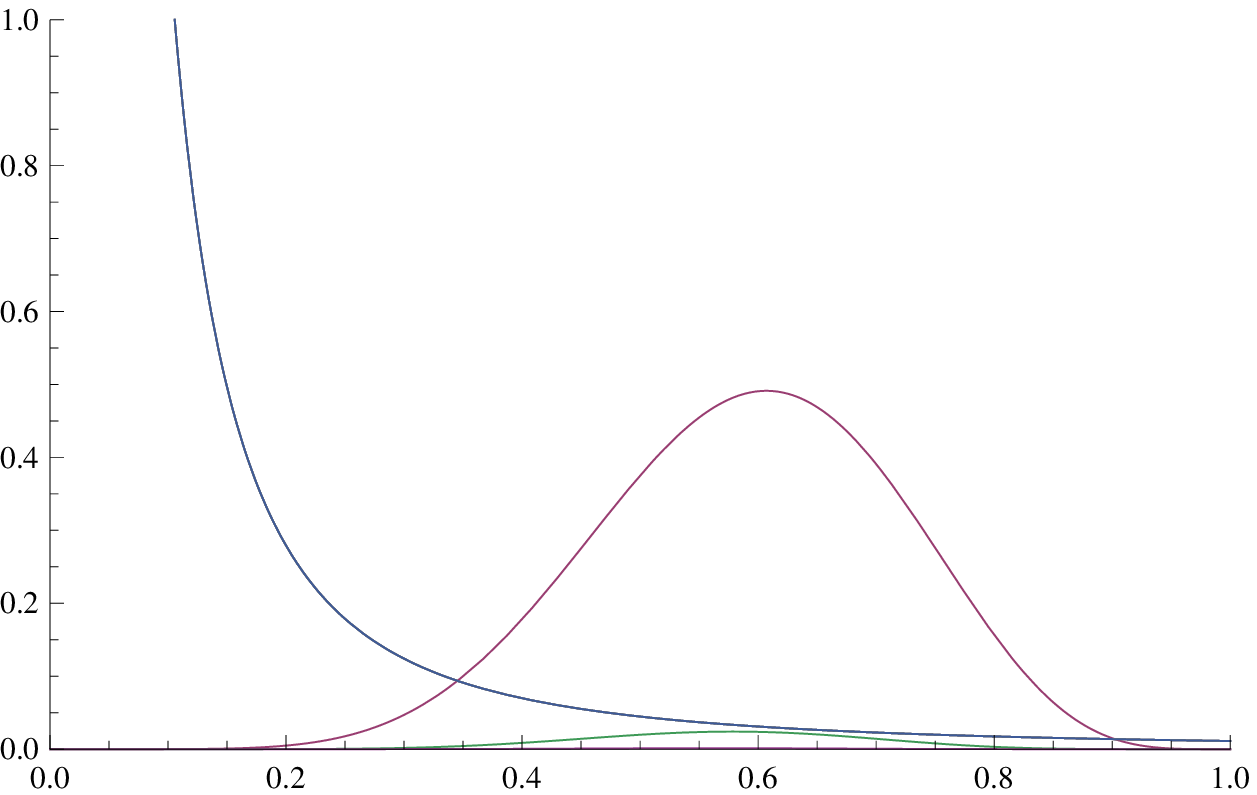}
\label{} } \subfigure[$F_2(x)$ function]{
\includegraphics[scale=0.45]{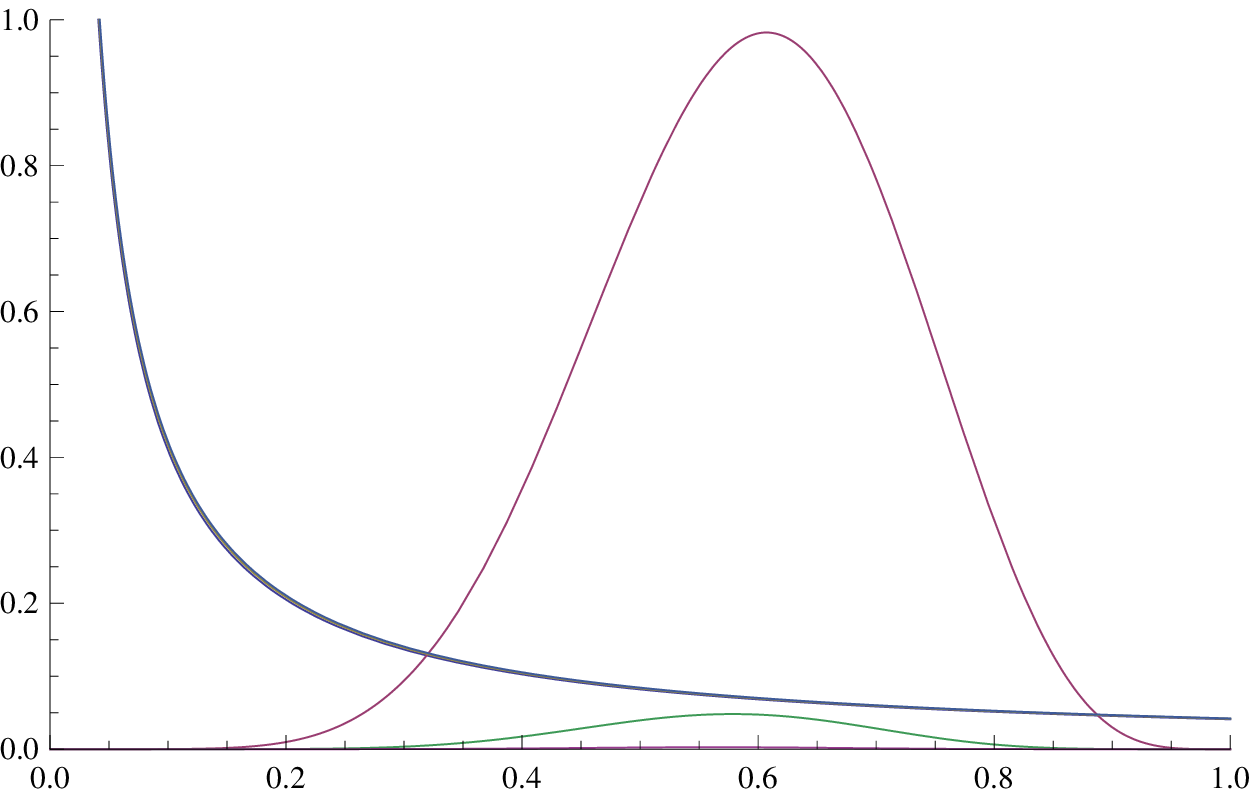}
\label{} } \subfigure[$b_1(x)$ function]{
\includegraphics[scale=0.45]{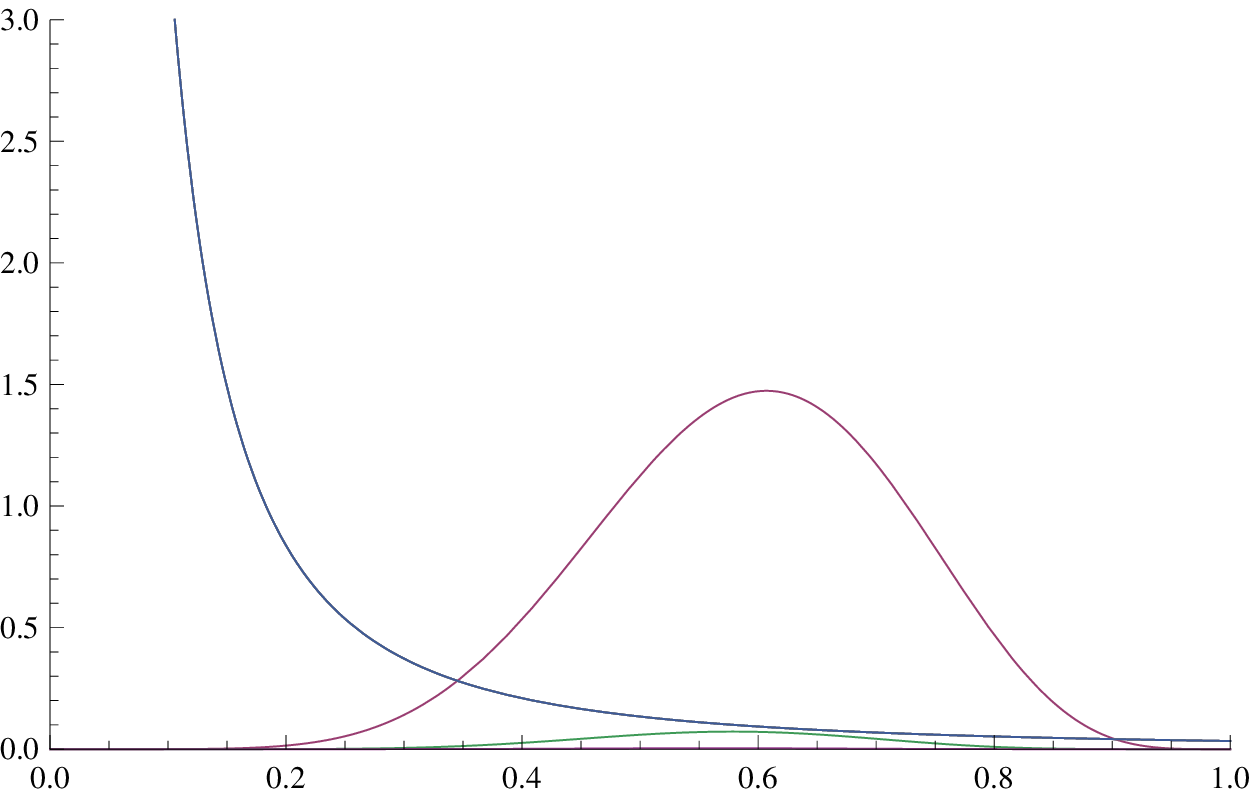}
\label{} } \subfigure[$b_2(x)$ function]{
\includegraphics[scale=0.45]{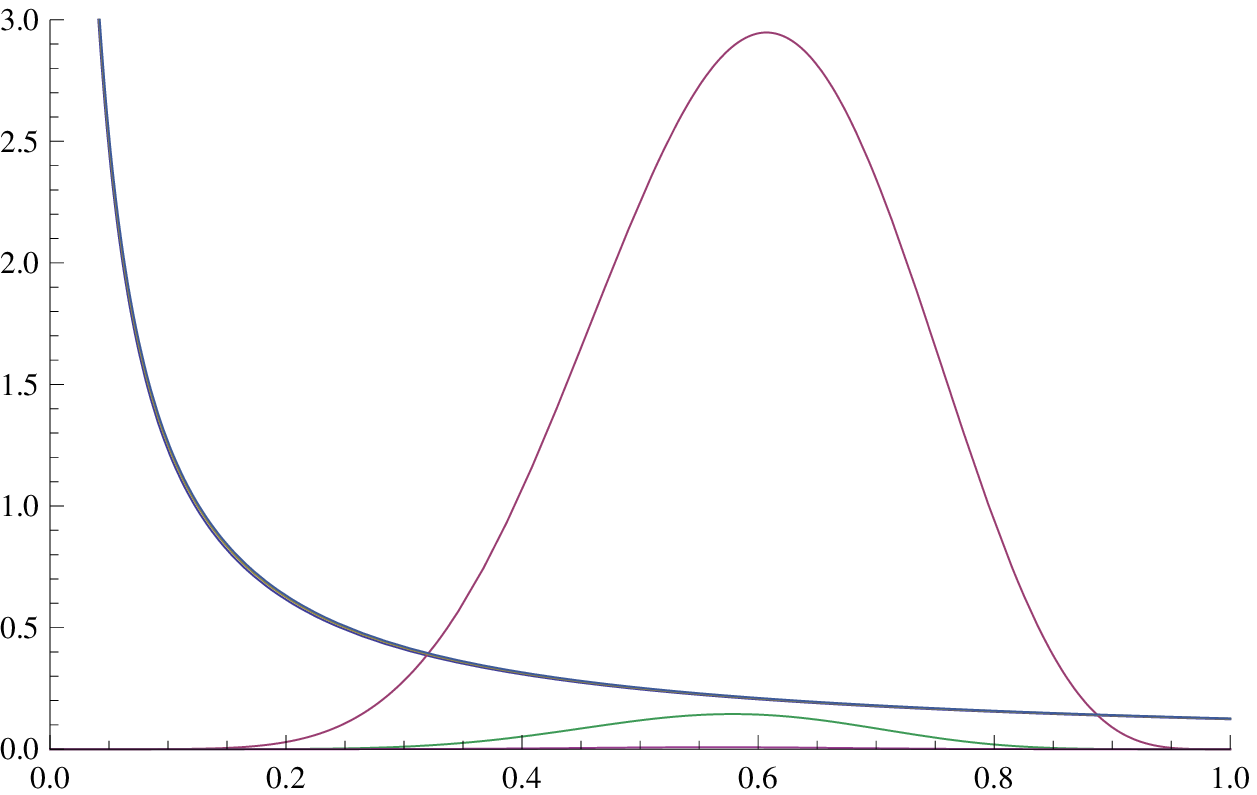}
\label{} } \caption{\small Structure functions $F_1$, $F_2$, $b_1$
and $b_2$ in terms of the Bjorken parameter $x$ for the $\ell=1, 2$
and $3$ vector mesons in the D4D8$\mathrm{\overline{D8}}$-brane
model. Bell-shaped curves correspond to the large-$x$ region (height
decreases as $\ell$ increases) while the hyperbolic curves
correspond to the small-$x$ region.} \label{functions2}
\end{figure}

\begin{figure}
\centering \subfigure[$F_1(x)$ function]{
\includegraphics[scale=0.45]{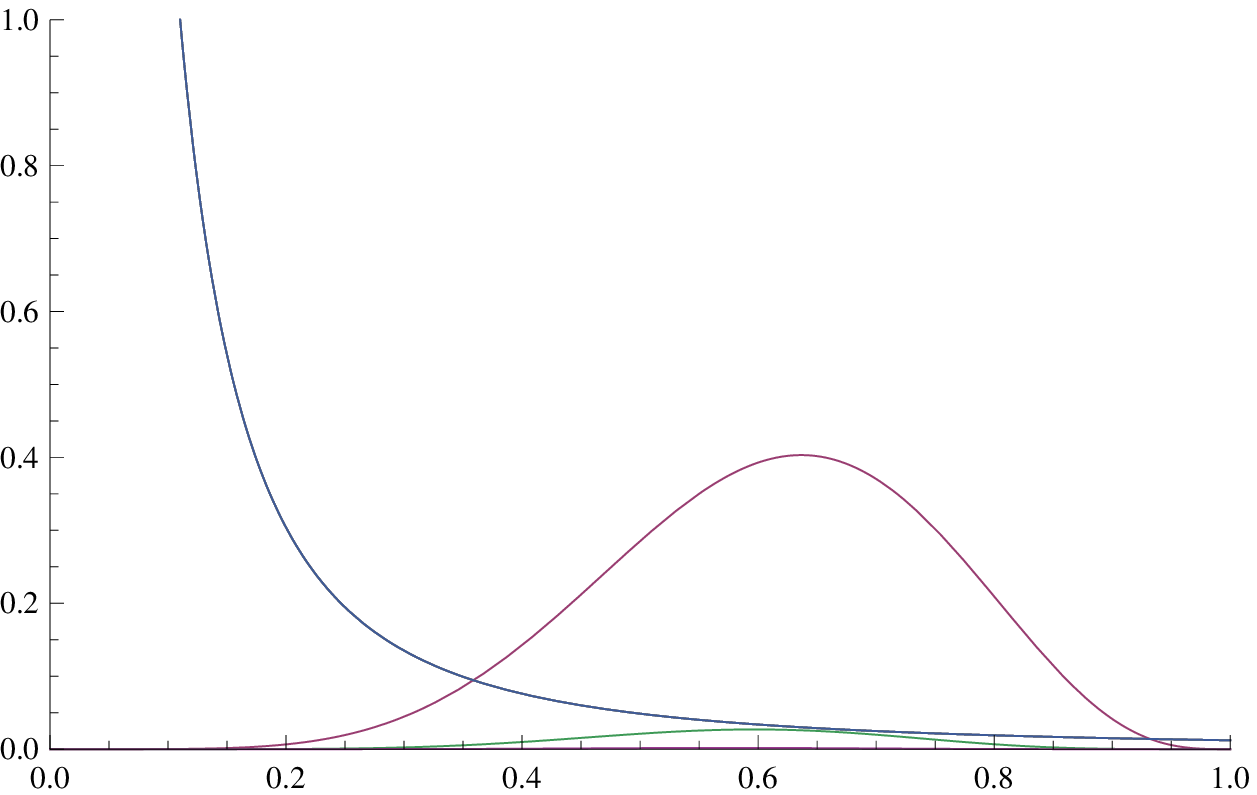}
\label{} } \subfigure[$F_2(x)$ function]{
\includegraphics[scale=0.45]{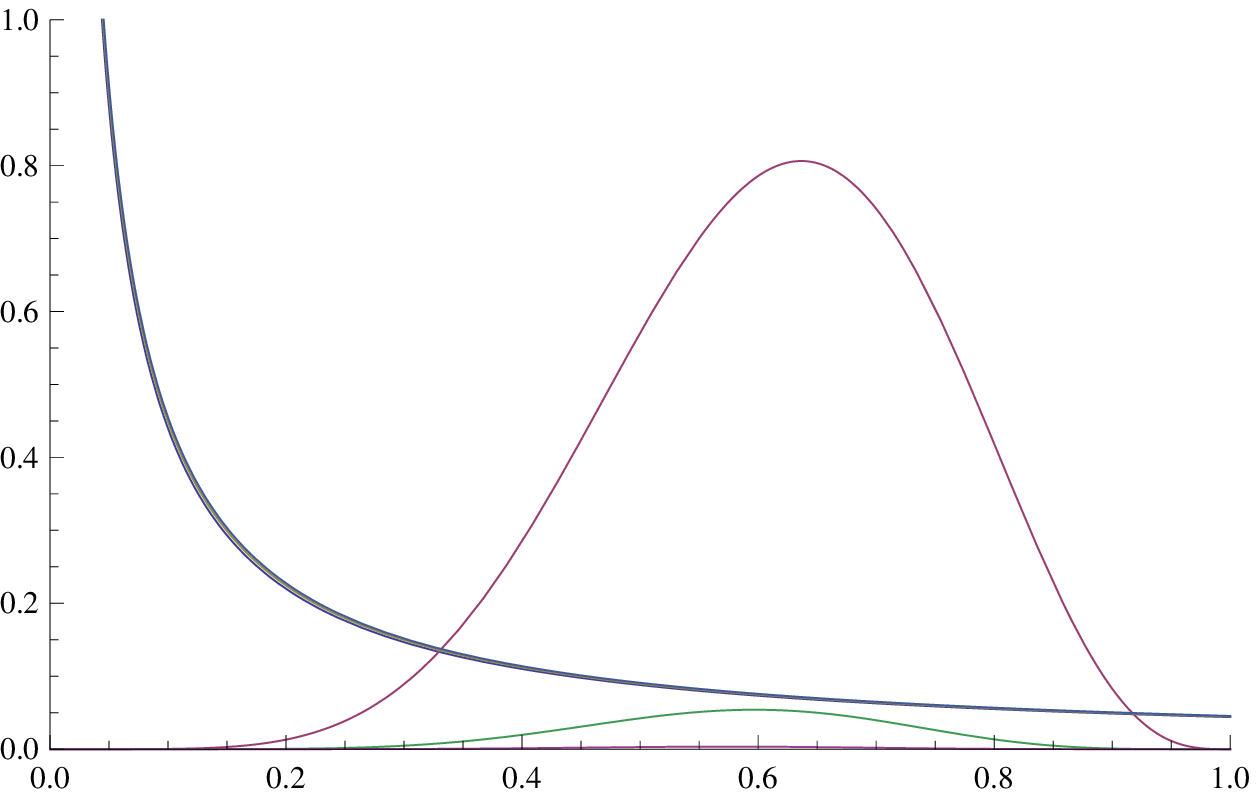}
\label{} } \subfigure[$b_1(x)$ function]{
\includegraphics[scale=0.45]{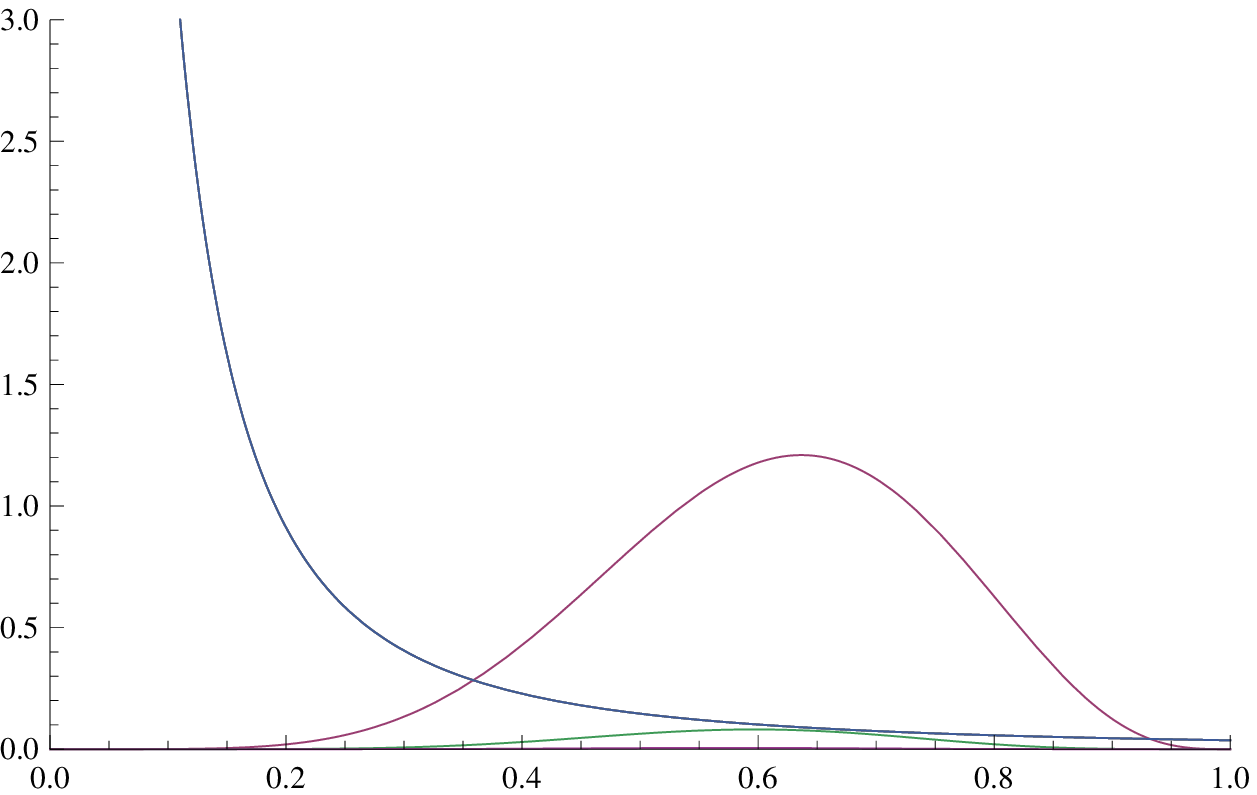}
\label{} } \subfigure[$b_2(x)$ function]{
\includegraphics[scale=0.45]{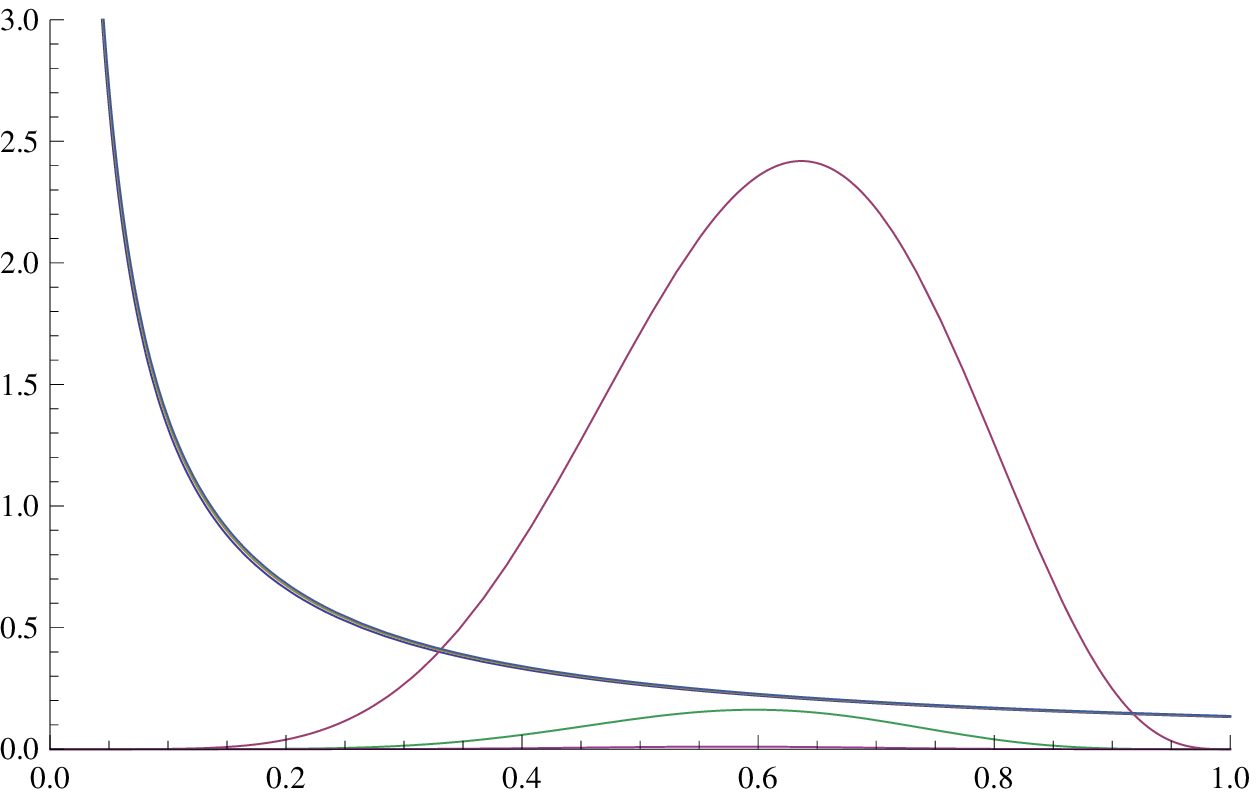}
\label{} } \caption{\small Structure functions $F_1$, $F_2$, $b_1$
and $b_2$ in terms of the Bjorken parameter $x$ for the $\ell=1, 2$
and $3$ vector mesons in the D4D6$\mathrm{\overline{D6}}$-brane
model. Bell-shaped curves correspond to the large-$x$ region (height
decreases as $\ell$ increases) while the hyperbolic curves
correspond to the small-$x$ region.} \label{functions3}
\end{figure}

Let us begin with the case of vector mesons. Figures
\ref{functions1}, \ref{functions2} and \ref{functions3} show the
structure functions $F_1$, $F_2$, $b_1$ and $b_2$ for polarized
vector mesons as a function of the Bjorken parameter. We consider
the D3D7-, D4D8$\mathrm{\overline{D8}}$- and
D4D6$\mathrm{\overline{D6}}$-brane models, respectively. In each
case we focus on the lightest particles given by $\ell=1, 2, 3$. For
the small-$x$ region the structure functions $F_1$, $F_2$, $b_1$ and
$b_2$ are represented by a continuously decreasing curve, which has
very little dependence on the values of $\ell=1, 2, 3$. On the other
hand, for large $x$-values all the figures show a set of bell-shaped
curves with maxima around $x \simeq 0.6$. The height of these curves
decreases as $\ell$ increases. The normalization constants are
specified in the next section and, they have been chosen in order to
have the best fitting to the lattice QCD results of the first three
moments of the $\rho$ meson\footnote{We omit the possible dependence
of these constants with $\ell$ since we compare with the moments of
the structure functions for $\ell=1$, {\it i.e.} the moments
corresponding to the $\rho$ meson, so the curves for $\ell > 1$ have
been drawn for the same values of the constants as for $\ell=1$.}.
These moments are shown in table 5 and discussed below. Recall that
in the small-$x$ parameter region there is a model-dependent and
$\ell$-dependent factor in the structure functions.
All the structure functions behave similarly. On the one hand, the
$x>0.2$ region is dominated by a smooth bell-shaped function with
maximum around $x \simeq 0.6$ (the specific value depends on $\ell$,
as it can be seen from equations (\ref{strfunctabel}) and
(\ref{Axgrande})). It approaches zero as $x \rightarrow 1$. On the
other hand, as $x$ decreases these bell-shaped functions also
decrease, and the small-$x$ structure functions defined in equations
(\ref{funcxchico}), (\ref{I1}) and (\ref{I2}) dominate. The $x<0.1$
region is described by functions $x^{-1}$ or $x^{-2}$ (depending on
which structure function one considers). The two type of functions
(the hyperbolic and the bell-shaped ones) overlap for $0.1<x<0.2$,
which is a transition region between both regimes of the Bjorken
parameter. This behavior is understood from the Callan-Gross type
relations in each range of the Bjorken parameter, as well as from
the fact that $b_i = 3F_i$ in both cases. Also, remember that when
$x$ becomes exponentially small this description fails because the
approximation that strings interact locally enough in order to use a
flat space amplitudes breaks down. In this case the correct behavior
should be obtained by considering a diffusion operator, and
ultimately the description of full string theory on a curved
background. The analysis in this parameter region is very difficult.
However, there is evidence that the divergence disappears (at least
for the glueball DIS \cite{Polchinski:2002jw}), leading to finite
moments of the structure functions, but this is beyond the scope of
this work. In the next section we will deal with this by introducing
a low-$x$ cutoff. We can see that for $\ell=1$ it gives similar
curves for the three models we consider, even if the normalization
constants are different for each model. It is interesting to see
that the structure functions at small $x$ have similar behavior for
$\ell=1, 2, 3$, while at large $x$ they get smaller as the parameter
$\ell$ increases. This is more evident in the
D4D8$\mathrm{\overline{D8}}$ and D4D6$\mathrm{\overline{D6}}$-brane
models than in the D3D7-brane model. In those cases, for $\ell=3$
the corresponding curve is one order of magnitude smaller than the
corresponding one for $\ell=1$ of the large-$x$ structure function.

We can also compare with the existing literature. In particular, at
large $x$ the curves for $F_1$ and $F_2$ have a similar bell-shaped
behavior as those presented in \cite{BallonBayona:2010ae}, although
the maxima in that reference occurs for $x \simeq 0.8$. In
\cite{BallonBayona:2010ae} the DIS structure functions for the
lightest vector mesons (they only consider the unpolarized case)
have been obtained by using the D4D8$\mathrm{\overline{D8}}$-brane
model, {\it i.e.} the $\rho$ meson and the $a_1$ axial-vector meson.
In fact, in that reference $F_1$ and $F_2$ structure functions have
been obtained in the range $0.2 < x < 1$.  We recall that the
calculations in that reference are different from ours since they
considered a four-dimensional effective Lagrangian only describing
the $1/\sqrt{\lambda} \ll x < 1$ range. Obviously, this describes
only a part of the problem. Instead, in
\cite{Koile:2011aa,Koile:2013hba} we have derived the corresponding
Lagrangians for the three holographic dual models from first
principles from type IIA and type IIB supergravities within this
parametric region, while for small $x$ in our paper
\cite{Koile:2014vca} we have obtained the structure functions by
using type IIA and type IIB string theories. From the structure
functions obtained in \cite{BallonBayona:2010ae} they found the
ratio $F_2/(2 F_1) \approx 1$, satisfying this Callan-Gross type
relation (as expected from the supergravity calculation
\cite{Polchinski:2002jw} within this parametric range, note that it
has not the factor $x$ multiplying $F_1$ which agrees with our
results) within the interval $0.4 < x < 0.6$ and momentum transfer
10 GeV$^2 < |q^2| < 80$ GeV$^2$. However, in
\cite{Koile:2011aa,Koile:2013hba,Koile:2014vca} we have proved that
a modified version of the Callan-Gross relation holds within the
whole range of the Bjorken parameter and for any value of the
momentum transfer, in fact there appear two Callan-Gross type
relations: $F_2=2 F_1$ for $1/\sqrt{\lambda} \ll x < 1$
(supergravity) and $F_2=2 x F_1 \delta$ for
$\exp(-\sqrt{\lambda})\ll x \ll 1/\sqrt{\lambda}$ (from superstring
theory, see section 4.2). Note that $0.5 < \delta <1$ for all mesons
described by the models we consider. Moreover, we have also obtained
new exact relations among the structure functions for scalar and for
polarized vector mesons in the planar limit at strong 't Hooft
coupling. In addition, in reference \cite{BallonBayona:2010ae} only
structure functions for the lowest states of vector and axial vector
mesons were studied, while our results hold for any scalar or
polarized vector meson.

\begin{figure}
\centering \subfigure[D3D7 $\&$ $\ell=1$]{
\includegraphics[scale=0.3]{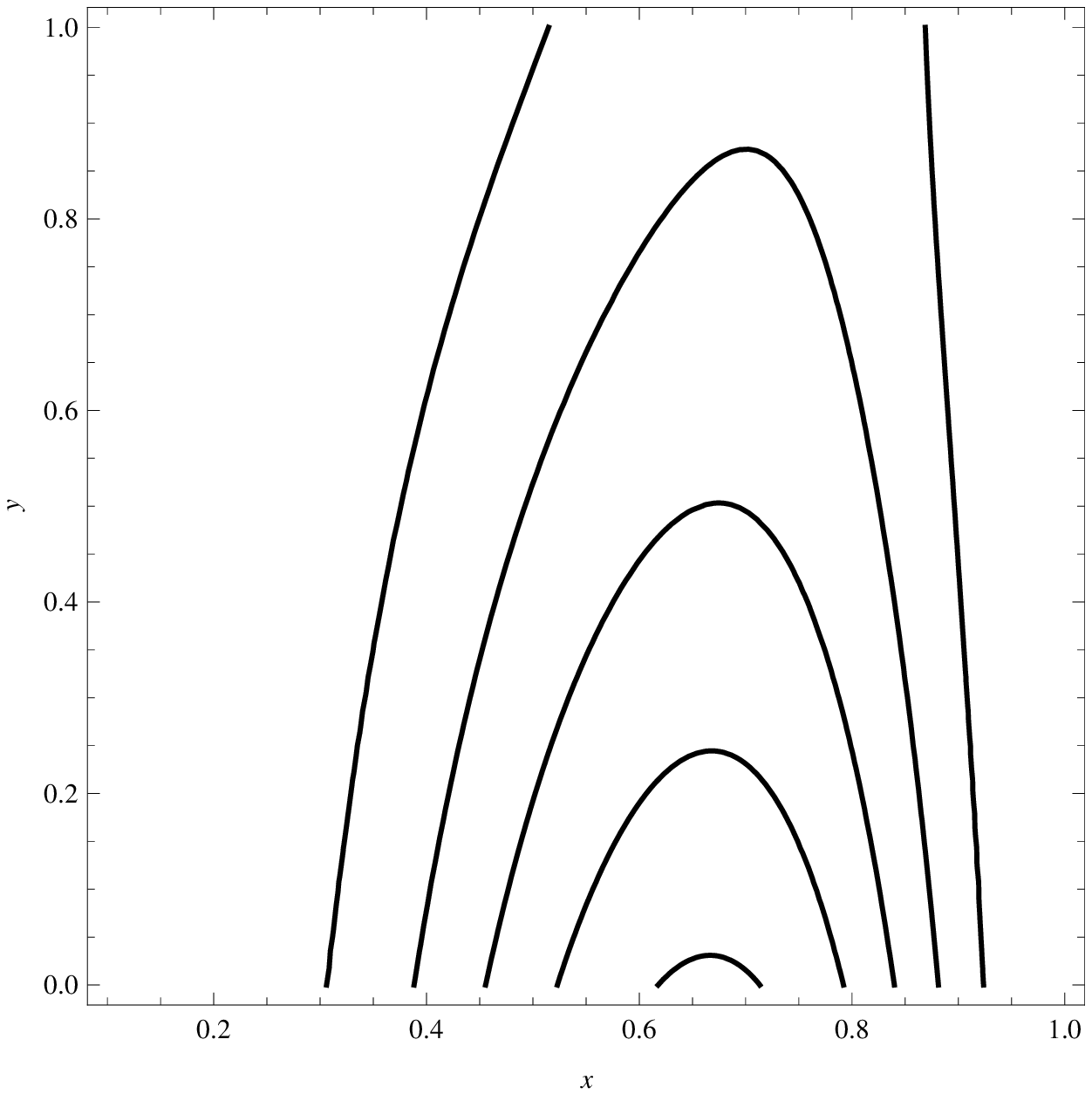}
\label{} } \subfigure[D4D6$\mathrm{\overline{D6}}$ $\&$ $\ell=1$]{
\includegraphics[scale=0.3]{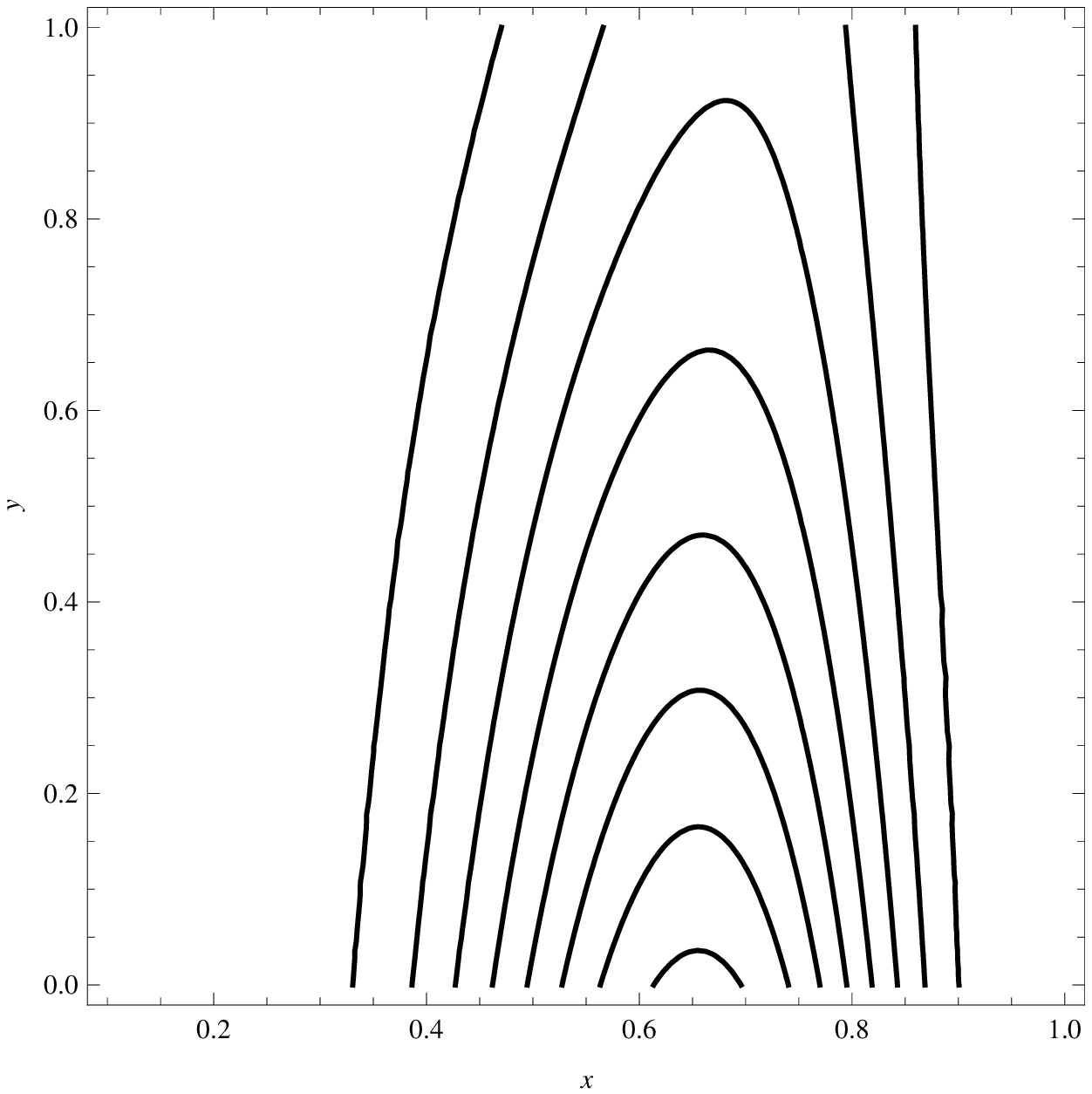}
\label{} } \subfigure[D4D8$\mathrm{\overline{D8}}$ $\&$ $\ell=1$]{
\includegraphics[scale=0.3]{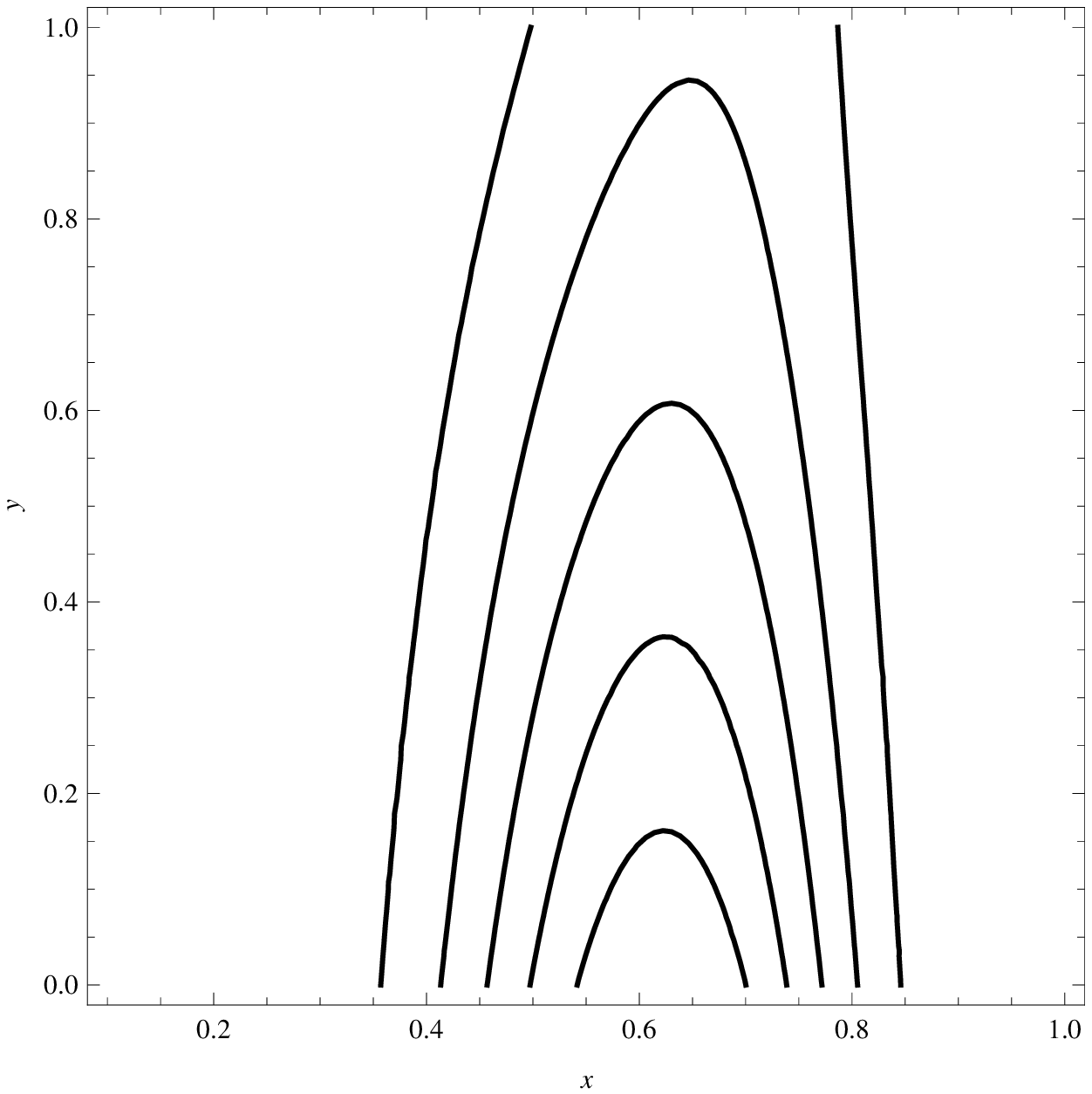}
\label{} } \subfigure[D3D7 $\&$ $\ell=2$]{
\includegraphics[scale=0.3]{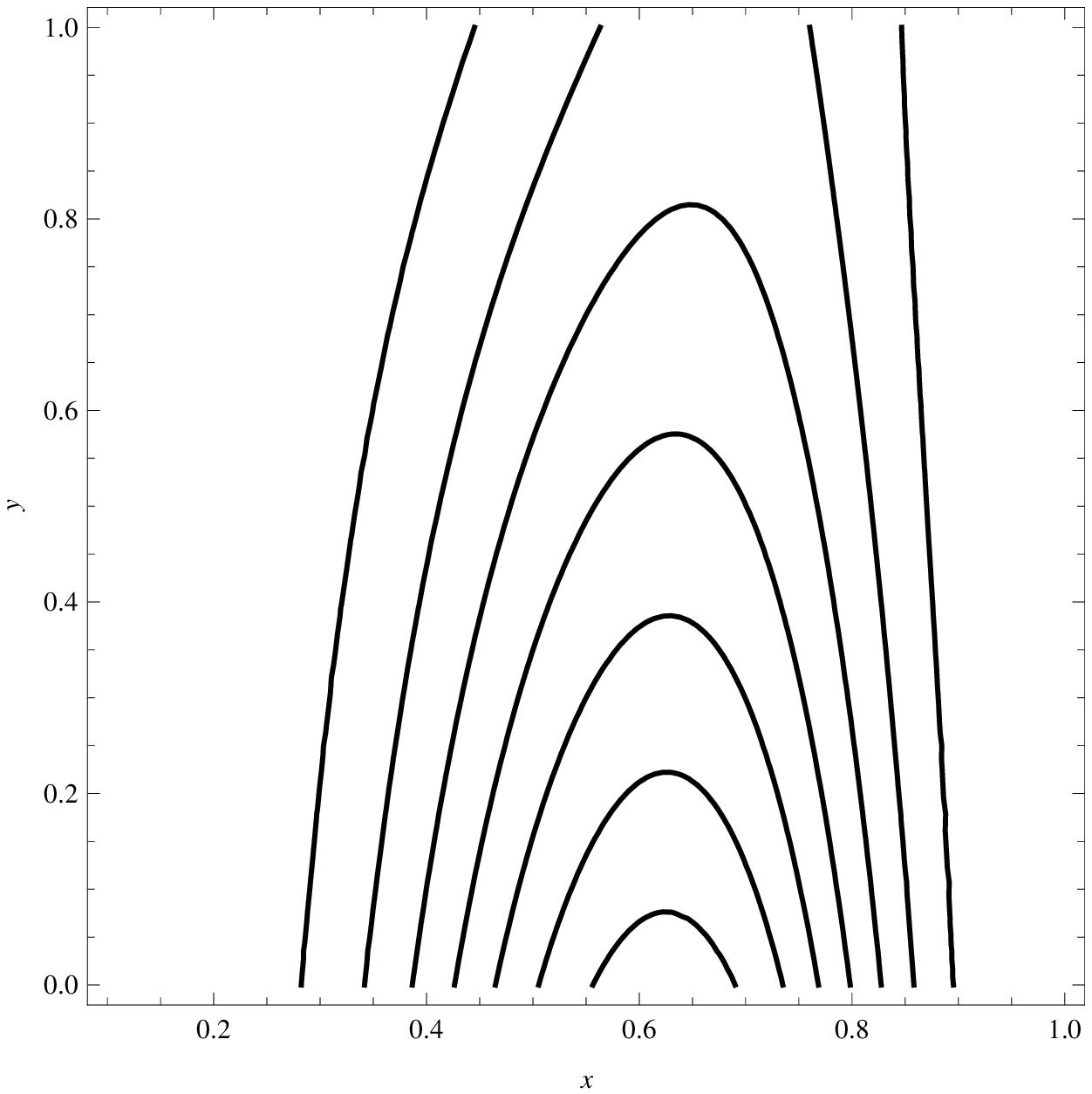}
\label{} } \subfigure[D4D6$\mathrm{\overline{D6}}$ $\&$ $\ell=2$]{
\includegraphics[scale=0.3]{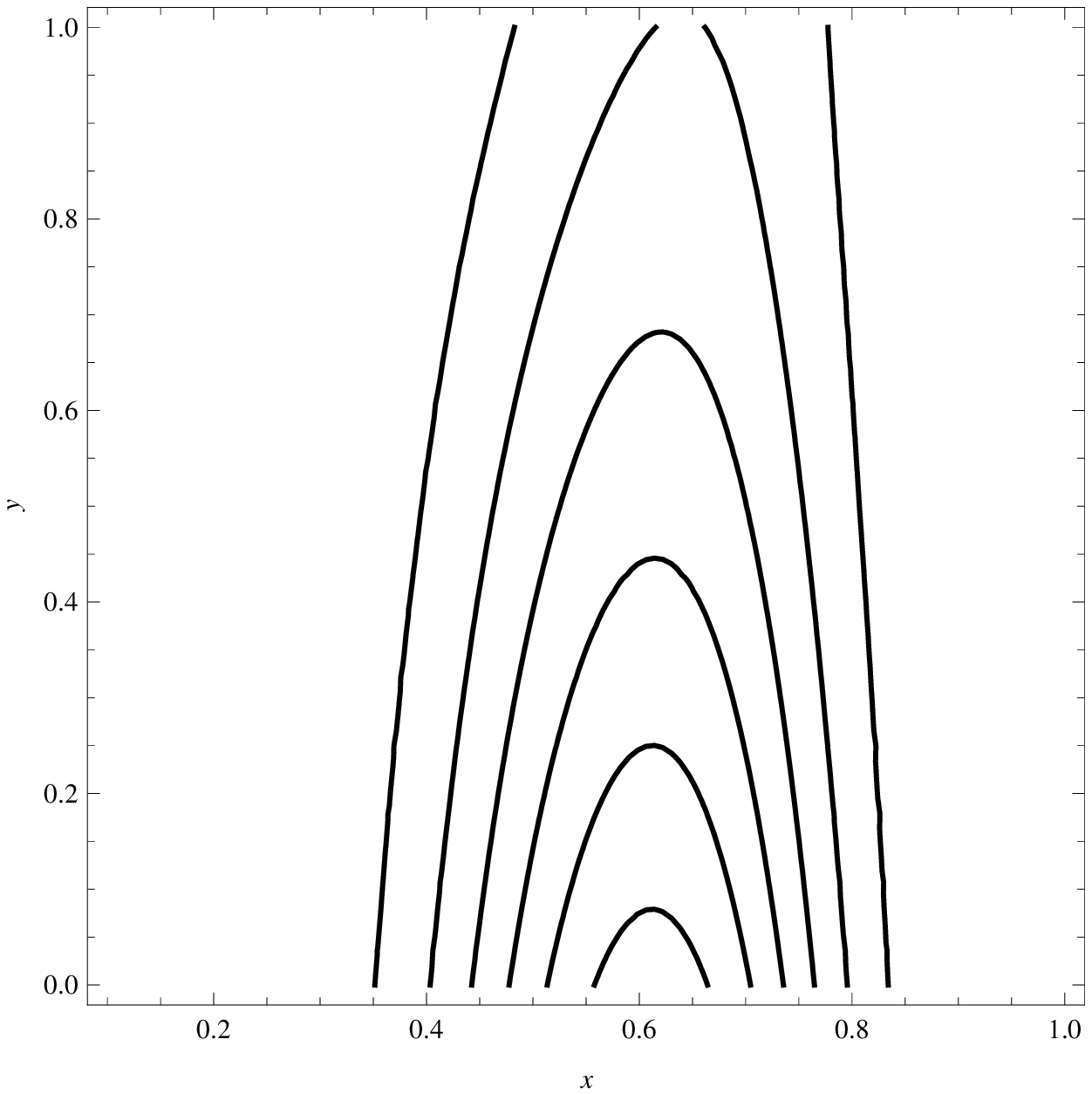}
\label{} } \subfigure[D4D8$\mathrm{\overline{D8}}$ $\&$ $\ell=2$]{
\includegraphics[scale=0.3]{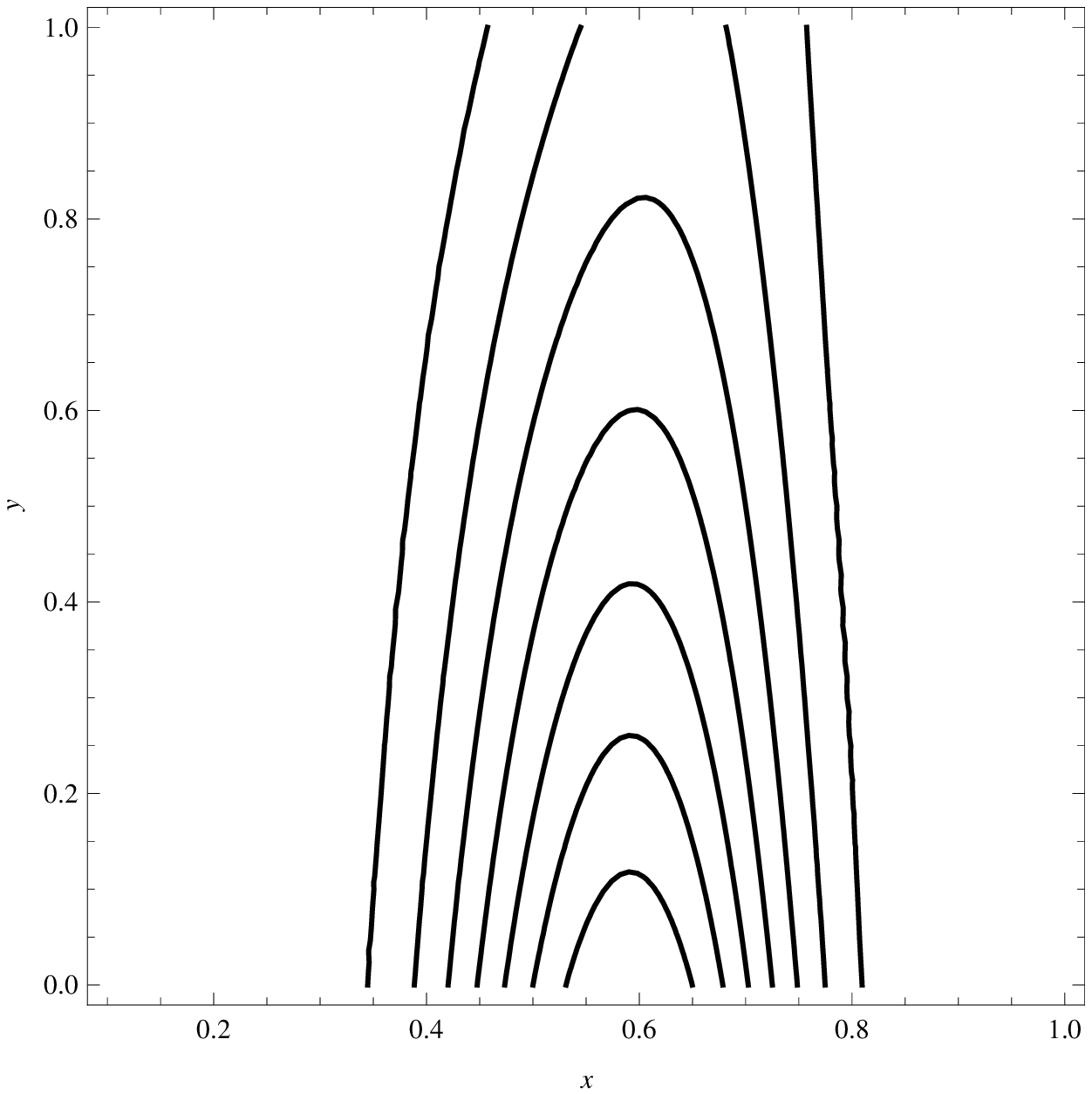}
\label{} } \subfigure[D3D7 $\&$ $\ell=3$]{
\includegraphics[scale=0.3]{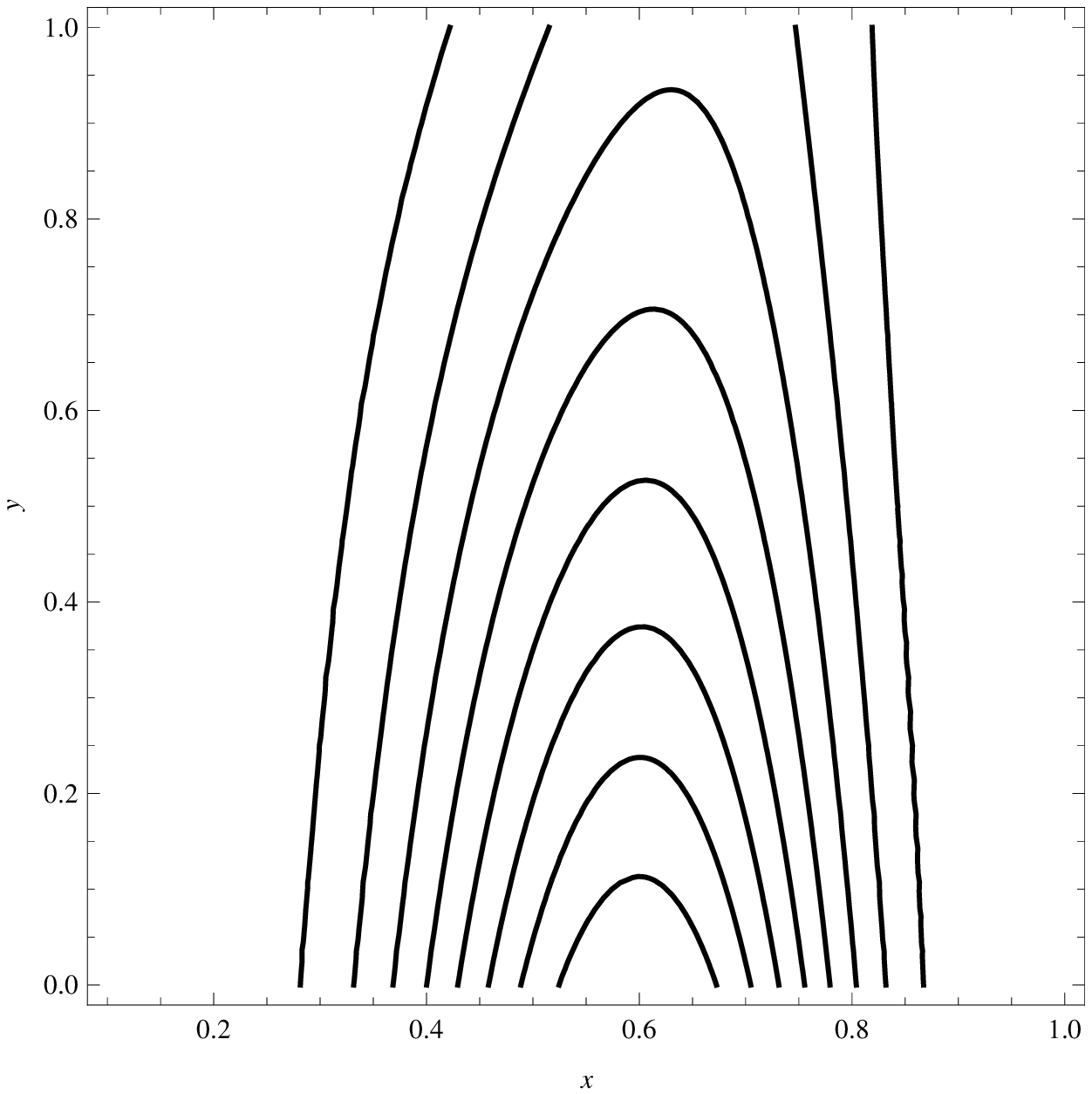}
\label{} } \subfigure[D4D6$\mathrm{\overline{D6}}$ $\&$ $\ell=3$]{
\includegraphics[scale=0.3]{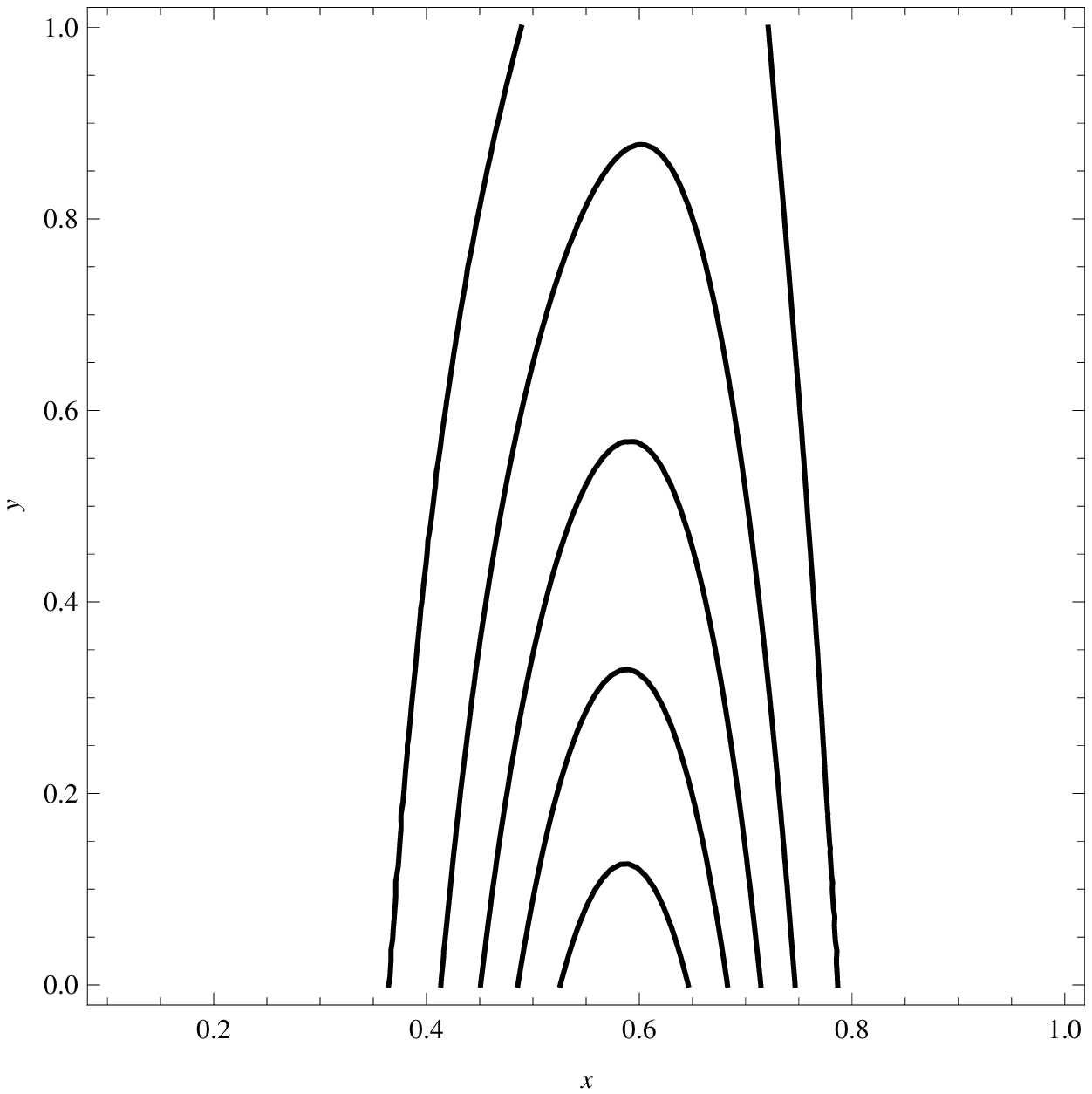}
\label{} } \subfigure[D4D8$\mathrm{\overline{D8}}$ $\&$ $\ell=3$]{
\includegraphics[scale=0.3]{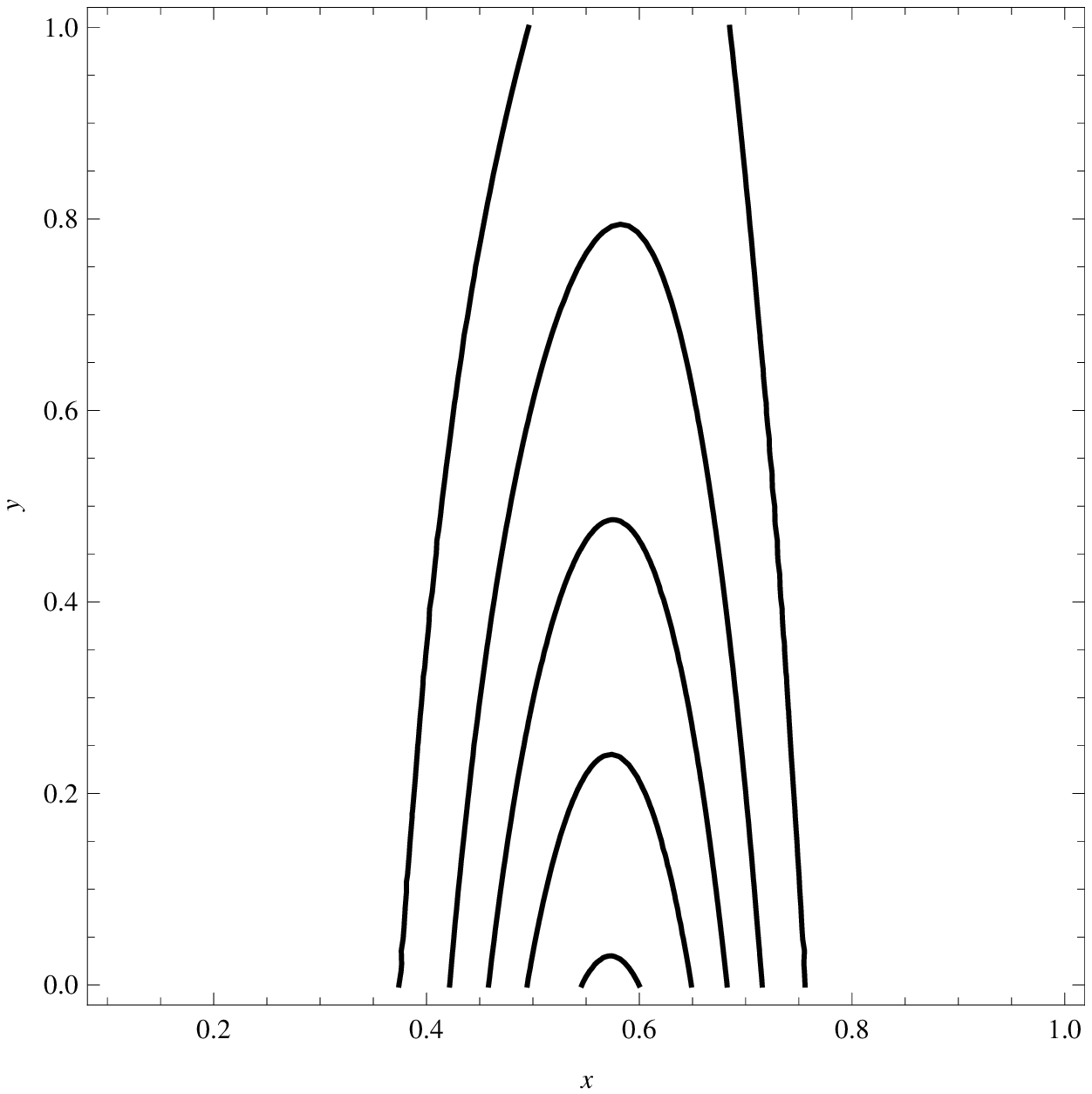}
\label{} } \subfigure[D3D7 $\&$ $\ell=10$]{
\includegraphics[scale=0.3]{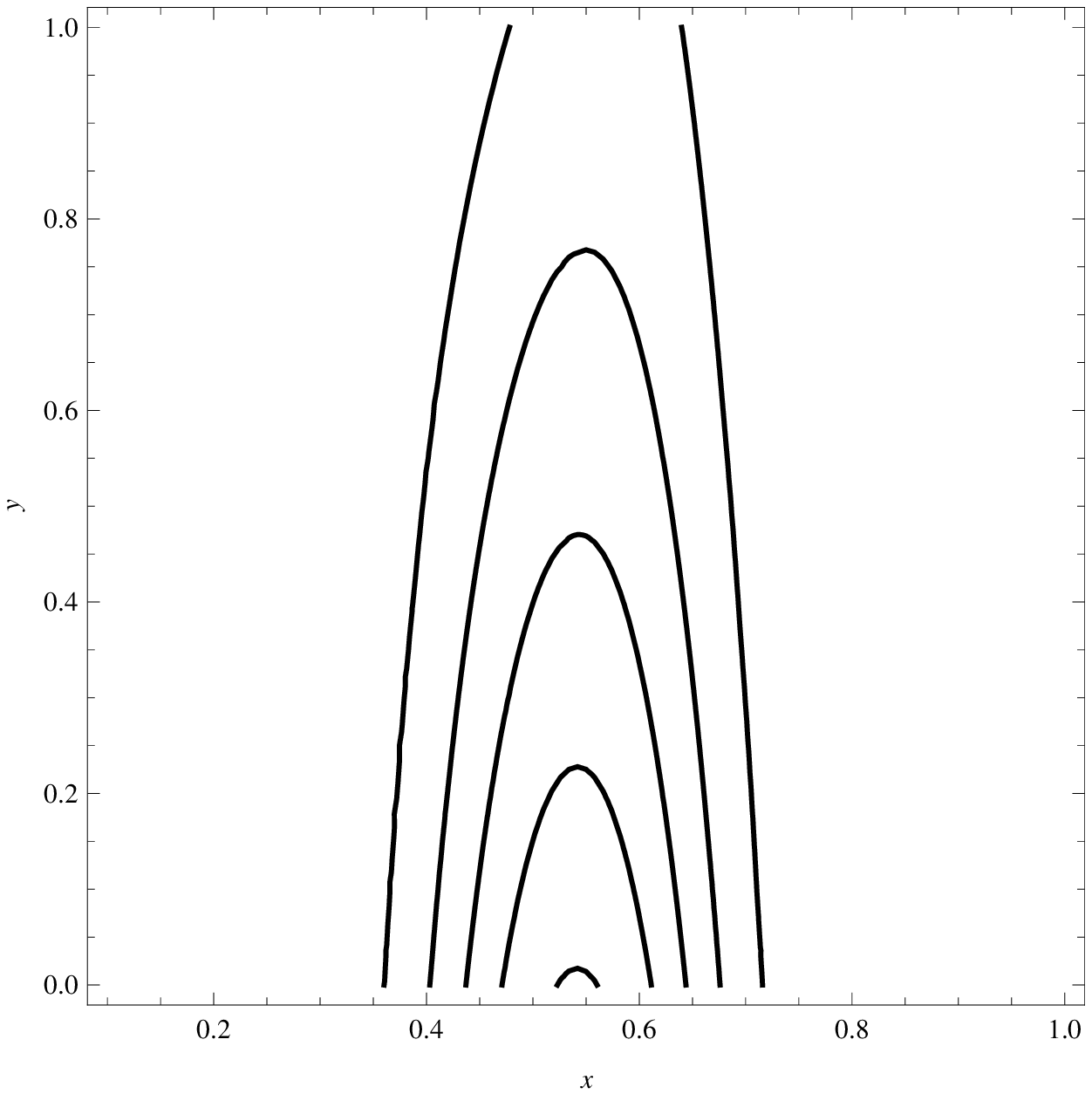}
\label{} } \subfigure[D4D6$\mathrm{\overline{D6}}$ $\&$ $\ell=10$]{
\includegraphics[scale=0.3]{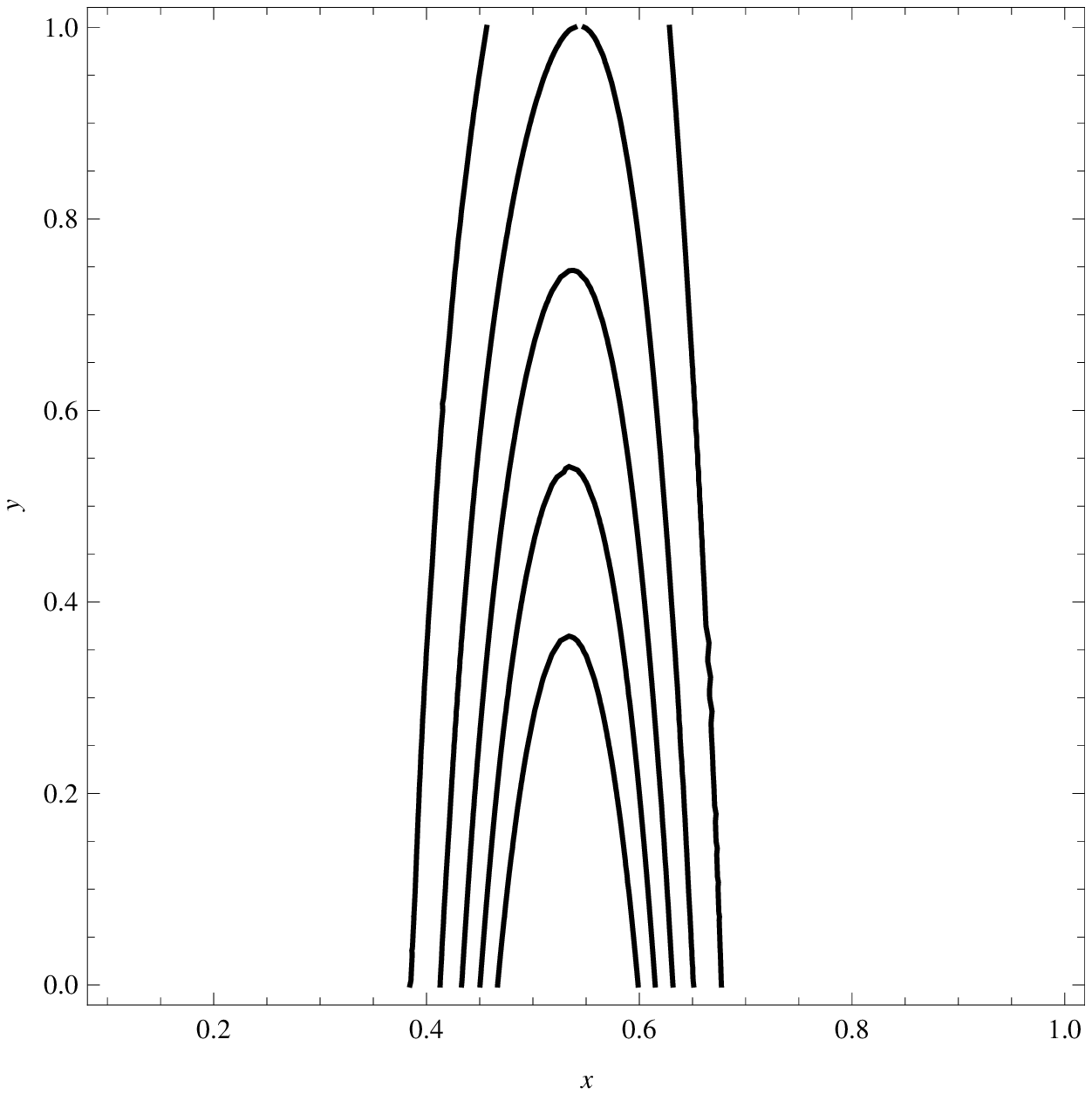}
\label{} } \subfigure[D4D8$\mathrm{\overline{D8}}$ $\&$ $\ell=10$]{
\includegraphics[scale=0.3]{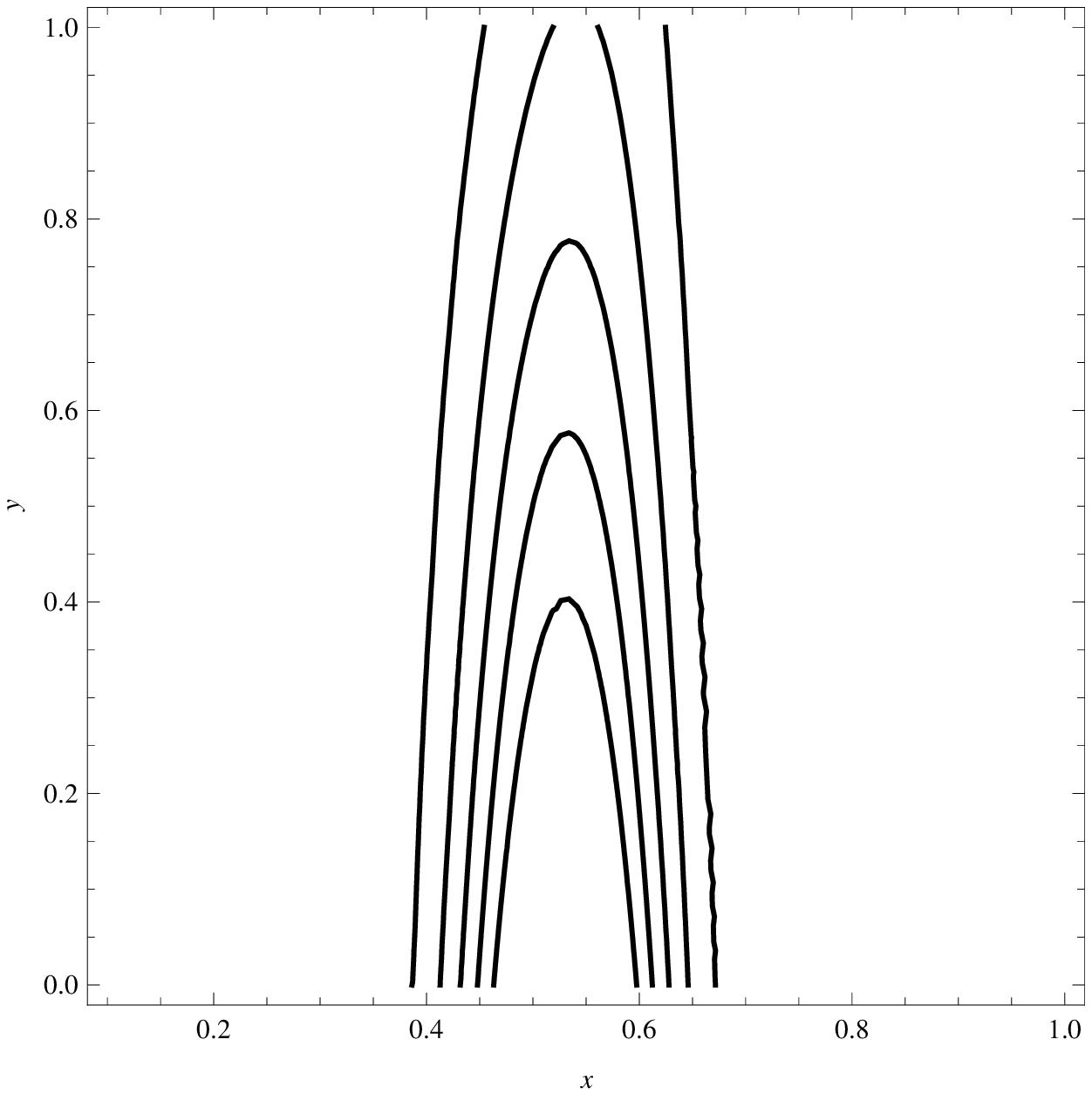}
\label{} } \caption{\small Contour line maps of DIS differential
cross section for unpolarized holographic vector mesons for the
three flavor-brane models we consider, in the region
$1/\sqrt{\lambda} \ll x < 1$. Different mesons are labeled with
$\ell=1, 2, 3$ and $10$ (displayed on the four rows). The
differential cross sections have been normalized as explained in the
main text. The horizontal axis represents the Bjorken parameter
within the range $[0.1, \, 1]$, while the vertical one corresponds
to the variable $y$, $[0, \, 1]$. Broader curves correspond to lower
values of the DIS cross section.} \label{Crosssection1}
\end{figure}

\begin{figure}
\centering \subfigure[D3D7 $\&$ $\ell=1$]{
\includegraphics[scale=0.3]{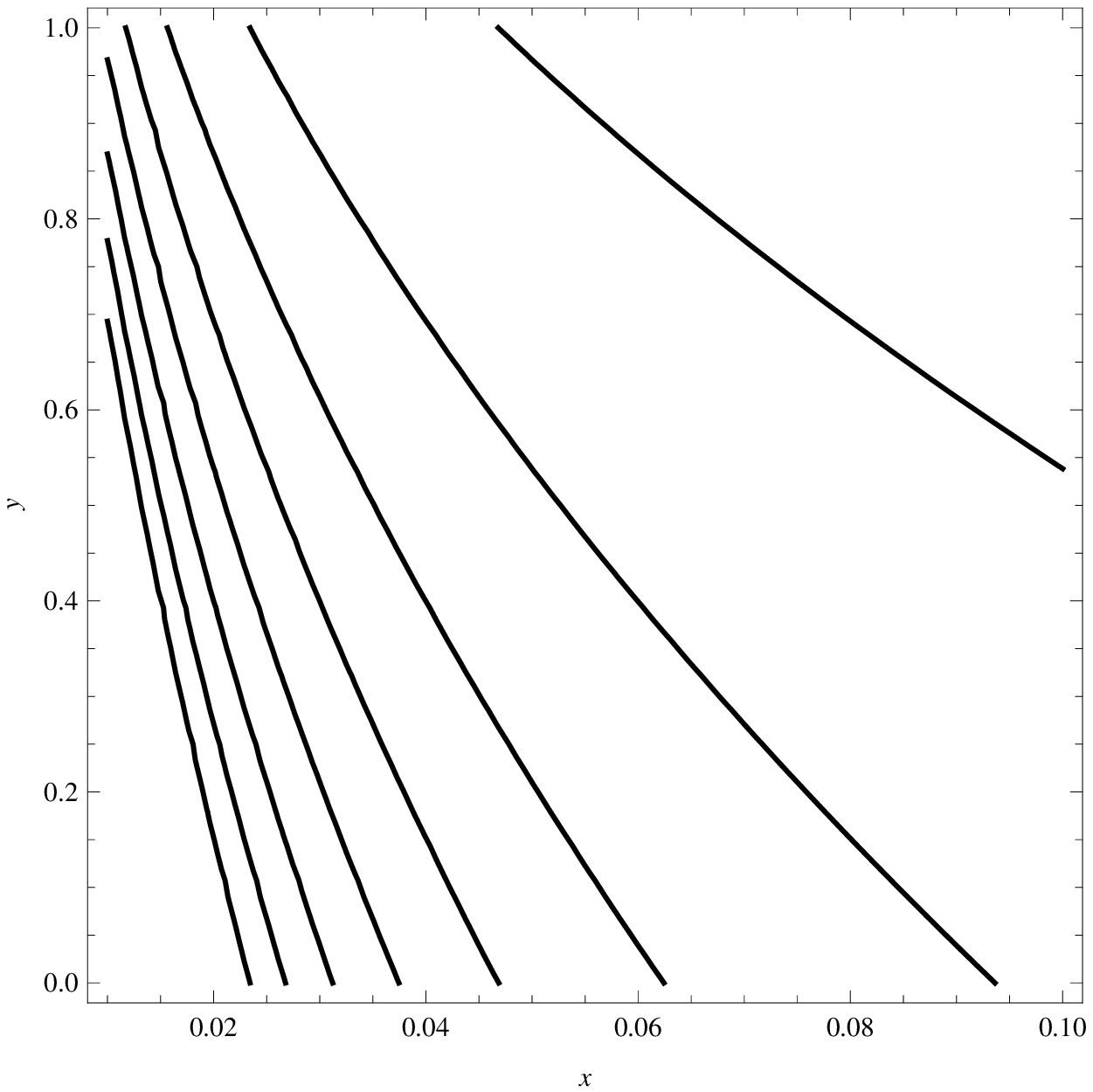}
\label{} } \subfigure[D4D6$\mathrm{\overline{D6}}$ $\&$ $\ell=1$]{
\includegraphics[scale=0.3]{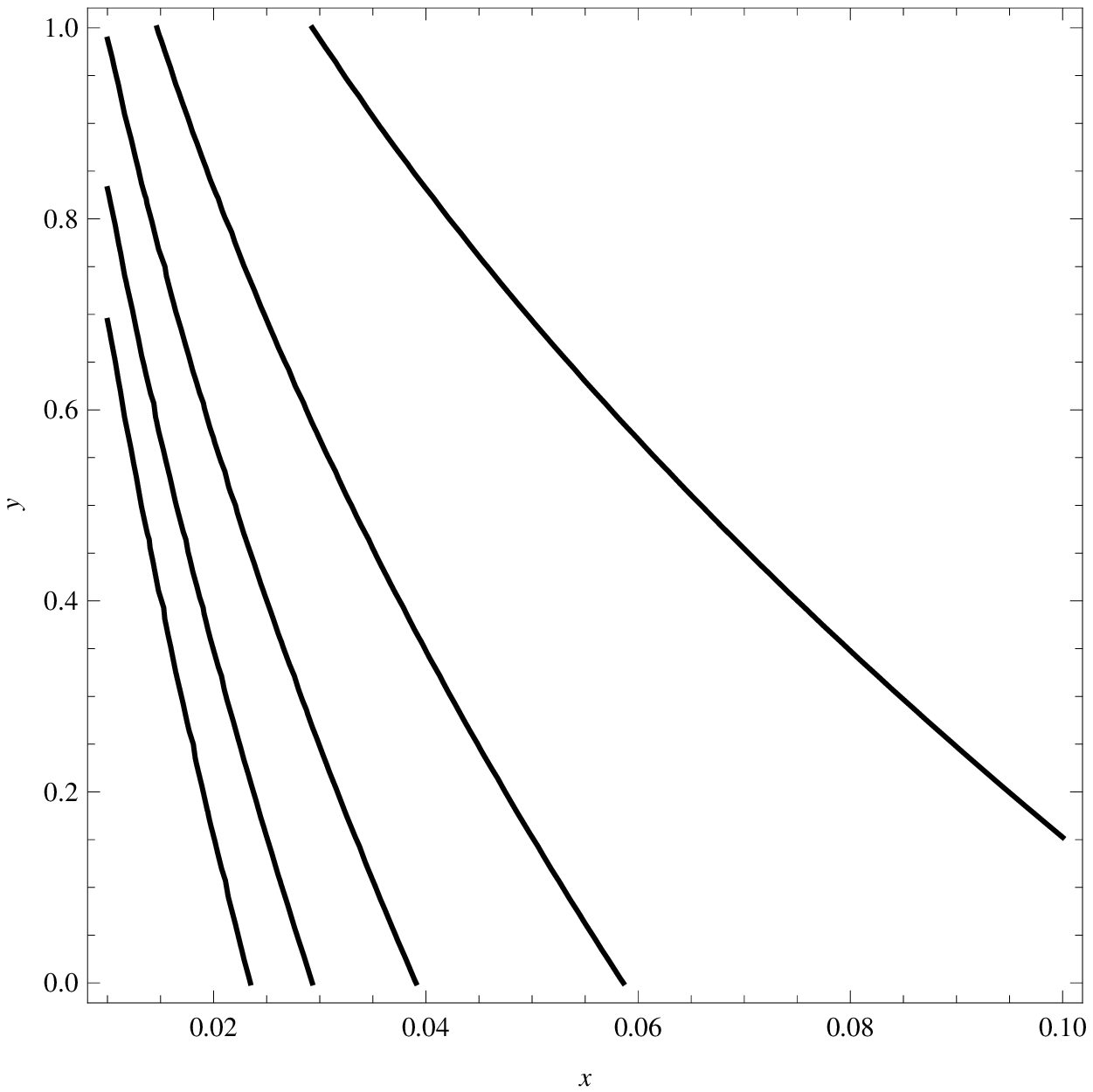}
\label{} } \subfigure[D4D8$\mathrm{\overline{D8}}$ $\&$ $\ell=1$]{
\includegraphics[scale=0.3]{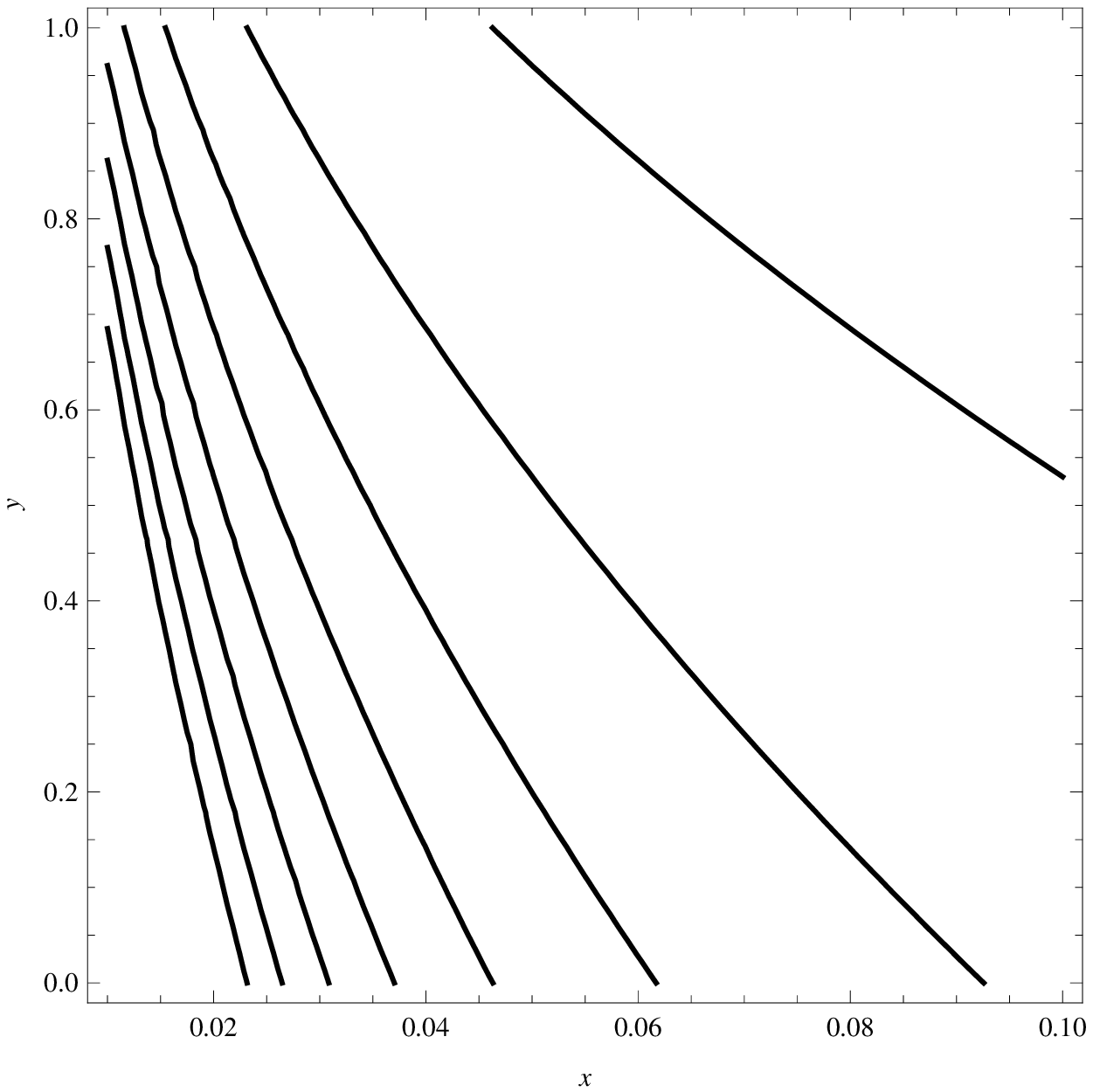}
\label{} } \subfigure[D3D7 $\&$ $\ell=2$]{
\includegraphics[scale=0.3]{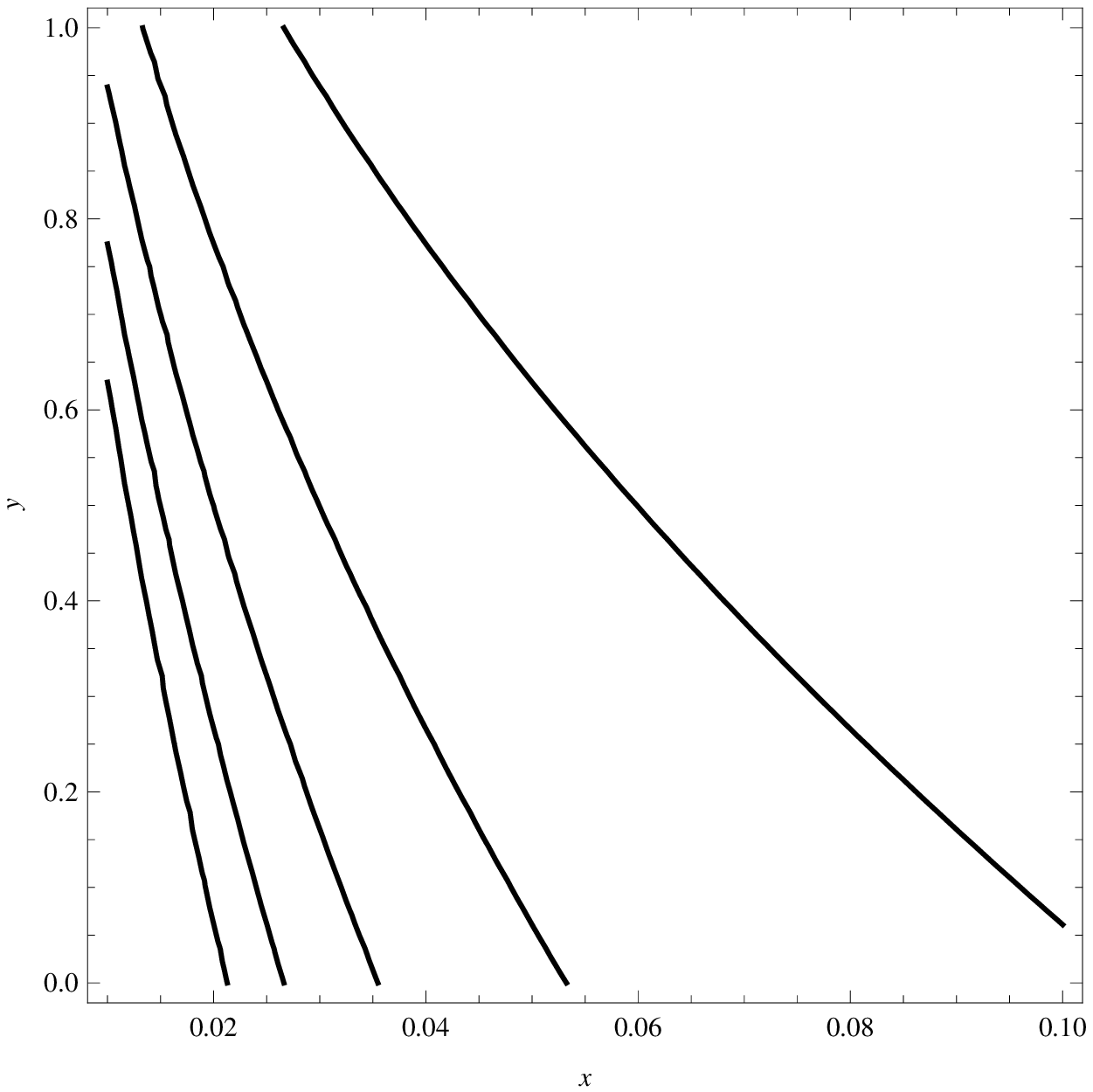}
\label{} } \subfigure[D4D6$\mathrm{\overline{D6}}$ $\&$ $\ell=2$]{
\includegraphics[scale=0.3]{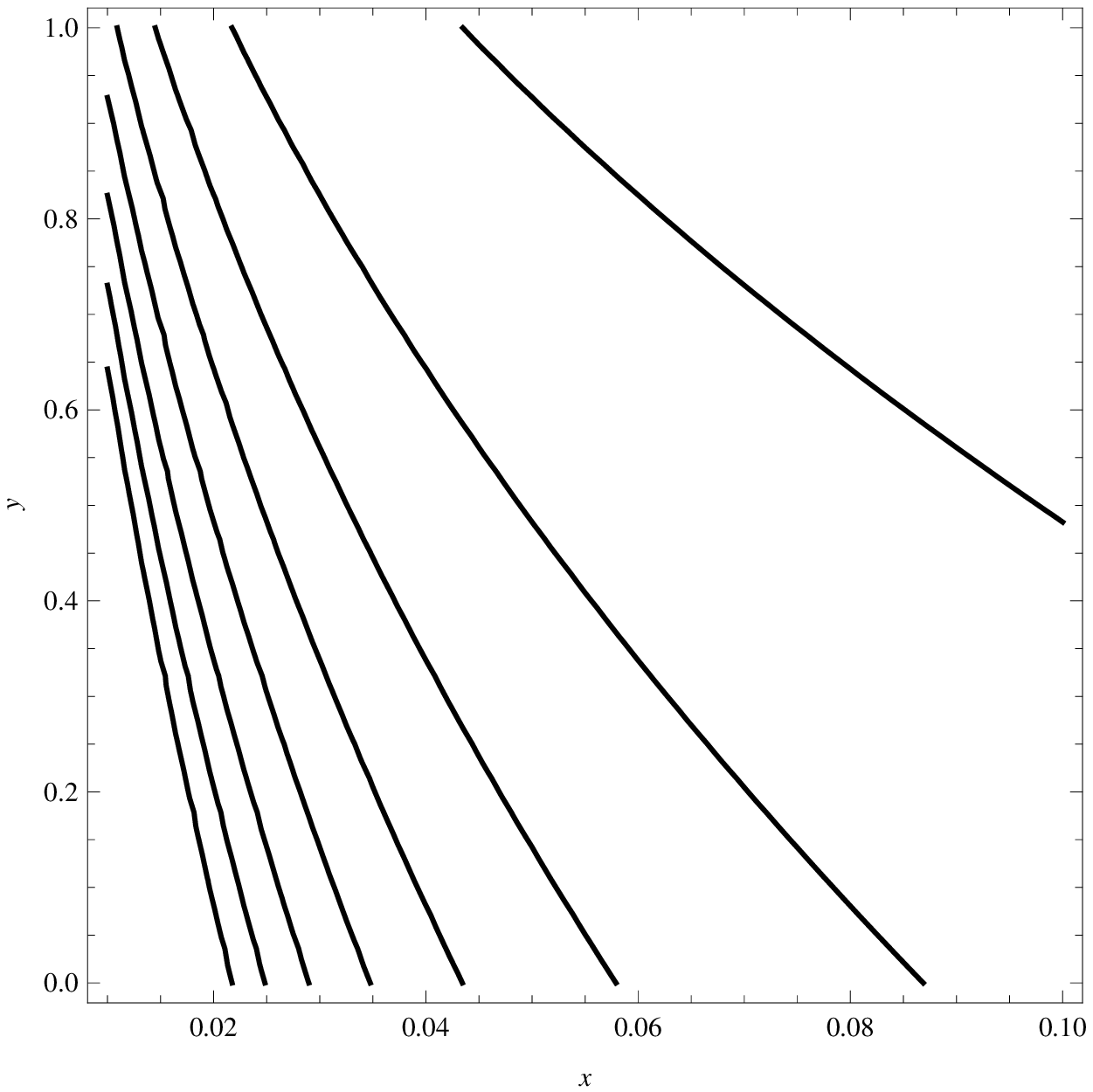}
\label{} } \subfigure[D4D8$\mathrm{\overline{D8}}$ $\&$ $\ell=2$]{
\includegraphics[scale=0.3]{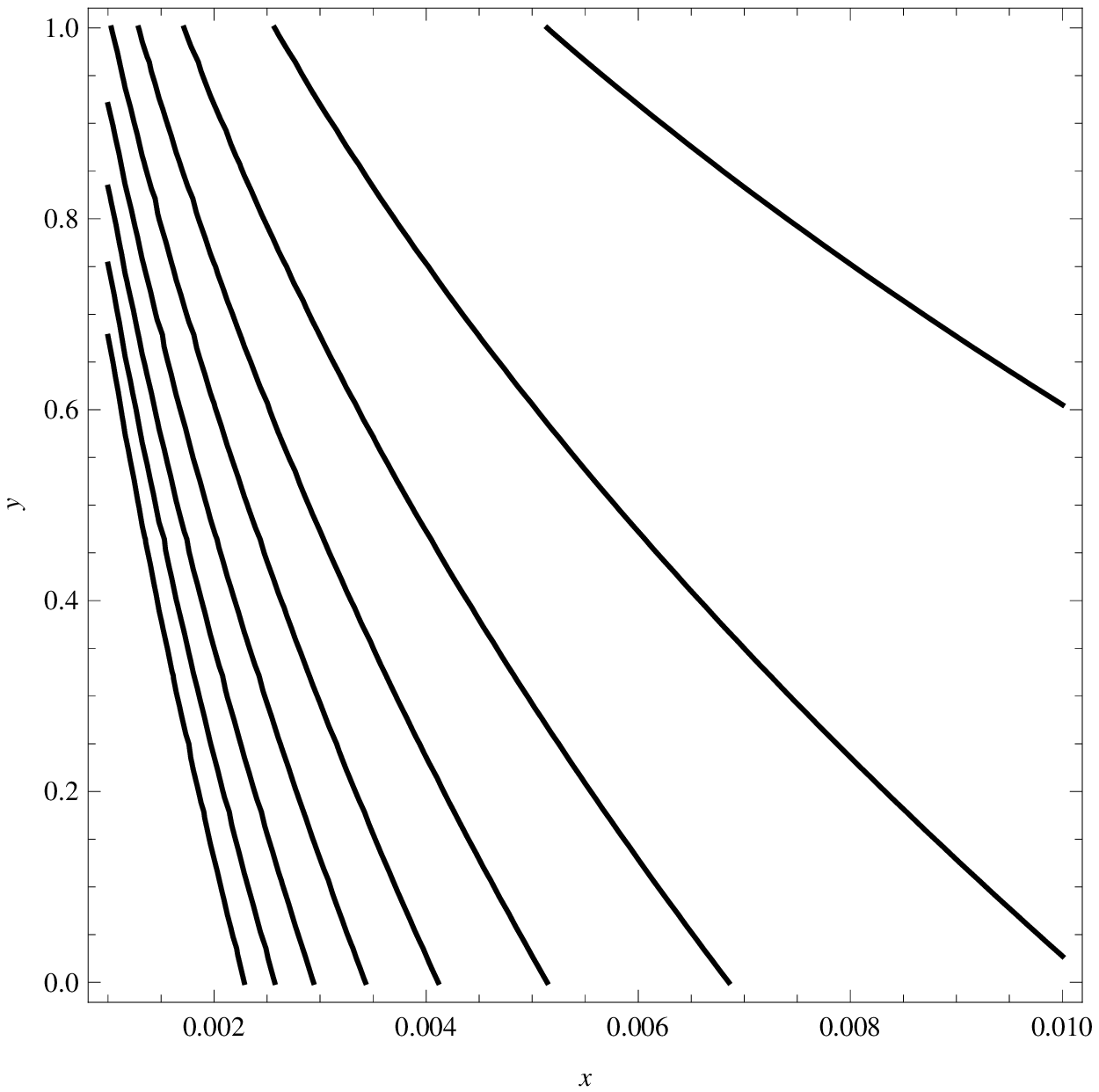}
\label{} } \subfigure[D3D7 $\&$ $\ell=3$]{
\includegraphics[scale=0.3]{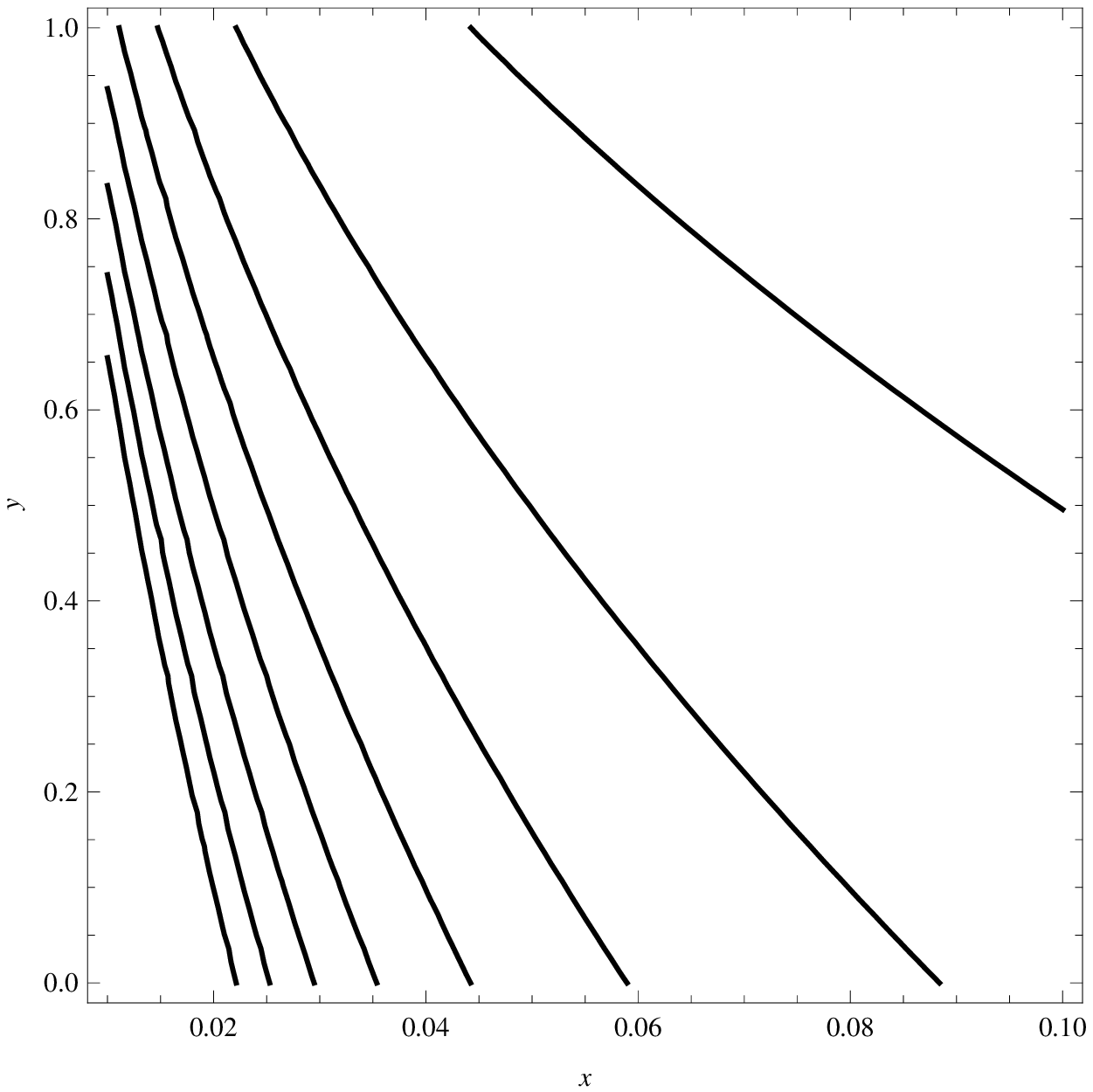}
\label{} } \subfigure[D4D6$\mathrm{\overline{D6}}$ $\&$ $\ell=3$]{
\includegraphics[scale=0.3]{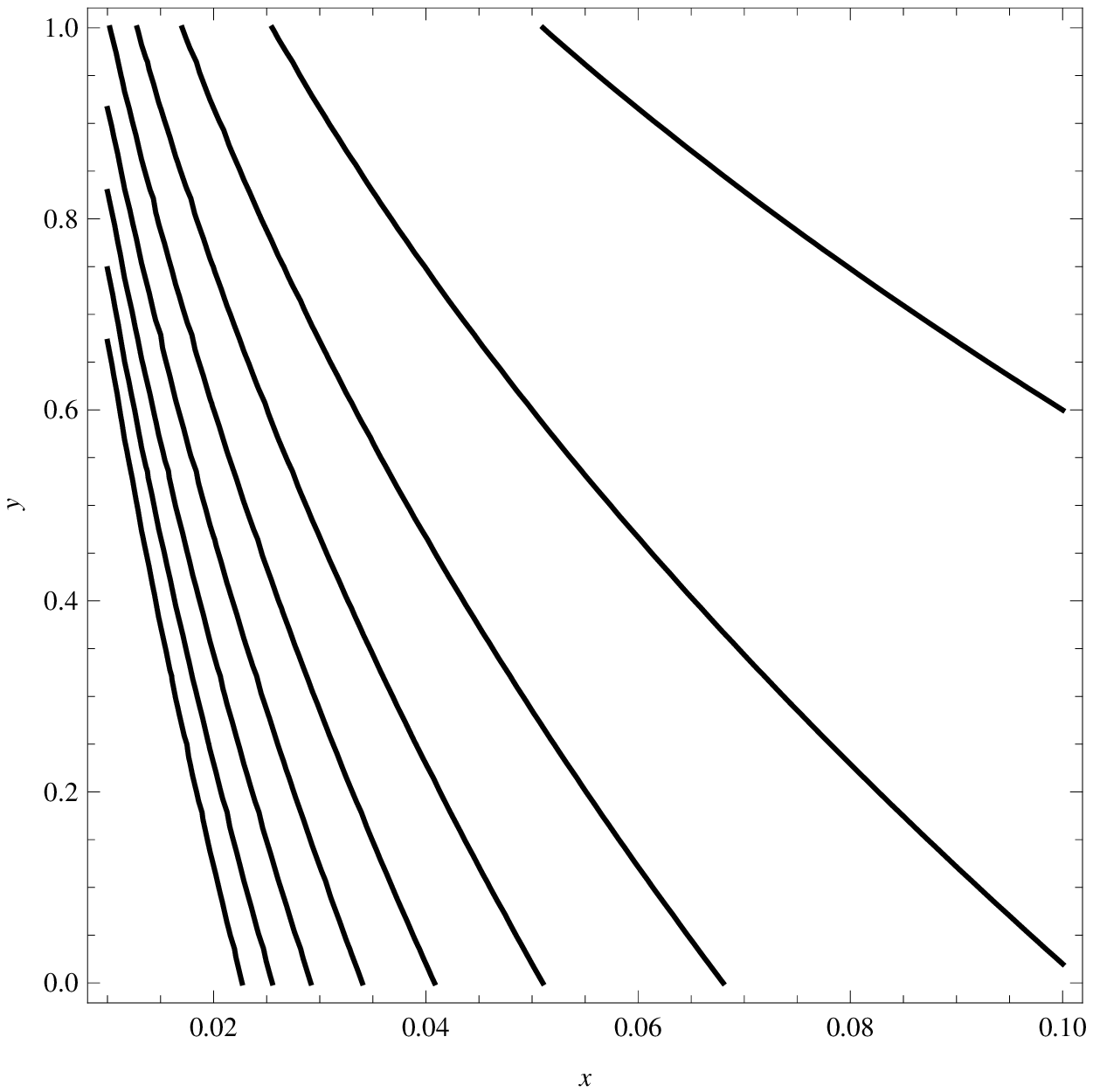}
\label{} } \subfigure[D4D8$\mathrm{\overline{D8}}$ $\&$ $\ell=3$]{
\includegraphics[scale=0.3]{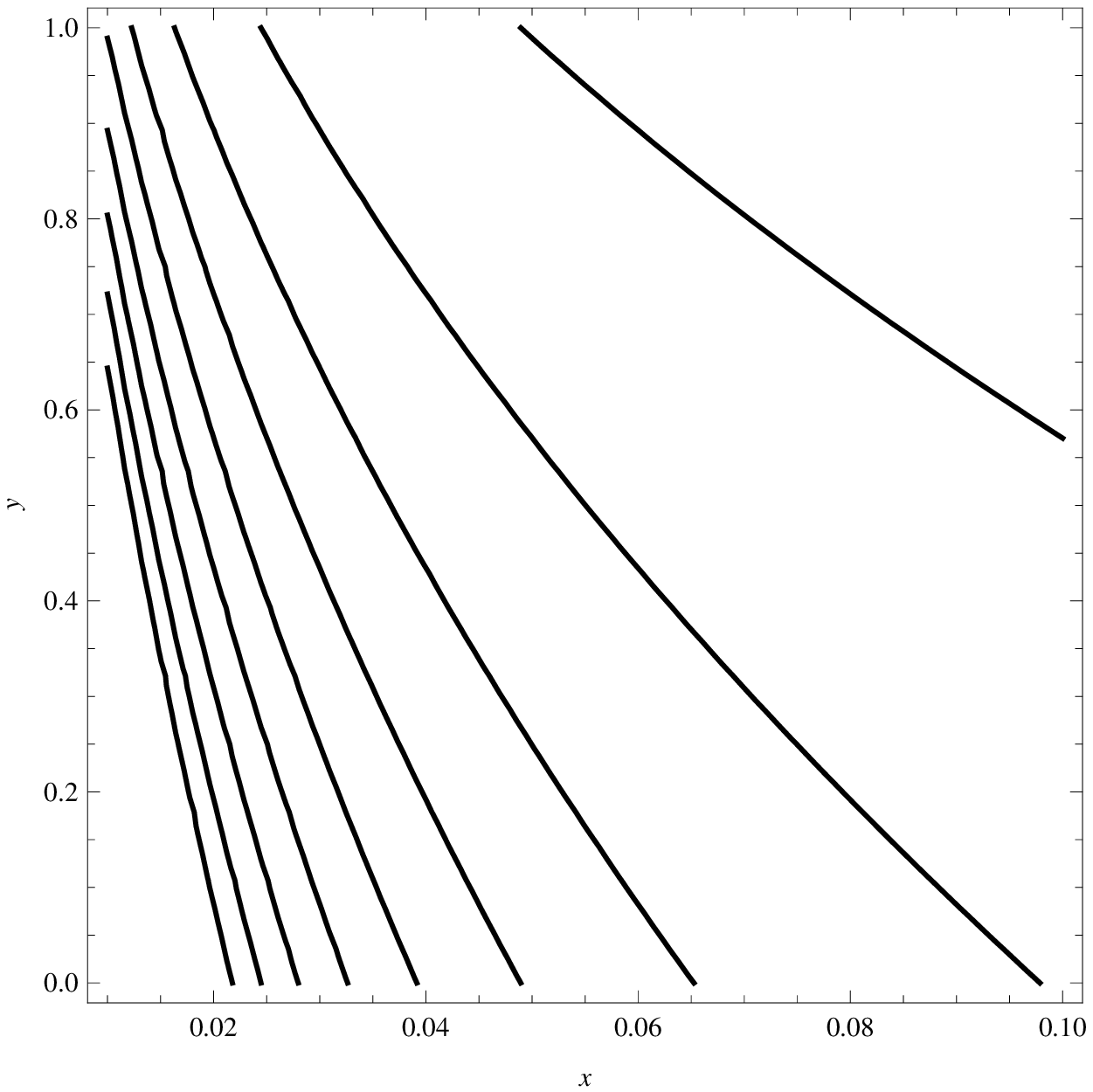}
\label{} } \subfigure[D3D7 $\&$ $\ell=10$]{
\includegraphics[scale=0.3]{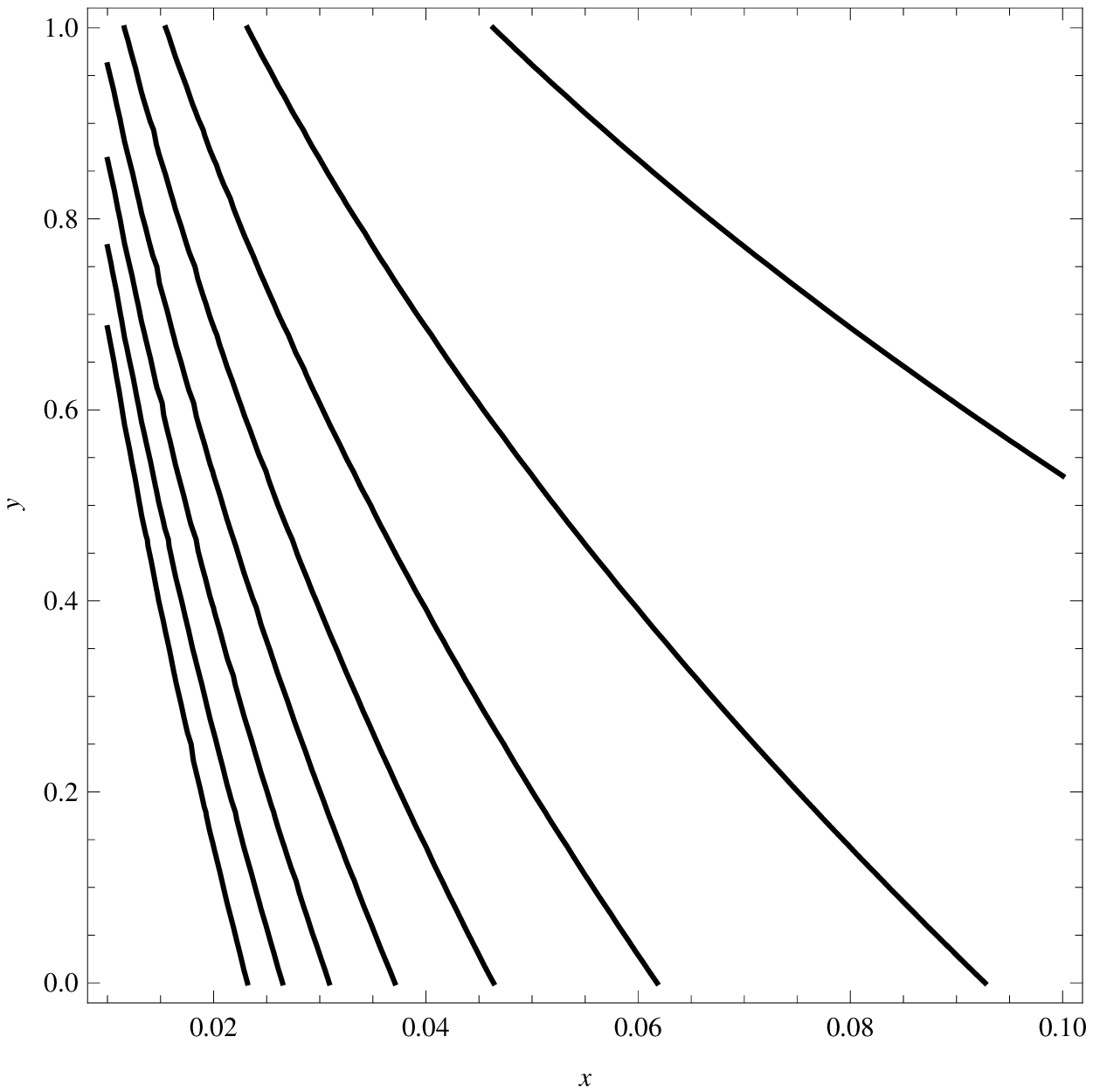}
\label{} } \subfigure[D4D6$\mathrm{\overline{D6}}$ $\&$ $\ell=10$]{
\includegraphics[scale=0.3]{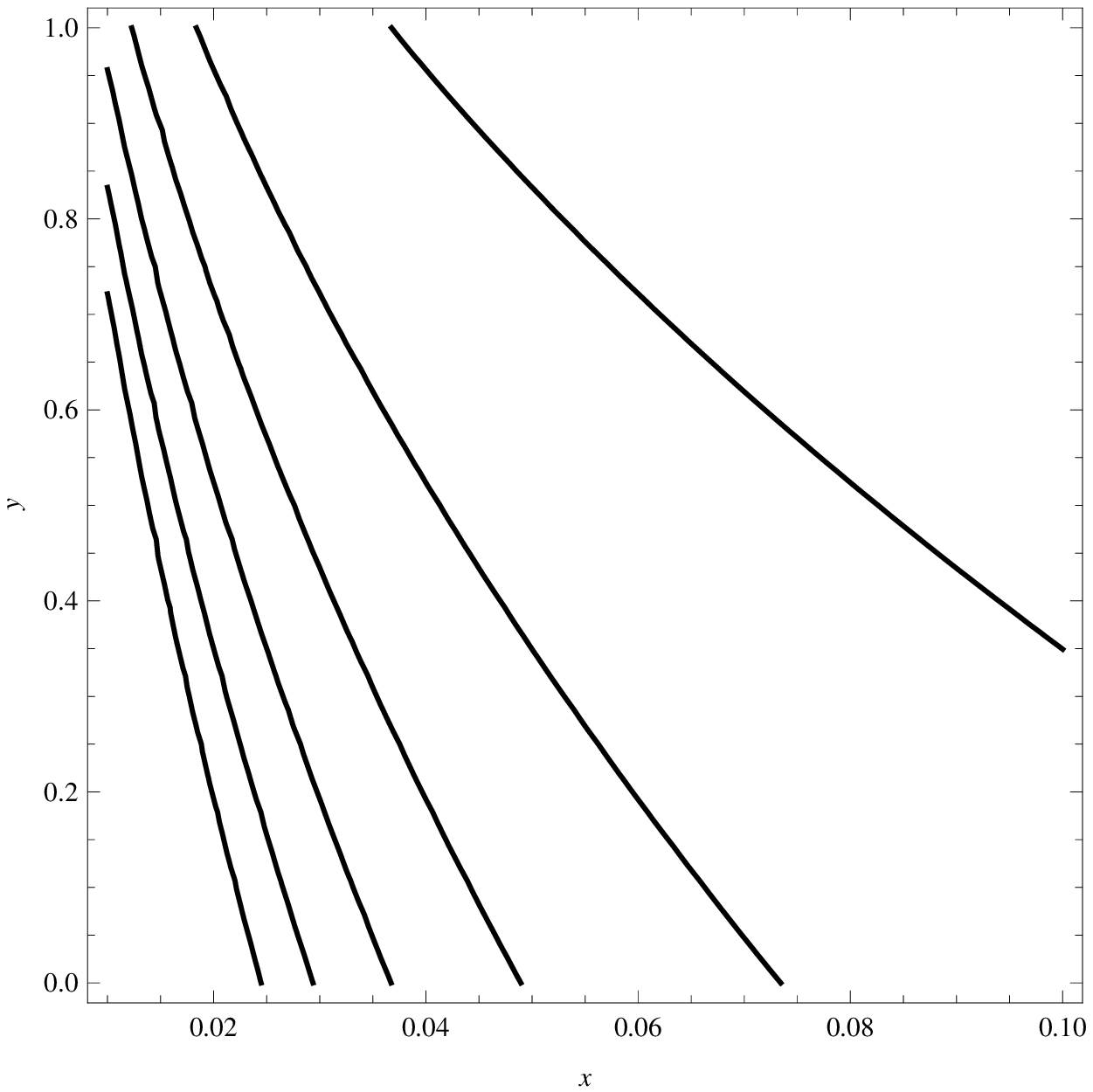}
\label{} } \subfigure[D4D8$\mathrm{\overline{D8}}$ $\&$ $\ell=10$]{
\includegraphics[scale=0.3]{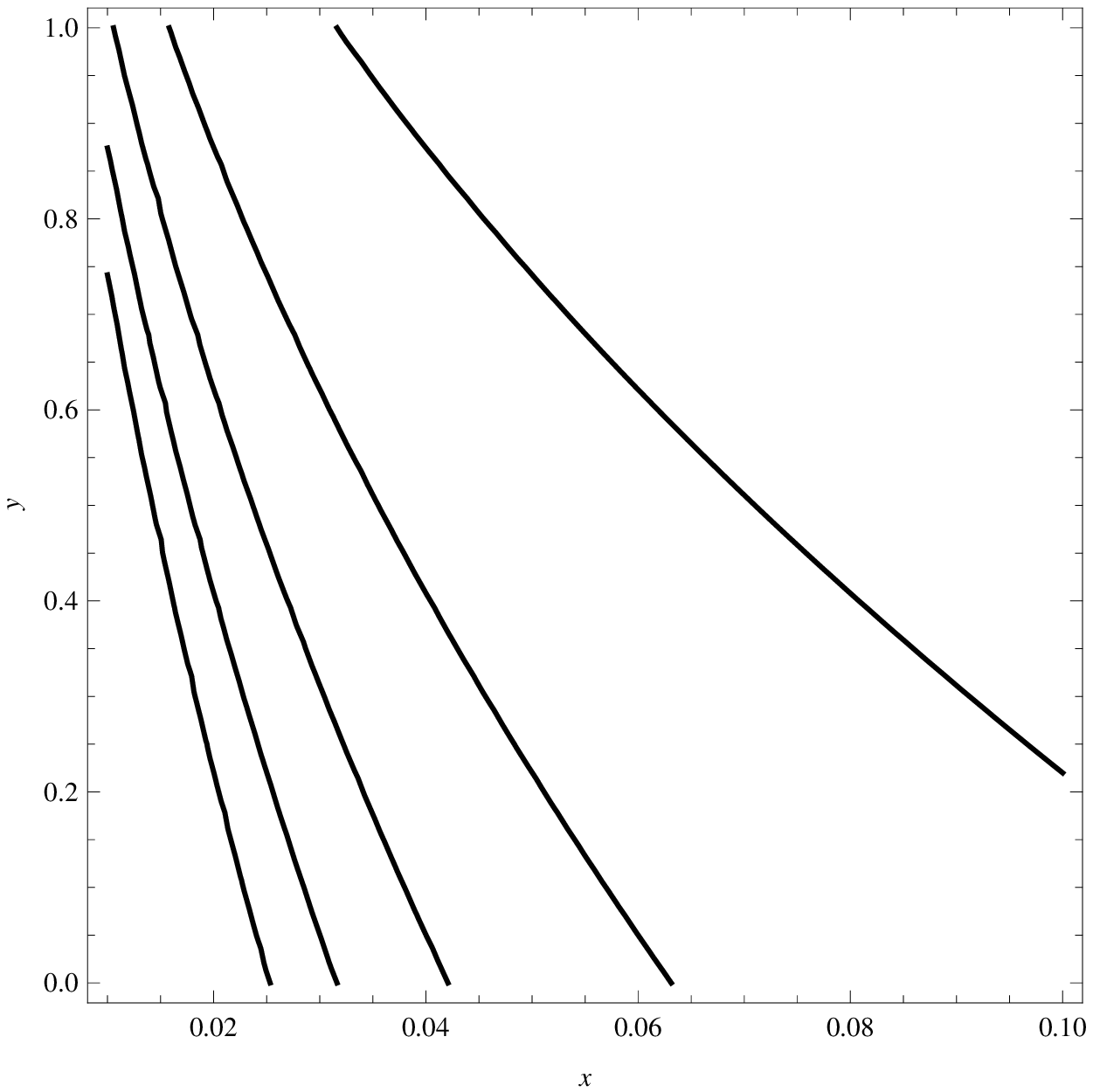}
\label{} } \caption{\small Contour line maps of DIS differential
cross section for unpolarized holographic vector mesons for the
three flavor-brane models we consider, in the region
$\exp{(-\sqrt{\lambda})}\ll x \ll 1/\sqrt{\lambda}$. Different
mesons are labeled with $\ell=1, 2, 3$ and $10$ (displayed on the
four rows). The differential cross sections have been normalized as
explained in the main text. The horizontal axis represents the
Bjorken parameter within the range $[0.01, \, 0.1]$, while the
vertical one corresponds to the variable $y$, $[0, \, 1]$. Curves
with larger slope correspond to larger values of the DIS cross
section.} \label{Crosssection2}
\end{figure}

Now, we present our results for the DIS differential cross section
of unpolarized vector mesons depicted as contour line maps in terms
of $x$ and $y$ variables. We show that for large $x$ in figure
\ref{Crosssection1} and for small $x$ in figure \ref{Crosssection2},
and for the three different Dp-brane models we have considered.
Broader curves in figure \ref{Crosssection1} correspond to lower
values of the DIS differential cross section. On the other hand, in
figure \ref{Crosssection2} curves with larger slope correspond to
larger values of the DIS differential cross section. In figure
\ref{Crosssection1} we can observe that the differential cross
section of unpolarized holographic vector mesons is larger around $x
\approx 0.5$, and also it becomes less spread for the Sakai-Sugimoto
model. Another feature is that this differential cross section is
larger for smaller values of the fractional energy loss of the
charged lepton $y$, having a maximum when the energy of the incident
lepton and the one of the scattered lepton are the same. On the
other hand, curves in figure \ref{Crosssection2} show contour line
maps of DIS differential cross section for unpolarized holographic
vector mesons for the three flavor-brane models we describe, in the
region $\exp{(-\sqrt{\lambda})}\ll x \ll 1/\sqrt{\lambda}$.  The
horizontal axis represents the Bjorken parameter within the range
$[0.01, \, 0.1]$, while the vertical one corresponds to the variable
$y$, $[0, \, 1]$.

Next, let us focus on the comparison with phenomenology of the meson
structure functions. This turns out to be difficult because the
experimental data of these functions is actually much more limited
than in the case of baryons. Moreover, it is more difficult to carry
out such a comparison when the Bjorken parameter $x$ is small. In
the case of pions there have been important phenomenological studies
\cite{Wijesooriya:2005ir,Holt:2010vj, Reimer:2011,Chang:2014gga},
however all these investigations focus on the valence structure
function related to fixed-target pion Drell-Yan experiments. The
corresponding data are restricted to the parametric region $x \geq
0.2$. Notice that in our calculations the pion corresponds to the
$\ell=1$ meson in the D3D7-model, since it is the lightest
pseudoscalar particle. These phenomenological studies could in
principle be compared with the results obtained in our previous
papers studying the parametric range $1/\sqrt{\lambda} \ll x < 1$
\cite{Koile:2011aa,Koile:2013hba}. The small-$x$ behavior we find is
difficult to compare with the existing phenomenological models.
However, in \cite{Altarelli:1995mu} it was proposed a model in which
the parton distribution functions have a sea term that diverges as
$x \rightarrow 0$, as it occurs for our structure functions. In
addition, this type of divergences beyond the valence term was shown
to appear on the pion contribution to the deuteron structure
function $b_1$ (see for instance figure 3 in reference
\cite{Miller:2013hla}).

In general terms, our approach shares fundamental concepts with
references \cite{Brower:2006ea, Brower:2007qh, Brower:2010wf,
Brower:2012sg, Brower:2014sxa} as well as it involves somehow
similar techniques to the ones used by them. Very interestingly, in
those references there is agreement with baryon DIS data from H1 and
ZEUS experiments at HERA. In fact in these references the objects of
study are the glueball structure functions. Since the structure
functions of scalar mesons behave similarly as those of glueballs
for all values of the Bjorken parameter, we regard it as an
indication suggesting that our predictions should be correct.
Keeping in mind that the results for glueballs can be compared with
experimental information from HERA, we would expect that our results
for mesons (which at least for scalar ones are closely related to
the glueball results of \cite{Polchinski:2002jw}) could be
confirmed.

\subsection{Comparison with lattice QCD}

Now, let us focus on the comparison with lattice QCD results.
Particularly, we can compare the lower moments of DIS structure
functions of the pion and the $\rho$ meson obtained from lattice QCD
\cite{Best:1997qp} with our own predictions. The $n$-th moment of a
generic function $h(x, q^2)$ is given by
\be
M_n [h] \equiv \int_0^1 dx \, x^{n-1} h(x, q^2) \, .
\ee
We consider the spin-dependent structure functions of the pion and
$\rho$ mesons. In particular, the lattice QCD calculations in
\cite{Best:1997qp} have considered Wilson fermions and they have
been carried out for three values of the quark mass. Therefore, it
allows one to carry out an extrapolation to the chiral limit. The
way we can compare our results with the lattice QCD ones for the
pion is by using our expressions of the $F_2$ structure
function\footnote{For the pion we use $F_2$ and not $F_1$ because
one of the features of the large $N$ limit is that the virtual
photon strikes the entire hadron, thus $F_1=0$ in the large-$x$
region.} for small and large $x$ values, and integrate them with
appropriate lower and upper cut-offs. The former is integrated from
$x=0.0001$ to $x=0.1$, while the latter is integrated from $x=0.1$
to $x=1$. Notice that within each structure function there is an
undetermined constant containing $\Lambda$, $R$, $q^2$ and some
other numbers. In fact, there is a different constant for each
parametric regime of the Bjorken variable. Thus, for each model
there are two undetermined constants. They are effectively two free
parameters for each flavor-brane model, which are determined by
carrying out the best fitting to the lattice QCD data for the
moments $M_i(F_2)$ with $i=1, 2, 3$. The rest of the fittings are
performed in the same way.

Our results and the lattice QCD data as well as the deviations from
each other are listed in table 3. The constants of the small-$x$
regime which leads to the best fitting to the lattice QCD data of
reference \cite{Best:1997qp} are $0.025$, $0.020$ and $0.022$, while
for the large-$x$ regime we found $24$, $2020$ and $267$, for the
D3D7-, the D4D8$\mathrm{\overline{D8}}$- and the
D4D6$\mathrm{\overline{D6}}$-brane models, respectively. We find
discrepancies with respect to the lattice QCD calculations up to $18
\%$ for our results in the case of the pion, $\ell=1$. Recall that
in the $1/N$ expansion discrepancies are of order $30 \%$, thus our
results are within the expectations. We should keep in mind that
when analyzing discrepancies between gauge/string duality
predictions, phenomenology and experimental data, for instance
bottom-up models in five dimensions (the AdS/QCD model) give results
related to masses and decay constants of mesons with discrepancies
of order $5 \%$ \cite{Erlich:2005qh, Da Rold:2005zs}. On the other
hand, for the case of more involved calculations leading to the
$\Delta I=1/2$ rule describing the kaon decays, related to the
calculation of four-point correlation functions, discrepancies
become of order $30 \%$ \cite{Hambye:2006av,Hambye:2005up}.
\begin{table}
\def\arraystretch{1.5}
\begin{center}
\begin{tabular}{|c|c|c|c|}
\hline
Model / Moment & $M_1(F_2)$ & $M_2(F_2)$ & $M_3(F_2)$ \\
\hline
Lattice QCD & 0.3047 & 0.1180 & 0.0583 \\
\hline
D3D7 & 0.3067 & 0.0962 & 0.0658 \\
\hline
Percentage error & -0.6 & 18.5 & -12.8 \\
\hline
D4D8$\mathrm{\overline{D8}}$ & 0.3061 & 0.1018 & 0.0643 \\
\hline
Percentage error & -0.5 & 13.7 & -10.3 \\
\hline
D4D6$\mathrm{\overline{D6}}$ & 0.3064 & 0.0990 & 0.0650 \\
\hline
Percentage error & -0.6 & 16.1 & -11.6 \\
\hline
\end{tabular}
\caption{\small Comparison of our results for the first moments of
the structure function $F_2$ of the pion for a suitable choice of
the normalization constants with respect to the average results of
the lattice QCD computations in \cite{Best:1997qp}. Uncertainties in
the lattice computations are omitted.}\label{momentstablepi}
\end{center}
\end{table}

There are more recent results in the lattice QCD literature to
compare with. For instance we can consider data from references
\cite{Brommel:2006zz, Chang:2014lva}. The results for these fittings
are shown in table 4. One can see that all the results are improved
being the deviation smaller than $10\%$.  In this case, the
constants we find are $0.014$, $0.011$ and $0.012$ for the small-$x$
regime, and $29$, $2450$ and $325$ for the large-$x$ regime, for the
D3D7-, the D4D8$\mathrm{\overline{D8}}$- and the
D4D6$\mathrm{\overline{D6}}$-brane models, respectively.

It is possible to understand a bit better the comparison between
lattice QCD data and our holographic results by isolating the
contributions from the small-$x$ and large-$x$ structure functions
to each moment. We are tempted to associate the small-$x$ region
with the concept of {\it sea}-quark distribution function, which
describes the possibility of finding a quark or an anti-quark
(generated by gluon splitting) carrying a very small fraction of the
hadron momentum. In this picture the large-$x$ structure function is
then related to the valence distribution function associated with
quarks that carry a considerable fraction of the hadron
momentum\footnote{These are not formal definitions of these
concepts.}. Table I in \cite{Detmold:2003tm} shows that both for
lattice and for phenomenological results one finds that the
contribution of the sea of quarks is important only for the first
moment of the structure functions, since it yields a considerable
fraction of the final result. In the second and third moment its
contribution is substantially reduced, and the contribution from the
valence distribution function gives almost the full result. This is
similar to what happens with our structure functions: the integral
for the small-$x$ region is important in order to fit the first
moment $M_1$, but one can almost ignore it for $M_2$ and $M_3$.
\begin{table}
\def\arraystretch{1.5}
\begin{center}
\begin{tabular}{|c|c|c|c|}
\hline
Model / Moment & $M_1(F_2)$ & $M_2(F_2)$ & $M_3(F_2)$ \\
\hline
Lattice QCD & 0.27 & 0.13 & 0.074 \\
\hline
D3D7 & 0.2708 & 0.1161 & 0.0803 \\
\hline
Percentage error & -0.3 & 10.7 & -8.5 \\
\hline
D4D8$\mathrm{\overline{D8}}$ & 0.2705 & 0.1221 & 0.0779 \\
\hline
Percentage error & -0.2 & 6.1 & -5.2 \\
\hline
D4D6$\mathrm{\overline{D6}}$ & 0.2706 & 0.1191 & 0.0791 \\
\hline
Percentage error & -0.2 & 8.4 & -6.9 \\
\hline
\end{tabular}
\caption{\small Comparison of our results for the first moments of
the structure function $F_2$ for the pion for a suitable choice of
the normalization constants with respect to the average results of
the lattice QCD computations in \cite{Brommel:2006zz,
Chang:2014lva}. Uncertainties in the lattice computations are
omitted.}\label{momentstablepi}
\end{center}
\end{table}

\begin{figure}
\centering
\includegraphics[scale=1.4]{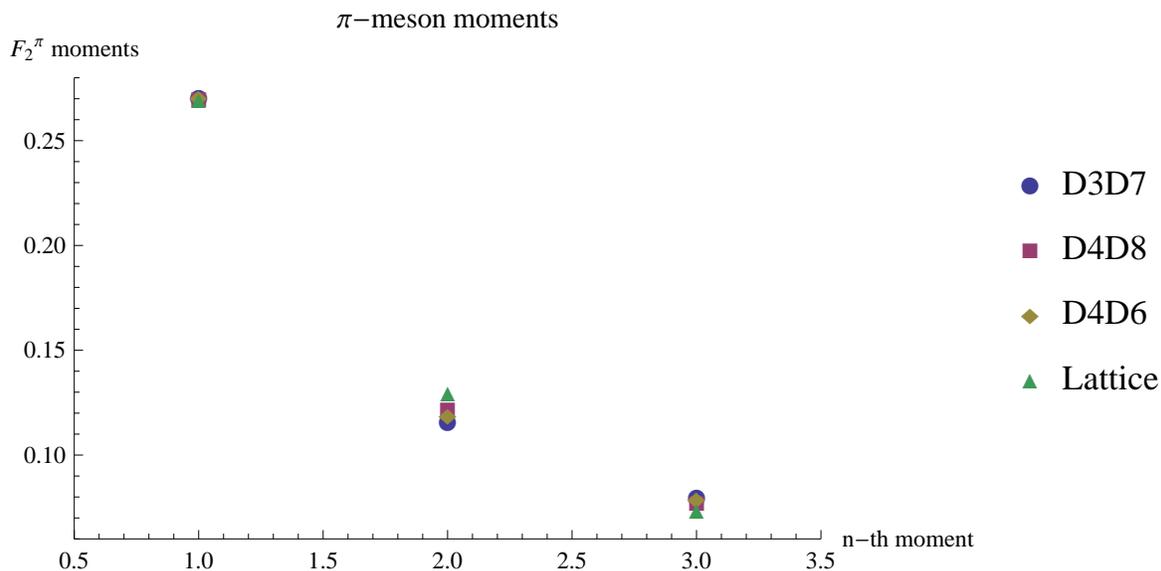}
\caption{\small The first three moments of $F_2$ are shown for the
$\pi$ meson. The free parameters for each dual holographic model are
chosen in order to fit the results in \cite{Best:1997qp}, obtained
with lattice QCD as explained before. The three holographic dual
models are shown, along with the average value obtained in the
mentioned reference. Discrepancies with respect to the reference are
under $20 \%$.} \label{pi}
\end{figure}

\begin{figure}
\centering
\includegraphics[scale=1.4]{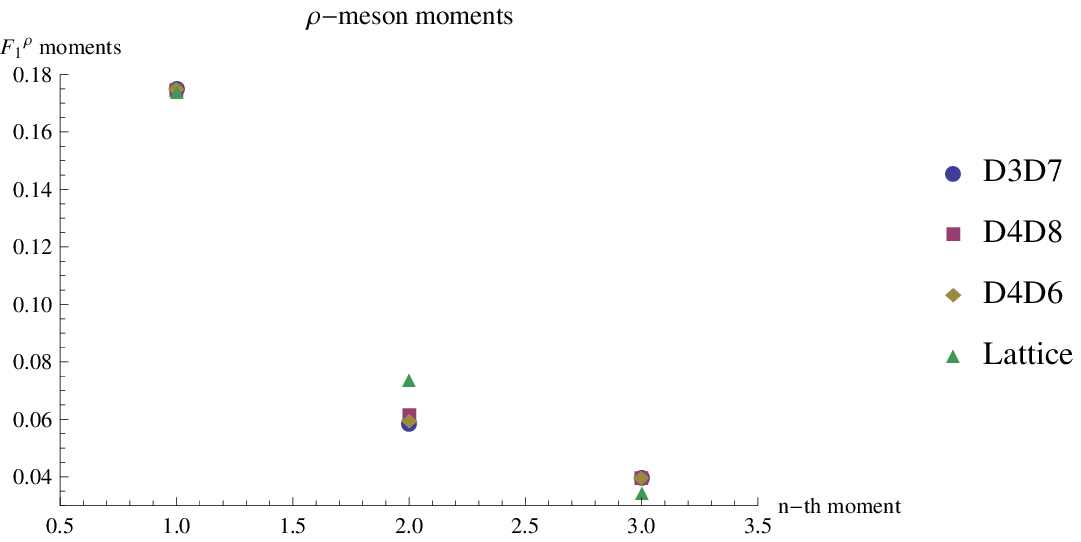}
\caption{\small  The first three moments of $F_1$ are shown for the
$\rho$ meson. The free parameters for each holographic dual model
are chosen in order to fit the results in \cite{Best:1997qp},
obtained with lattice QCD. The three holographic dual models are
shown, along with the average value obtained in this reference.
Discrepancies with respect to the reference are under $21 \%$ in all
cases.} \label{rho}
\end{figure}

Table 5 shows the comparison of our results for the first moments of
the structure function $F_1$ for the $\rho$ meson for a suitable
choice of the normalization constants with respect to the average
results of the lattice QCD computations in \cite{Best:1997qp}. The
constants in this case are $0.013$, $0.011$ and $0.012$ for the
small-$x$ regime, and $14$, $1229$ and $162$ for the large-$x$
regime, for the D3D7-, the D4D8$\mathrm{\overline{D8}}$- and the
D4D6$\mathrm{\overline{D6}}$-brane models, respectively. As it was
commented for table 3, it is very interesting that the relative
discrepancies are smaller than 21 percent.
\begin{table}
\def\arraystretch{1.5}
\begin{center}
\begin{tabular}{|c|c|c|c|}
\hline
Model / Moment & $M_2(F_1)$ & $M_3(F_1)$ & $M_4(F_1)$ \\
\hline
Lattice QCD & 0.1743 & 0.074 & 0.035 \\
\hline
D3D7 & 0.1755 & 0.059 & 0.040 \\
\hline
Percentage error & -0.7 & 21.3 & -14.1 \\
\hline
D4D8$\mathrm{\overline{D8}}$ & 0.1752 & 0.062 & 0.040 \\
\hline
Percentage error & -0.5 & 16.4 & -11.8 \\
\hline
D4D6$\mathrm{\overline{D6}}$ & 0.1754 & 0.060 & 0.040 \\
\hline
Percentage error & -0.6 & 18.8 & -13.0 \\
\hline
\end{tabular}
\caption{\small Comparison of our results for the first moments of
the structure function $F_1$ for the $\rho$ meson for a suitable
choice of the normalization constants with respect to the average
results of the lattice QCD computations in \cite{Best:1997qp}.
Uncertainties in the lattice computations are
omitted.}\label{momentstablerho}
\end{center}
\end{table}
Now, by using tables 4 and 5 we can draw figures \ref{pi} and
\ref{rho}.
In figure \ref{pi} the first three moments for $F_2$ are displayed
for the pion. The free parameters for each dual holographic model
are chosen in order to fit the results in \cite{Best:1997qp}
obtained with lattice QCD as explained before. The three studied
holographic dual models are shown, together with the value obtained
in the reference. We find good agreement with lattice QCD data with
discrepancies lower than $20 \%$. By using more recent lattice QCD
results of references \cite{Brommel:2006zz, Chang:2014lva} the
fittings improve with discrepancies lower than $10 \%$.
In addition, in figure \ref{rho} the first three moments of $F_1$
are shown for the $\rho$ meson. The three holographic dual models
are presented in a similar way as in figure \ref{pi}. Discrepancies
with respect to that reference are under $21 \%$ in all cases.
Therefore, this also shows good agreement with lattice QCD data.

Notice that in all our fittings the constant of the structure
function corresponding to the small-$x$ regime is always a few
orders of magnitude smaller than the one associated with the
large-$x $ structure function. This is in full agreement with our
calculations, at least in the regime where the holographic dual
model represents an accurate description of the dual gauge theory.
This could have been directly predicted from our analytical results.
By examining the difference between the constants in front of
(\ref{PiLargex}) and (\ref{PiSmallx1}, \ref{PiSmallx2}) one can see
that for the $\ell=1$ scalar meson the numerical coefficients and
normalization constants are of the same order, and the main
difference is given by the $\lambda^{-1/2}$ factor that appears in
the small-$x$ case. This is small at large 't Hooft coupling, and it
is directly related to the value of $x=0.1$ which we use in order to
separate the first two regimes (A and B) in our calculations.

Another important point comes also from the comparison between our
analytical results and the different models formulated in the
literature in order to fit the phenomenological data for the
measurement of the pion valence distribution functions. As it is
explained in \cite{Aicher:2011ai,Detmold:2003tm} there have been
different attempts to describe the available data, and while there
are some differences between them, they all agree on the fact that
the distribution functions should behave as $(1-x)^2$ in the
$x\rightarrow 1$ limit. In \cite{Aicher:2010cb} it has been found a
fall-off $(\sim (1-x)^{2\pm 0.1})$ for the valence distribution
function consistent with Drell-Yan data. This is consistent with
theoretical predictions based on perturbative QCD (see for instance
\cite{Yuan:2003fs} and references therein) as well as with
calculations using Dyson-Schwinger equations \cite{Hecht:2000xa}.
This is exactly what happens with our structure functions
(\ref{PiLargex}) in the D3D7-brane model for the $\ell=1$ scalar
meson, {\it i.e.} the pion. This would suggest that the D3D7-brane
model gives better predictions than the other two ones studied in
this work. In the D4D8$\mathrm{\overline{D8}}$-brane and
D4D6$\mathrm{\overline{D6}}$-brane models the exponents are $4.59$
and $3.33$, respectively.

A very interesting review of lattice QCD calculations on moments of
hadron distribution functions, as well as other important hadron
structure observables is given in \cite{Hagler:2009ni}. The earliest
calculations of the lowest two moments of the pion quark
distribution function have been done in
\cite{Martinelli:1987zd,Martinelli:1987si,Martinelli:1987bh} in
quenched lattice QCD. An extensive study of several moments of the
pion and the $\rho$ meson have been done in \cite{Best:1997qp}. In
fact, we have compared our results with those in that reference in
our tables 3 and 5 and in our figures 8 and 9, while in our table 4
we compare with references \cite{Brommel:2006zz, Chang:2014lva}. The
earliest results of \cite{Best:1997qp} have been obtained by using
quenched lattice QCD simulations, which means that the fermion
determinant is set to one, thus neglecting all quark-antiquark
quantum fluctuations. This introduces an error which can only be
quantified by comparison with dynamical simulations with physical
quark masses, which are only computationally possible since a few
years\footnote{We thank Andreas Sch\"afer for this comment.}. In
fact for instance in
\cite{Brommel:2006zz,Aicher:2011ai,Chang:2014lva} dynamical quarks
have been considered.

We may also try to use this kind of fittings to compare the first
few moments corresponding to our $b_i$ functions with the ones
calculated in \cite{Best:1997qp}. However, in that paper the results
found for these moments are smaller than the $F_i$
moments\footnote{In fact, even if we actually tried to do this the
very small values of these moments and the considerable errors
associated with them in \cite{Best:1997qp} would probably make the
fittings unreliable.}. Since the structure functions we obtained
from string theory satisfy the relations $b_i = 3F_i$, the numerical
comparison becomes meaningless. It seems that there is a conceptual
difference, and it would be very interesting to study where it comes
from or to be able to resolve this discrepancy from new experimental
data.

Once again we would like to emphasize that the discussion presented
in this last section has to be understood qualitatively. One should
keep in mind that the gauge theories that we are able to study in
terms of string theory dual models do not look exactly like QCD.
What makes them interesting though is that they have some important
similarities related to the meson spectrum and the concepts of color
confinement, chiral symmetry breaking and other properties. Thus, it
is really interesting that these holographic dual models give
predictions of the moments of the structure functions which are
comparable to lattice QCD results.

The good agreement we have obtained with lattice QCD data is somehow
not unexpected taking into account several examples where the
supergravity dual models give results which compare reasonably well
with lattice QCD data. For instance, in the case of SYM plasmas good
level of agreement has been found between lattice QCD results and
the gauge/gravity duality approach by including string theory
corrections to the supergravity calculations as it has been shown
for the mass and electric charge transport coefficients. In the case
of the mass transport the quantitative agreement between lattice QFT
calculations for $\eta/s$ \cite{Meyer:2007ic} and the gauge/string
duality \cite{Kovtun:2004de,Myers:2008yi} is remarkable ($\eta$ is
the ${\cal {N}}=4$ SYM plasma shear viscosity and $s$ the entropy
density). Also for electric charge transport the DC conductivity has
been calculated using lattice QCD \cite{Aarts:2007wj} and the
gauge/string duality \cite{CaronHuot:2006te}, including ${\cal
 {O}}(\alpha'^3)$ corrections, leading to good level of agreement
\cite{Hassanain:2011ce,Hassanain:2011fn,Hassanain:2010fv,Hassanain:2012uj}.
Of course, also in those cases the gauge/string dual calculations
strictly hold for large $N$ and the strong coupling expansion
($1/\lambda$), but if one insists on using $N=3$ and $\lambda
\approx 15$ it leads to the results we mention in this paragraph.

Also, the fact that the methods developed for the holographic dual
description of DIS can be extended to other strongly coupled $SU(N)$
gauge theories is very interesting.

~

~

\centerline{\large{\bf Acknowledgments}}

~

We would like to thank Sergio Iguri, Jos\'e Goity, Carlos N\'u\~nez
and Andreas Sch\"afer for very valuable comments on the manuscript.
The work of E.K., N.K. and M.S. is supported by the CONICET. This
work has been partially supported by the CONICET-PIP 0595/13 the
grant and UNLP grant 11/X648.


\newpage

\begin{thebibliography}{99}



\bibitem{Koile:2011aa}
  E.~Koile, S.~Macaluso and M.~Schvellinger,
  ``Deep Inelastic Scattering from Holographic Spin-One Hadrons,''
  JHEP {\bf 1202} (2012) 103
  [arXiv:1112.1459 [hep-th]].

\bibitem{Koile:2013hba}
  E.~Koile, S.~Macaluso and M.~Schvellinger,
  ``Deep inelastic scattering structure functions of holographic spin-1 hadrons with $N_f \geq 1$,''
  JHEP {\bf 1401} (2014) 166
  [arXiv:1311.2601 [hep-th]].

\bibitem{Koile:2014vca}
  E.~Koile, N.~Kovensky and M.~Schvellinger,
  ``Hadron structure functions at small $x$ from string theory,''
  JHEP {\bf 1505} (2015) 001
  [arXiv:1412.6509 [hep-th]].

\bibitem{Kruczenski:2003be}
  M.~Kruczenski, D.~Mateos, R.~C.~Myers, D.~J.~Winters,
  ``Meson spectroscopy in AdS / CFT with flavor,''
  JHEP {\bf 0307 } (2003)  049.
  [arXiv:hep-th/0304032 [hep-th]].

\bibitem{Kruczenski:2003uq}
  M.~Kruczenski, D.~Mateos, R.~C.~Myers and D.~J.~Winters,
  ``Towards a holographic dual of large N(c) QCD,''
  JHEP {\bf 0405} (2004) 041
  [hep-th/0311270].

\bibitem{Sakai:2004cn}
  T.~Sakai, S.~Sugimoto,
  ``Low energy hadron physics in holographic QCD,''
  Prog.\ Theor.\ Phys.\  {\bf 113 } (2005)  843-882.
  [arXiv:hep-th/0412141 [hep-th]].

\bibitem{Polchinski:2002jw}
  J.~Polchinski, M.~J.~Strassler,
  ``Deep inelastic scattering and gauge / string duality,''
  JHEP {\bf 0305 } (2003)  012.
  [hep-th/0209211].

\bibitem{Polchinski:2000uf}
  J.~Polchinski and M.~J.~Strassler,
  ``The String dual of a confining four-dimensional gauge theory,''
  hep-th/0003136.

\bibitem{Brower:2006ea}
  R.~C.~Brower, J.~Polchinski, M.~J.~Strassler and C.~I.~Tan,
  ``The Pomeron and gauge/string duality,''
  JHEP {\bf 0712}, 005 (2007)
  [hep-th/0603115].

\bibitem{Manohar:1992tz}
A.~V.~Manohar, ``An Introduction to spin dependent deep inelastic
scattering,'' In *Lake Louise 1992, Symmetry and spin in the
standard model* 1-46 [hep-ph/9204208].

\bibitem{Hoodbhoy:1988am}
  P.~Hoodbhoy, R.~L.~Jaffe and A.~Manohar,
  ``Novel Effects in Deep Inelastic Scattering from Spin 1 Hadrons,''
  Nucl.\ Phys.\ B {\bf 312}, 571 (1989).

\bibitem{BallonBayona:2010ae}
  C.~A.~Ballon Bayona, H.~Boschi-Filho, N.~R.~F.~Braga and M.~A.~C.~Torres,
  ``Deep inelastic scattering for vector mesons in holographic D4-D8 model,''
  JHEP {\bf 1010} (2010) 055
  [arXiv:1007.2448 [hep-th]].

\bibitem{Wijesooriya:2005ir}
  K.~Wijesooriya, P.~E.~Reimer and R.~J.~Holt,
  ``The pion parton distribution function in the valence region,''
  Phys.\ Rev.\ C {\bf 72}, 065203 (2005)
  [nucl-ex/0509012].

\bibitem{Holt:2010vj}
  R.~J.~Holt and C.~D.~Roberts,
  ``Distribution Functions of the Nucleon and Pion in the Valence Region,''
  Rev.\ Mod.\ Phys.\  {\bf 82}, 2991 (2010)
  [arXiv:1002.4666 [nucl-th]].

\bibitem{Reimer:2011}
P.~Reimer, R.~Holt and K.~Wijesooriya,
 ``The Partonic Structure of the Pion at Large-x,''
AIP Conference Proceedings 1369, 153 (2011); doi: 10.1063/1.3631531

\bibitem{Chang:2014gga}
  L.~Chang and A.~W.~Thomas,
  ``Pion Valence-quark Parton Distribution Function,''
  arXiv:1410.8250 [nucl-th].

\bibitem{Altarelli:1995mu}
  G.~Altarelli, S.~Petrarca and F.~Rapuano,
  Phys.\ Lett.\ B {\bf 373}, 200 (1996)
  [hep-ph/9510346].

\bibitem{Miller:2013hla}
  G.~A.~Miller,
  ``Pionic and Hidden-Color, Six-Quark Contributions to the Deuteron b1 Structure Function,''
  Phys.\ Rev.\ C {\bf 89}, no. 4, 045203 (2014)
  [arXiv:1311.4561 [nucl-th]].

\bibitem{Brower:2007qh}
  R.~C.~Brower, M.~J.~Strassler and C.~I.~Tan,
  ``On the eikonal approximation in AdS space,''
  JHEP {\bf 0903}, 050 (2009)
  [arXiv:0707.2408 [hep-th]].

\bibitem{Brower:2010wf}
  R.~C.~Brower, M.~Djuric, I.~Sarcevic and C.~I.~Tan,
  ``String-Gauge Dual Description of Deep Inelastic Scattering at Small-$x$,''
  JHEP {\bf 1011}, 051 (2010)
  [arXiv:1007.2259 [hep-ph]].

\bibitem{Brower:2012sg}
  R.~C.~Brower, M.~Djuric, I.~Sarcevic and C.~I.~Tan,
  ``Small-x Deep Inelastic Scattering via the Pomeron in AdS,''
  arXiv:1204.0472 [hep-ph].

\bibitem{Brower:2014sxa}
  R.~Brower, R.~C.~Brower, M.~Djuriat, T.~Raben and C.~I.~Tan,
  ``Towards holographic QCD: AdS/CFT, confinement deformation, and DIS at small-x,''
  arXiv:1412.3443 [hep-ph].

\bibitem{Best:1997qp}
  C.~Best, M.~Gockeler, R.~Horsley, E.~M.~Ilgenfritz, H.~Perlt, P.~E.~L.~Rakow, A.~Schafer and G.~Schierholz {\it et al.},
  Phys.\ Rev.\ D {\bf 56}, 2743 (1997)
  [hep-lat/9703014].

\bibitem{Erlich:2005qh}
  J.~Erlich, E.~Katz, D.~T.~Son and M.~A.~Stephanov,
  ``QCD and a holographic model of hadrons,''
  Phys.\ Rev.\ Lett.\  {\bf 95} (2005) 261602
  [hep-ph/0501128].

\bibitem{Da Rold:2005zs}
  L.~Da Rold and A.~Pomarol,
  ``Chiral symmetry breaking from five dimensional spaces,''
  Nucl.\ Phys.\ B {\bf 721} (2005) 79
  [hep-ph/0501218].

\bibitem{Hambye:2006av}
  T.~Hambye, B.~Hassanain, J.~March-Russell and M.~Schvellinger,
  ``Four-point functions and Kaon decays in a minimal AdS/QCD model,''
  Phys.\ Rev.\ D {\bf 76} (2007) 125017
  [hep-ph/0612010].

\bibitem{Hambye:2005up}
  T.~Hambye, B.~Hassanain, J.~March-Russell and M.~Schvellinger,
  ``On the Delta I = 1/2 rule in holographic QCD,''
  Phys.\ Rev.\ D {\bf 74} (2006) 026003
  [hep-ph/0512089].

\bibitem{Brommel:2006zz}
  D.~Brommel {\it et al.}  [QCDSF-UKQCD Collaboration],
  ``Quark distributions in the pion,''
  PoS LAT {\bf 2007}, 140 (2007).

\bibitem{Chang:2014lva}
  L.~Chang, C.~Mezrag, H.~Moutarde, C.~D.~Roberts, J.~Rodriguez-Quintero and P.~C.~Tandy,
  ``Basic features of the pion valence-quark distribution function,''
  Phys.\ Lett.\ B {\bf 737}, 23 (2014)
  [arXiv:1406.5450 [nucl-th]].

\bibitem{Detmold:2003tm}
  W.~Detmold, W.~Melnitchouk and A.~W.~Thomas,
  ``Parton distribution functions in the pion from lattice QCD,''
  Phys.\ Rev.\ D {\bf 68}, 034025 (2003)
  [hep-lat/0303015].

\bibitem{Aicher:2011ai}
  M.~Aicher, A.~Schafer and W.~Vogelsang,
  ``Threshold-Resummed Cross Section for the Drell-Yan Process in Pion-Nucleon Collisions at COMPASS,''
  Phys.\ Rev.\ D {\bf 83}, 114023 (2011)
  [arXiv:1104.3512 [hep-ph]].

\bibitem{Aicher:2010cb}
  M.~Aicher, A.~Schafer and W.~Vogelsang,
  ``Soft-gluon resummation and the valence parton distribution function of the pion,''
  Phys.\ Rev.\ Lett.\  {\bf 105}, 252003 (2010)
  [arXiv:1009.2481 [hep-ph]].

\bibitem{Yuan:2003fs}
  F.~Yuan,
  ``Generalized parton distributions at $x \rightarrow 1$,''
  Phys.\ Rev.\ D {\bf 69}, 051501 (2004)
  [hep-ph/0311288].

\bibitem{Hecht:2000xa}
  M.~B.~Hecht, C.~D.~Roberts and S.~M.~Schmidt,
  ``Valence quark distributions in the pion,''
  Phys.\ Rev.\ C {\bf 63}, 025213 (2001)
  [nucl-th/0008049].

\bibitem{Hagler:2009ni}
  P.~Hagler,
  ``Hadron structure from lattice quantum chromodynamics,''
  Phys.\ Rept.\  {\bf 490}, 49 (2010)
  [arXiv:0912.5483 [hep-lat]].

\bibitem{Martinelli:1987zd}
  G.~Martinelli and C.~T.~Sachrajda,
  ``Pion Structure Functions From Lattice {QCD},''
  Phys.\ Lett.\ B {\bf 196} (1987) 184.

\bibitem{Martinelli:1987si}
  G.~Martinelli and C.~T.~Sachrajda,
  ``A Lattice Calculation of the Second Moment of the Pion's Distribution Amplitude,''
  Phys.\ Lett.\ B {\bf 190} (1987) 151.

\bibitem{Martinelli:1987bh}
  G.~Martinelli and C.~T.~Sachrajda,
  ``A Lattice Calculation of the Pion's Form-Factor and Structure Function,''
  Nucl.\ Phys.\ B {\bf 306} (1988) 865.

\bibitem{Meyer:2007ic}
  H.~B.~Meyer,
  ``A Calculation of the shear viscosity in SU(3) gluodynamics,''
  Phys.\ Rev.\ D {\bf 76} (2007) 101701
  [arXiv:0704.1801 [hep-lat]].

\bibitem{Kovtun:2004de}
  P.~Kovtun, D.~T.~Son and A.~O.~Starinets,
  ``Viscosity in strongly interacting quantum field theories from black hole physics,''
  Phys.\ Rev.\ Lett.\  {\bf 94} (2005) 111601
  [hep-th/0405231].

\bibitem{Myers:2008yi}
  R.~C.~Myers, M.~F.~Paulos and A.~Sinha,
  ``Quantum corrections to eta/s,''
  Phys.\ Rev.\ D {\bf 79} (2009) 041901
  [arXiv:0806.2156 [hep-th]].

\bibitem{Aarts:2007wj}
  G.~Aarts, C.~Allton, J.~Foley, S.~Hands and S.~Kim,
  ``Spectral functions at small energies and the electrical conductivity in hot, quenched lattice QCD,''
  Phys.\ Rev.\ Lett.\  {\bf 99} (2007) 022002
  [hep-lat/0703008 [HEP-LAT]].

\bibitem{CaronHuot:2006te}
  S.~Caron-Huot, P.~Kovtun, G.~D.~Moore, A.~Starinets and L.~G.~Yaffe,
  ``Photon and dilepton production in supersymmetric Yang-Mills plasma,''
  JHEP {\bf 0612} (2006) 015
  [arXiv:hep-th/0607237].

\bibitem{Hassanain:2011ce}
  B.~Hassanain and M.~Schvellinger,
  ``Diagnostics of plasma photoemission at strong coupling,''
  Phys.\ Rev.\ D {\bf 85} (2012) 086007
  [arXiv:1110.0526 [hep-th]].

\bibitem{Hassanain:2011fn}
  B.~Hassanain and M.~Schvellinger,
  ``Plasma conductivity at finite coupling,''
  JHEP {\bf 1201} (2012) 114
  [arXiv:1108.6306 [hep-th]].

\bibitem{Hassanain:2010fv}
  B.~Hassanain, M.~Schvellinger,
  ``Towards 't Hooft parameter corrections to charge transport in strongly-coupled plasma,''
  JHEP {\bf 1010 } (2010)  068
  [arXiv:1006.5480 [hep-th]].

\bibitem{Hassanain:2012uj}
  B.~Hassanain and M.~Schvellinger,
  ``Plasma photoemission from string theory,''
  JHEP {\bf 1212} (2012) 095
  [arXiv:1209.0427 [hep-th]].




\end{thebibliography}
\end{document}